\documentclass[11pt,a4paper]{article}
\usepackage{jstyle}
\usepackage[vcentermath]{youngtab}
\usepackage{ytableau}
\usepackage{tikz}
\usetikzlibrary{decorations.pathreplacing, decorations.pathmorphing}
\tikzset{snake it/.style={decorate, decoration=snake}}

\usepackage{hyperref}
\usepackage{caption}
\usepackage[all]{hypcap}
\usepackage{float}
\usepackage{subcaption}

\usepackage[colorinlistoftodos,draft]{todonotes}

\newcommand{\dd}{{\rm d}}
\newcommand{\cd}{{\rm D}}

\author{\quad Euihun Joung}
\author{\quad Min-gi Kim}
\author{\quad Yujin Kim}

\affiliation{Department of Physics, Kyung Hee University \\ Seoul 02447, Korea}

\emailAdd{euihun.joung@khu.ac.kr}
\emailAdd{xp2425@khu.ac.kr}
\emailAdd{4happiness@khu.ac.kr}

\title{\centering Unfolding Conformal Geometry}

\abstract{Conformal geometry is studied using the unfolded formulation \`a la Vasiliev.
Analyzing the first-order consistency of the unfolded equations, 
we identify the content of zero-forms as the spin-two off-shell Fradkin-Tseytlin module of $\mathfrak{so}(2,d)$.
We sketch the nonlinear structure of the equations and explain
how Weyl invariant densities, which Type-B Weyl anomaly consist of,
could be systematically computed within the unfolded formulation. 
The unfolded equation for conformal geometry is also shown to be reduced to various on-shell gravitational systems
by requiring additional algebraic constraints.
}

\begin{document}

\maketitle

\section{Introduction}

Conformal geometry plays an important role in many areas of gravitational and high energy physics
as well as in certain fields of geometry.
It extends the diffeomorphism invariant Riemannian geometry with local rescaling symmtery, namely the Weyl (rescaling) symmetry.
Conformal geometry can be also viewed as the geometry of the asymptotic boundary of the bulk spacetime with negative cosmological constant \cite{Fefferman:2007rka}, and hence it can be used in the AdS/CFT correspondence \cite{Henningson:1998gx}.
Conformal gravity in four dimensions is an alternative gravitational theory enjoying Weyl symmetry besides  diffeomorphism.
Since its introduction by Weyl and Bach, many studies were devoted to it (see e.g. 
\cite{Wheeler:2013ora,Scholz:2017pfo,Hobson:2021vzg} and references therein).
As a four-derivative gravitational theory in four dimensions, it is power-counting renormalizable as opposed to Einstein gravity
and has many interesting features which might be relevant in phenomenological models of gravity:
notably, it has the conformal symmetry which the early universe seems to exhibit,
and hence conformal gravity or its variant may replace Einstein gravity in the very early time of the universe 
(see e.g. recent works \cite{Hobson:2021vzg,Kehagias:2021smx}
and references therein).
More generally, when we consider various modifications of gravity, conformal geometry also plays a distinguished role
since the diffeomorphism plus Weyl rescaling is the maximum gauge symmetry
that a theory of symmetric rank-two tensor field can afford. 

Another prominent use of conformal geometry is in conformal field theories 
where it appears as Weyl anomaly (see e.g. \cite{Duff:1993wm} for an historical overview).
Like the other anomalies, Weyl anomaly is subjected to the Wess-Zumino consistency condition,
and the classification of Weyl anomalies by the relevant cohomological analysis has been innitiated in 
\cite{Bonora:1983ff, Bonora:1985cq} with results up to dimension six.
The structure in general dimensions, postulated already in \cite{Bonora:1985cq},
was confirmed  first by using the technics of dimensional regularization on the effective gravitational action \cite{Deser:1993yx}
and later by a cohomological analysis \cite{Boulanger:2007ab,Boulanger:2007st}.
According to these results, a Weyl anomaly in $d=2n$ dimensions is the spacetime integral of 
a linear combination of the following densities multiplied by the Weyl rescaling parameter $\sigma$.
\begin{itemize}
\item Type-A anomaly, associated with $a$-coefficient: Euler density,
$\cE=\e_{a_1\cdots a_{d}}\,R^{a_1a_2}\wedge \cdots\wedge R^{a_{d-1}a_{d}}$\,,
where $R^{ab}$ is the Riemann curvature two-forms.
\item Type-B anomaly, associated with $c$-coefficient: strictly Weyl invariant density,
which is a specific contraction of (covariant derivatives of) Riemann tensors,
including any full contraction of $n$ Weyl tensors.
\end{itemize}
The explicit form or even the number of the non-trivial Weyl invariant densities ---
by ``non-trivial Weyl invariant densities'', we mean Weyl invariant densities which are not   contractions of Weyl tensors ---
are not known in general dimensions
but up to dimensions eight.
In six dimensions, it was shown in \cite{Bonora:1985cq,Parker:1987,Deser:1993yx} that there exists only one non-trivial Weyl invariant density (see also  \cite{Henningson:1998gx}).
In eight dimensions, the authors of \cite{Boulanger:2004zf} showed,
by employing purely algebraic methods based on a Weyl covariant tensor
calculus \cite{Boulanger:2004eh} and with a help of computer program, 
that there exists five non-trivial Weyl invariant densities.
The covariant derivative and the calculus used in \cite{Boulanger:2004eh} is 
closely related (see \cite{Francois:2015pga} for the relation) to the Thomas D operator and the tractor calculus for conformal geometry
\cite{Curry:2014yoa}. See e.g. \cite{Erdmenger:1997gy} for more background in this topic.
The current work originated from the attempt of 
understanding the above structure from a different  angle.

As Riemaniann geometry can be understood as the gauge theory of an isometry group, such as Poincar\'e or 
(Anti) de Sitter group,
conformal geometry can be viewed as the gauge theory of conformal group $SO(2,d)$.
Such algebraic approaches to geometries have been pioneered by Cartan and Weyl, and known as Cartan geometry
 (see e.g. \cite{sharpe}).
It incorporates not only Riemaniann, but also parabolic geometry of which conformal geometry 
is an example. 
In physics, there have been several attempts
to obtain conformal gravity actions using $\mathfrak{so}(2,d)$-gauge fields. 
Analogously to the re-derivation of Einstein action using the isometry-algebra-valued gauge field  \cite{MacDowell:1977jt}
(see also \cite{Stelle:1979aj}),
four-dimensional conformal gravity action has been expressed as an action of $\mathfrak{so}(2,4)$-gauge field subject to 
certain off-shell constraints \cite{CrispimRomao:1977hj,Kaku:1977pa,Crispim-Romao:1978zlo}
(see also the review \cite{Fradkin:1985am}).
In three dimensions, the Chern-Simons theory of $\mathfrak{so}(2,3)$-gauge field provides a Weyl invariant theory with
all necessary constraints integrated in it \cite{Horne:1988jf}. It has been also shown that conformal gravity actions can be obtained from an ambient space formulation \cite{Preitschopf:1998ei}
and from a  dimensional reduction \cite{Aros:2013yaa}.

Yet another framework intimately related to the Cartan geometry is the Vasiliev's unfolded formulation 
\cite{Vasiliev:1986td,Vasiliev:1987hv,Lopatin:1987hz,Vasiliev:1988xc,Vasiliev:1988sa}.
It is basically the zero-form extension of free differential algebra introduced in the context of supergravity 
\cite{DAuria:1982mkx}.
Thanks to the zero-forms, dynamical equations can be expressed in an integrable first-derivative form,
and the consistency of dynamical system becomes purely algebraic.
The unfolded formulation has been originally introduced for higher spin gravity \cite{Vasiliev:1988sa,Vasiliev:1990en,Vasiliev:2003ev} but can be applied to any dynamical system. 
In particular, a huge variety of free dynamical systems
governed by the conformal algebra $\mathfrak{so}(2,d)$ have been analyzed in the unfolded formulation \cite{Shaynkman:2004vu,Vasiliev:2009ck},
and it has been shown that the structure of the representations appearing there matches 
that of the Bernstein-Gelfand-Gelfand resolution (see e.g. \cite{BGG}).
In \cite{Bekaert:2017bpy}, Fefferman-Graham ambient construction
has been integrated into the unfolded system as Hamiltonian constraints.

In this work, we unfold conformal geometry starting from the $\mathfrak{so}(2,d)$ gauge connection.
As mentioned above, the merit of the unfolding procedure lies in the zero-form sector 
and we closely look into this part and fill the details which are not delineated in the literature. 
By explicitly analyzing the first-order consistency of the unfolded equation,
we find that the content of zero-forms corresponds to the spin-two off-shell 
Fradkin-Tseyltin module \cite{Shaynkman:2004vu,Beccaria:2014jxa} (see also \cite{Basile:2018eac}),
that is the module associated with the off-shell spin-two Fradkin-Tseytlin field \cite{Fradkin:1985am}.
In this paper, we are primarily interested in the unfolding of the off-shell system. 
Indeed, the unfolding formulation  can be applied not only to dynamical systems but also systems without prescribed dynamics,
namely, off-shell systems \cite{Vasiliev:2009ck}. See \cite{Misuna:2019ijn,Misuna:2020fck} for recent works on the off-shell extension of higher spin gravity.  
An interesting point of off-shell unfolding is that it can be posteriorly reduced to an on-shell system
by imposing additional algebraic constraints on the zero-form sector.
In this way, once we unfold conformal geometry, we can reduce it
to the on-shell conformal gravity, namely Bach-flat geometry, by eliminating some zero-forms with additional algebraic constraints.
Moreover, by imposing additional constraints on both of the one-form and zero-form sectors, 
one can reduce it even to Einstein gravity or various modifications of it.
If we restrict to the linear equation, then
the unfolded off-shell spin-two Fradkin-Tseytlin system can be reduced 
to on-shell Fradkin-Tseytlin system,
as well as partially massless or massive spin-two systems \cite{Skvortsov:2006at,Ponomarev:2010st}.
In this paper, we mostly consider the spin-two cases, but clearly, many of the above reductions can be generalized to higher spins. 
Let us note --- to avoid a potential confusion of interpreting the above statements  in the AdS/CFT context --- that all these theories are defined in the same $d$ dimensional spacetime.

Another merit of the unfolding of conformal geometry 
is that it allows to revisit the classification of Weyl anomalies, which was our original motivation. 
As we shall show below, the classification of Weyl anomalies \`a la unfolding 
is essentially the same as the method proposed in \cite{Boulanger:2004zf}.
The latter makes use of Weyl-covariant derivatives of Weyl tensors, which
carry reducible Lorentz representations, to make an ansatz for Weyl invariant density.
Instead, the unfolded equation is equipped with the zero-forms carrying irreducible Lorentz representations,
and the ansatz for a Weyl invariant density is made by these zero-forms. 
Unfortunately, the unfolding does not gain sizable computational efficiency, and 
to revisit the eight dimensional classification or even to attack 
the ten dimensional problem, 
it would be inevitable to use a computer programming,  like in \cite{Boulanger:2004zf}, 
which was beyond the scope of the current work.
Despite this limitation, we still find it useful in understanding the general structure of  unfolding conformal geometry 
and  the essence of the Weyl anomaly classification problem. 
The explicit knowledge of the zero-form content and their first-order gauge transformation 
allows us to do a few preliminary assessments of the classification.

The organization of paper is as follows.
In Section \ref{sec: gauge}, we begin with a review of the $\mathfrak{so}(2,d)$ gauge formulation of conformal gravity
and set the problem of unfolding conformal geometry,
which can be approached perturbatively in the power of zero-forms.
In Section \ref{sec: linear}, we explain how the linear part of the unfolded equation
in the zero-form expansion
defines a $\mathfrak{so}(2,d)$ representation in the space of zero-form fields.
This differs from the analysis of linearized equations around a specific background
as the latter captures only the isometry part of $\mathfrak{so}(2,d)$\,.
We solve the consistency conditions of the linear part
by using cell operators and 
find explicit form of the equation up to linear order in the zero-form.
In Section \ref{sec: representation}, we 
review the spin-two off-shell Fradkin-Tseyltin module,
and show its $\mathfrak{so}(2)\oplus \mathfrak{so}(d)$ decomposition
coincides with the content of the zero-form fields.
We comment also on the subtle points on the active and passive actions
and the role of dual representation.
In Section \ref{sec: nonlinear}, we sketch the structure of the unfolding at nonlinear orders
and show how the nonlinear equations can be systematically determined by moving up from 
the zero-form field equation of the lowest conformal weight.
In Section \ref{sec: Weyl}, 
we review the gauge symmetry of an unfolded system,
and show how the non-trivial Weyl invariants could be determined 
as gauge invariant $d$-forms.
In particular, we determine the unique quadratic part of the Weyl invariants in eight and ten dimensions.
We also provide discussions on higher order parts.
In Section \ref{sec: reduction}, 
we show how various on-shell systems can be obtained 
by requiring certain algebraic constraints which are invariant under special conformal transformations.
We also discuss how the off-shell conformal system itself can be viewed 
as a constrained system.

\section{Unfolding conformal geometry}
\label{sec: gauge}

In this section, first  we review the $\mathfrak{so}(2,d)$ gauge formulation
of conformal geometry, which has been known since \cite{CrispimRomao:1977hj,Kaku:1977pa,Crispim-Romao:1978zlo} (see also \cite{Kuzenko:2019ill} for a recent use of it).
Then, we introduce the unfolding scheme to conformal geometry 
with a rather pedagogical account.

\subsection{Gauge formulation of conformal geometry}

Let us begin with setting the convention for the conformal algebra: 
the Lie algebra $\mathfrak{so}(2,d)$ is generated by
anti-Hermitian generators $\hat M_{AB}$
with the Lie bracket,
\be
	[\hat M_{AB},\hat M_{CD}]= \eta_{AD}\hat M_{BC}+\eta_{BC}\hat M_{AD}-\eta_{AC}\hat M_{BD}
	-\eta_{BD}\hat M_{AC}\,,
\ee
where $\eta_{AB}$ is the flat metric with signature $(2,d)$\,.
Taking the basis with indices $A=+,-,a$ and $a=0,1,\ldots, d-1$
where $\eta_{+-}=1$ and $\eta^{ab}$ is the $d$-dimensional flat metric with signature 
$(1,d-1)$, the
Lie bracket of
$\hat M_{ab}=\hat J_{ab}$, 
$\hat M_{a+}=\hat P_a$, $\hat M_{a-}=\hat K_a$ and $\hat M_{+-}=\hat D$
read 
\ba
	&[\hat J_{ab},\hat J_{cd}]= \eta_{ad}\hat J_{bc}+\eta_{bc}\hat J_{ad}-\eta_{ac}\hat J_{bd}-\eta_{bd}\hat J_{ac}\,, \qquad 
	&[\hat J_{ab},\hat D]=0\,,\nn
	&[\hat J_{ab},\hat P_c]= \eta_{bc}\hat P_a-\eta_{ac}\hat P_b\,,\qquad 
	&[\hat J_{ab},\hat K_c,]= \eta_{bc}\hat K_a-\eta_{ac}\hat K_b\,,\nn
	& [\hat D, \hat P_a]=\hat P_a\,, \qquad 
	[\hat D,\hat K_a]= -\hat K_a \,,\qquad 
	& [\hat K_a,\hat P_b]=\eta_{ab}\,\hat D-\hat J_{ab}\,.
\ea
From now on, all the Latin indices $a,b,c,d,\ldots$ are lowered and raised by $\eta_{ab}$ and $\eta^{ab}$.

Let us review now the gauge formulation of conformal geometry.
We consider the gauge one-form taking value in $\mathfrak{so}(2,d)$ algebra,
\be
	\hat A=e^a\,\hat P_a+\frac12\,\omega^{ab}\,\hat J_{ab}+f^a\,\hat K_a+b\,\hat D\,,
	\label{1form}
\ee
where $e^a, \omega^{ab}, f^a, b$ are one-form fields
which are all independent at this stage.
For geometric interpretation, we assume the one-form $e^a$ has components $e^a_\mu$
which are invertible, and 
the inverse 
is denoted by $E_a^\mu$, which  define vector fields $E_a=E_a^\mu\,\partial_\mu$\,.
The curvature two form,
\be
	\hat F= \dd\hat A+\hat A\wedge \hat A 
	= F_{\hat P}^a\,\hat P_a+\frac12\,F_{\hat J}^{ab}\,\hat J_{ab}+	
	F_{\hat K}^a\,\hat K_a+F_{\hat D}\,\hat D\,.
\ee
has the components, 
\ba
	&F_{\hat P}^a= ({\cd^L}+b)\,e^a\,,\qquad 
	&F_{\hat J}^{ab} = R^{ab}-2\,f^{[a}\wedge e^{b]}\,,\nn
	&F_{\hat K}^a= ({\cd^L}-b)\,f^a\,,\qquad 
	&F_{\hat D}= \dd b+f^a\wedge e_a\,.
\ea
Here, ${\cd^L}$ is the Lorentz covariant differential,
\be
	{\cd^L}\,V^a=\dd\,V^a +\omega^a{}_b\,V^b\,,
\ee
and $R^{ab}$ is given by
\be
	R^{ab}=\dd\,\o^{ab}+\o^{ac}\wedge \o_{c}{}^b\,.
\ee
This system has $\mathfrak{so}(2,d)$ gauge symmetry,
\be
	\delta \hat A=\dd \hat \L+[\hat A,\hat \L]\,,
	\qquad 
	\delta \hat F=[\hat F,\hat \L]\,.
\ee
Labeling the components of the gauge parameter $\hat \L$ as
\be
	\hat \L=\e^a\,\hat P_a+\frac12\,\l^{ab}\,\hat J_{ab}+\k^a\,\hat K_a+\s\,\hat D\,,
\ee
the one-forms transform as
\ba
&& \delta e^a=({\cd^L}+b)\,\e^a -\l^a{}_b\,e^b-\s\,e^a\,,\\
&& \delta \omega^{ab}={\cd^L} \l^{ab}+2\,e^{[a}\,\k^{b]}+2\,f^{[a}\,\e^{b]}\,,
\\
&& \delta f^a=({\cd^L}-b)\,\k^a-\l^a{}_b\,f^b+\s\,f^a\,,\\
&& \delta b=\dd \s-e^a\,\k_a+\e^a\,f_a\,.
	\label{gauge sym}
\ea
We would need a partial gauge fixing as well as imposing constraints to recover the usual
geometry based on metric tensor.

Let us review the set of constraints which bring the $\mathfrak{so}(2,d)$-gauge theory
to conformal geometry.
First, we impose the torsionless constraint,
\be
	{\rm C}_{\hat P}\,: \qquad F_{\hat P}^a=({\cd^L}+b)\,e^a\overset{!}{=}0\,,
	\label{CP}
\ee
which is modified by the presence of $b$.
Here, we use the notation $\overset{!}{=}$ to emphasize that it is a constraint that we 
decided to impose.
We also impose the curvature for dilation $\hat D$ to vanish
\be
	{\rm C}_{\hat D}\,: \qquad F_{\hat D}=\dd\,b+f^a\wedge e_a\overset{!}{=}0\,.
	\label{CD}
\ee
From the Bianchi identity $\dd \hat F+[\hat A,\hat F]=0$, we find
\ba
	&{\rm C}_{\hat P}\ +\ {\rm BI}_{\hat P}\,:\qquad &F^{ab}_{\hat J}\wedge e_b= 0\,,\\
	&{\rm C}_{\hat D}\ +\ {\rm BI}_{\hat D}\,: \qquad & F^a_{\hat K}\wedge e_a= 0\,, \\
	&{\rm BI}_{\hat J}\,:\qquad & {\cd^L}\,F^{ab}_{\hat J}-2\,e^{[a}\wedge F_{\hat K}^{b]}=0\,,\label{dO BI}\\
	&{\rm BI}_{\hat K}\,:\qquad &({\cd^L}-b)\,F_{\hat K}^a-f_b\wedge F_{\hat J}^{ab}=0\,,\label{diff BI2}
\ea
where we implemented the constraints C$_{\hat P}$ \eqref{CP} and C$_{\hat D}$ \eqref{CD}.
Let us express
\be
	F_{\hat J\,ab}=\frac12\,C_{ab,cd}\,e^c\wedge e^d\,,
	\qquad 
	F_{\hat K\,a}= \frac12\,C_{a,bc}\,e^b\wedge e^c\,.
\ee
Then $C_{ab,cd}$ and $C_{a,bc}$ are given in terms of 
$R_{cd,ab}=i_a\,i_b\,R_{cd}$  and $f_{b,a}=i_a\,f_b$ 
with $i_a:=i_{E_a}$
(or  $R_{ab}=\frac12\,R_{ab,cd}\,e^{c}\wedge e^d$ and $f_a=f_{a,b}\,e^b$) as
\ba
	C_{ab,cd} \eq R_{ab,cd}-\eta_{ad}\,f_{b,c}+\eta_{bd}\,f_{a,c}+\eta_{ac}\,f_{b,d}-\eta_{cb}\,f_{a,d}
	\,,
	\label{W exp}
	\\
	C_{a,bc} \eq 2\left({\cd^L}_{[b|} f_{a,|c]}-2\,b_{[b|}\,f_{a,|c]}\right),
\ea
where ${\cd^L}=e^a\,{\cd^L}_a$\,.
The first two identities, namely, the algebraic Bianchi identities are equivalent to
\be
	{\rm C}_{\hat P}\ +\ {\rm BI}_{\hat P}
	\qquad \Longrightarrow \qquad C_{a[b,cd]}=0\,,
\ee
\be
	{\rm C}_{\hat D}\ +\ {\rm BI}_{\hat D} \qquad \Longrightarrow \qquad
	 C_{[ab,c]}=0\,,
\ee
so they are irreducible $GL_d$ tensors:
\be
	C_{ab,cd} \sim {\small \young(ac,bd)}\,,
	\qquad
	C_{c,ab} \sim {\small \young(ac,b)}\,.
\ee
Moreover if we impose the trace-free constraint on $C_{ab,cd}$\,,
\be
	{\rm C}_{\hat J}\,: \qquad \eta^{ac}\,C_{ab,cd}\overset{!}{=}0
	\quad \Longleftrightarrow\quad i_{a}\,F^{ab}_{\hat J}\overset{!}{=}0\,,
\ee
the trace of the differential Bianchi identity  \eqref{dO BI}  requires $C_{a,bc}$ to be trace-free as well: 
\be
	{\rm C}_{\hat J}\ +\ {\rm BI}_{\hat J}\qquad \Longrightarrow 
	\qquad \eta^{ab}\,C_{a,bc}=0\,.
\ee
In fact, the constraints ${\rm C}_{\hat D}$ and ${\rm C}_{\hat J}$ are not independent,
and the former is a consequence of the latter together with other constraints.
To recapitulate, we impose the following set of constraints,\footnote{In parabolic geometry \cite{sharpe},
these constraints are what define ``normal connection''.
See Section \ref{sec: reduction} for further discussions.}
\be
	F^a_{\hat P}\overset{!}{=}0\,,\qquad i_{a}\,F^{ab}_{\hat J}\overset{!}{=}0\,,
	\qquad (F_{\hat D}\overset{!}{=}0),
	\label{F constr}
\ee
and the resulting algebraic Bianchi identities are
\be 
	C_{a[b,cd]}=0\,,\qquad C_{[a,bc]}=0\,,
	\qquad \eta^{ab}\,C_{a,bc}=0\,.
\ee
The differential Bianchi identities \eqref{dO BI} and  \eqref{diff BI2} read
\ba
	&& ({\cd^L}-2\,b)_{[k}\,C^{ab,}{}_{cd]}-2\,\delta_{[k}^{[a}\,C^{b],}{}^{\phantom{]}}_{cd]} =0\,,\nn
	&&({\cd^L}-3\,b)_{[k}C^{a,}{}_{cd]}-\,f_{b,[k}\,C^{ab,}{}_{cd]}=0\,.
	\label{f BI}
\ea
The above is the starting point of the unfolding machinery. 
Before moving to that, let us review how the usual conformal geometry can be
recovered from this setting.

\subsection*{Reduction to metric formulation}

All the constraints can be solved algebraically:
\begin{itemize}
\item The constraint C$_{\hat P}$ determines $\o_{ab,c}=i_c\,\o_{ab}$ 
(or $\o_{ab}=\o_{ab,c}\,e^c$)
in terms of $e^a$ and $b$ as
\be
	\o_{ab,c}=
	E^\mu_{[b}\,E_{c]}^\nu\,\partial_{\mu}\,e_{a\nu}+
	E^\mu_{[c}\,E_{a]}^\nu\,\partial_{\mu}\,e_{b\nu}+E^\mu_{[b}\,E_{a]}^\nu\,\partial_{\mu}\,e_{c\nu}
	+2\,b_{[a}\,\eta_{b]c}\,.
\ee
\item The constraint C$_{\hat D}$ determines $f_{[a,b]}$ in terms of $b$\,:
\be
	f_{[a,b]}=\partial_{[a}\,b_{b]}\,,
\ee
where $\partial_a=E_a^\mu\,\partial_\mu=E_a$\,.
\item The constraint C$_{\hat J}$ determines $f_{(a,b)}$ in terms of $R_{ab}=R_{a}{}^{c}{}_{,bc}$\,:
from \eqref{W exp}, we find
\be
	f_{(a,b)}=\frac{1}{d-2}\left(R_{ab}-\frac{\eta_{ab}\,R}{2\,(d-1)}  \right).
\ee

\end{itemize}
After solving all the constraints, the $\mathfrak{so}(2,d)$-gauge symmetry reduces to\footnote{In fact,
the constraints are not invariant under the gauge transformation
\eqref{gauge sym} with the parameter $\e^a$:
\be
	\delta F^a_{\hat P}=F^{ab}_{\hat J}\,\e_b\,,
	\qquad
	\delta (i_a\,F^{ab}_{\hat J})=i_a\,F^{[a}_{\hat K}\,\e^{b]}\,,
	\qquad
	\delta F_{\hat D} =   - F^{a}_{\hat K}\, \epsilon_a\,.
\ee
However, the gauge symmetries can be properly modified 
by a ``non-geometrical" curvature term \cite{Kaku:1977pa}
so as to
leave all the constraints invariant. 
In fact, this modification naturally arises in the unfolded formulation.
See Section \ref{sec: sym} for the details.}
\ba
	&& \delta e^a=({\cd^L}+b)\,\e^a -\l^a{}_b\,e^b-\s\,e^a\,,\\
	&& \delta b=\dd \s-e^a\,\k_a+\e^a\,f_a\,.
\ea
We can fix the gauge symmetries associated with $\hat K_a$ and $\hat J_{ab}$ as follows.

\begin{itemize}

\item The gauge symmetry of $\hat K_a$  allows us to set $b_a$ to zero:
\be
	\delta_{\k}\,b_a=\k_a \qquad \Longrightarrow \qquad b_a=0\,.
\ee
Note that in this gauge, the $\hat K_a$ symmetry must transformation
under the $\hat D$ symmetry with the parameter,
\be
	\k_a=\partial_a\sigma\,.
\ee
\item The gauge symmetry of $\hat J_{ab}$ is
\be
	\delta_\l e^a_\mu= \l^{ab}\,e_{b\mu}\,,
\ee
which allows us to fix the degrees of freedom of $e^a_\mu$
besides those in
\be
	g_{\m\n}=\eta_{ab}\,e^a_\mu\,e^b_{\nu}\,.
\ee
\end{itemize}

The residual gauge symmetries are those of $\hat D$ and $\hat P$,
\begin{itemize}

\item The gauge symmetry of $\hat D$ gives the Weyl rescaling,
\be
	\delta_\s\,g_{\mu\nu}=2\,\s\,g_{\m\n}\,.
\ee

\item The gauge symmetry of $\hat P$ gives the diffeomorphism,
\be
	\delta_\e\,g_{\mu\nu}=\nabla_{\mu}\e_\nu+\nabla_{\nu}\e_\mu\,,
	\qquad \e_\mu=e^a_\mu\,\e_a\,.
\ee
\end{itemize}

After the reduction, we find that $R_{ab,cd}$ and $C_{ab,cd}$ 
coincide with the usual Riemann 
and Weyl tensor,
and 
$P_{ab}=f_{(a,b)}$ and $C_{a,bc}=\nabla_{b}\,P_{ac}-\nabla_{c}\,P_{ab}$ 
with the Schouten and Cotton tensors.

\subsection{Unfolding conformal geometry}

Let us now consider the unfolding of conformal geometry.
Remind that we have used the equations,
\ba
	&&\cd^Ke^a=0\,,\nn
	&&\cd^K \o^{ab}-2\,e^{[a}\wedge f^{b]}
	=\frac12\,e_c\wedge e_d\,C^{ab,cd}\,,\nn
	&&\cd^K b+e^a\wedge f_a=0\,,\nn
	&& \cd^K f^a=\frac12\,e_b\wedge e_c\,C^{a,bc} \,,
	\label{1f eq}
\ea
with 
\be 
	C_{a[b,cd]}=0\,,\qquad C_{[a,bc]}=0\,,
	\qquad \eta^{ab}\,C_{ab,cd}=0\,,\qquad \eta^{ab}\,C_{a,bc}=0\,.
\ee
In the former set of equations, 
we slightly simplified the expressions by introducing 
$K=SO(1,1)\times SO(1,d-1)$ or $
\mathfrak{k}=\mathfrak{so}(1,1)\oplus \mathfrak{so}(1,d-1)$\footnote{The maximal compact subalgebra
of $\mathfrak{so}(2,d)$ is not 
$\mathfrak{so}(1,1)\oplus \mathfrak{so}(1,d-1)$ but
$\mathfrak{so}(2)\oplus \mathfrak{so}(d)$.
However, the two subalgebras are intimately related as we shall comment later 
in Section \ref{sec: FT module}.}
covariant derivative $\cd^K$, which 
acts on a $\mathfrak{so}(1,d-1)$-tensor with conformal dimension $\Delta$ as
\be
	\cd^K W^{[\D] ab\cdots} =
	\cd^LW^{[\D] ab\cdots} -\D\,b\,W^{[\D]ab\cdots}\,,
	\label{conf dim}
\ee
and assigning the conformal dimensions 
$\D=-1,0, 0,  1$
to $e^a, \o^{ab}, b, f^a$, respectively.\footnote{Assigning
the conformal dimension $-1$ to $\dd x^\m$,
the fields
 $e^a_\mu, \o^{ab}_\m, b_\m, f^a_\mu$
have conformal dimensions $0, 1, 1, 2$ identical 
to the numbers of derivatives of the corresponding fields in the metric formulation.}
Note here that the eigenvalue of the dilation operator $\hat D$
is $-\Delta$.
This reflects that the fields $W^{[\D]ab\cdots}$ 
carry in fact a dual (or contragredient) representation
which is obtained by the anti-involution $(\hat P^a,\hat J^{ab},\hat D,\hat K^a)
\to (\hat P^a,-\hat J^{ab},-\hat D,\hat K^a)$\,.
See Section \ref{sec: rep} for more discussions on this point.

In the following,
we sketch the key reasoning of the unfolding scheme.
\begin{itemize}
\item 
The system \eqref{1f eq} can  be regarded  as a set of equations
for one-forms $e^a, \o^{ab}, b$ and $f^a$
as well as zero-forms $C^{ab,cd}$ and $C^{a,bc}$. 
Note that $C^{ab,cd}$ and $C^{a,bc}$ have
conformal dimensions $\D=2$ and $3$, respectively.
The zero-forms
are completely, namely algebraically, determined by the equations, 
and hence no new degrees of freedom are introduced by employing them.
About the one-forms, basically the equations tell how the (covariant) derivatives of the one-forms
are determined by the other fields without any derivatives.
Consequently, the one-forms are subject to certain conditions
which are necessary for the system to be
equivalent to conformal geometry. 

\item
Viewing \eqref{1f eq} as a dynamical system for the associated fields, that is,
the one-forms $e^a, \o^{ab}, b$ and $f^a$
and the zero-forms $C^{ab,cd}$ and $C^{a,bc}$,
it is more natural to introduce a new set of equations
which determine the evolution---that is, the (covariant) derivatives---of $C^{ab,cd}$ and $C^{a,bc}$:
\ba
	&& \cd^KC^{ab,cd}=(\cd^L-2\,b)C^{ab,cd}=e_f\,C^{ab,cd,e}+
	(\textrm{pre-existing fields})\,,
	\nn
	&&\cd^KC^{a,bc}=(\cd^L-3\,b)C^{a,bc}=e_d\,C^{a,bc,d}+(\textrm{pre-existing fields})\,.
	\label{new eq}
\ea
On the right hand side of the equations, we have introduced a new set of zero-form fields
$C^{ab,cd,e}$ and $C^{a,bc,d}$ besides
what can be expressed in terms of pre-existing fields.
The new fields $C^{ab,cd,e}$ and $C^{a,bc,d}$ should be subject to a proper set of conditions 
so that the new equations \eqref{new eq} with the new fields  neither introduce any new degrees of freedom
nor remove any pre-existing degrees of freedom.
For this, one need to examine the compatibility of the new equations \eqref{new eq} with
the Bianchi identities \eqref{f BI}. 

\item Viewing \eqref{1f eq} and \eqref{new eq} 
as a dynamical system for the one-forms $e^a, \o^{ab}, b$ and $f^a$, and the zero-forms
$C^{ab,cd},C^{ab,cd,e}, C^{a,bc}$ and  $C^{a,bc,d}$, 
we can again introduce `evolution equations' for 
$C^{ab,cd,e}$ and $C^{a,bc,d}$ in a similar manner
as we did for $C^{ab,cd}$ and $C^{a,bc}$\,.

\item 
This procedure can be continued iteratively, and introduces infinitely many  zero-form fields
with infinitely many  equations 
in a way that such an extension of the fields and equations does not alter the content
of degrees of the freedom of the system.

\end{itemize}

In order to work with an infinite number of additional zero-form fields,
we need to label them efficiently, and the subalgebra $\mathfrak{k}=\mathfrak{so}(1,1)\oplus \mathfrak{so}(1,d-1)$ can
provide such a good label:\footnote{At this stage, it is not clear whether the $K$-label would be sufficient
without necessitating an additional label to distinguish two fields of the same $K$-label.
As we will show below, the $K$-label is sufficient to describe all the fields in the system.
In other words, in the decomposition of the zero-form module into $K$-irreps,
there is no multiplicty.}
in the following any zero-form fields will be labeled as 
traceless fiberwise tensors with two groups of totally symmetric indices,
\be
	C^{[\D] a_1\cdots  a_m,b_1\cdots b_n},
	\label{0-form}
\ee
subject to the Young projection condition,
\be
	C^{[\D] (a_1\cdots a_m,b_1)b_2\cdots b_n}=0\,.
\ee
In this way, the fiberwise tensor carries an irrep under $\mathfrak{so}(1,d-1)$ corresponding
to a two-row Young diagram.
We adopt the following common short-hand notation,
\be
	C^{[\D]a(m),b(n)}=C^{[\D] a_1\cdots a_m,b_1\cdots b_n}\,.
	\label{0 form content}
\ee
Sometimes, it will be more convenient to use
what we will refer to as ``depth'' $\delta$, than the conformal dimension $\D$\,:
\be
	C^{\{\d\}a(m),b(n)}=C^{[\Delta]a(m),b(n)}\,,
	\qquad 
	\delta=\frac{\D-m+n}2\,.
	\label{depth}
\ee
The zero-form fields $C^{[2]a(2),b(2)}$ and $C^{[3]a(2),b}$
should be identified with
the usual Weyl and Cotton tensors:
\be
	C^{[2]a(2),b(2)}=C^{(a_1|b_1,|a_2)b_2}\,,
	\qquad
	C^{[3]a(2),b}=C^{(a_1,a_2)b}\,.
	\label{WC condition}
\ee
If we relax the above identification condition,  
the zero-form equations that we will elaborate below
has the capacity to describe a system of any integral spin.

The infinite amount of the $K$-covariant equations for zero-forms 
that we need to identify will take the following form,
\be
	\cd^K C^{[\Delta]a(m),b(n)}
	=e_c\,\cE^{[\Delta+1]a(m),b(n)|c}(C)+f_c\,\cF^{[\Delta-1]a(m),b(n)|c}(C)\,,
\ee
where $\cE^{[\Delta+1]a(m),b(n)|c}(C)$ and $\cF^{[\Delta-1]a(m),b(n)|c}(C)$
are functions of the zero-forms with total conformal weight $\D+1$ and $\D-1$.
Let us recall that the conformal dimensions of $e^a, f^a$ and $\cd^K$ 
are $-1,1$ and 0.
We can consider the Taylor expansion of $\cE^{[\Delta+1]a(m),b(n)|c}(C)$,
\ba
	&&\cE^{[\Delta+1]a(m),b(n)|c}(C)=
	\cE^{[\Delta+1]a(m),b(n)|c}_{d(p),e(q)}\,C^{[\Delta+1]d(p),e(q)}\nn
	&& \qquad +
	\sum_{\underset{\D_1+\D_2=\D+1}{\D_1,\D_2}} \cE^{[\D_1,\D_2]a(m),b(n)|c}_{d(p),e(q)|f(s),g(t)}\,
	C^{[\Delta_1]d(p),e(q)}\,C^{[\Delta_2]f(s),g(t)}+\cdots\,,
\ea
where the expansion coefficients
$\cE^{[\Delta+1]a(m),b(n)|c}_{d(p),e(q)}$
and $\cE^{[\D_1,\D_2]a(m),b(n)|c}_{d(p),e(q)|f(s),g(t)}$
are made only by Kronecker delta symbols so that
they only rearrange or contract indices.
The function $\cF^{[\Delta-1]a(m),b(n)|c}(C)$ can be expanded analogously.
Identifying the general 
form of $\cE^{[\Delta+1]a(m),b(n)|c}(C)$ and $\cF^{[\Delta-1]a(m),b(n)|c}(C)$
is a highly non-trivial task, and hence
we first identify  the linear parts,
which will determine the content of the zero-forms.
The identification of non-linear terms 
can be worked out, in principle,
order by order in $\D$.
Due to the boundness of $\D\ge 2$,
$\cd^K C^{[\D] a(m)b(n)}$ will involve
at most $[(\D+1)/2]$ order terms.

\section{First order unfolding of conformal geometry}
\label{sec: linear}

\subsection{First order unfolding}

Let us consider the system only up to the linear order,
\be
	\cd^K C^{[\Delta]a(m),b(n)}
	+
	e^c\,(\hat P_c\, C)^{[\Delta]a(m),b(n)}
	+f^c\,(\hat K_c \, C)^{[\Delta]a(m),b(n)}
	=\cO(C^2)\,,
	\label{0f W eq}
\ee
where we have used the notation,\footnote{Note that $(\cO\,T)^{a(m),b(n)}$ denotes
the $a(m),b(n)$ components of the tensor $\cO\,T$.
Here,  $T$ is not a tenor of type $(m,n)$ but $\cO\,T$ is.}
\ba
	&& (\hat P^c\, C)^{[\Delta]a(m),b(n)}=
	-\cE^{[\Delta+1]a(m),b(n)|c}_{d(p),e(q)}\,C^{[\Delta+1]d(p),e(q)}\,,\nn
	&& (\hat K^c\, C)^{[\Delta]a(m),b(n)}=
	-\cF^{[\Delta-1]a(m),b(n)|c}_{d(p),e(q)}\,C^{[\Delta-1]d(p),e(q)}\,.
\ea
Remark that the linear terms of $\cE^{[\D+1]a(m),b(n)|c}(C)$ and 
$\cF^{[\D-1]a(m),b(n)|c}(C)$ are 
denoted by the actions of $\hat P^c$ and $\hat K^c$, respectively.
It will become shortly clearly that they indeed correspond to
the action of translation and special conformal transformation.
Remind also that the action of $\hat P_a$ and $\hat K_a$ on
the space of zero-form fields is not yet defined.
The equation \eqref{0f W eq} suggests to combine
the linear terms with the $K$-covariant derivative as
\be
	\cd^GC=\cO(C^2)\,,
	\label{cov 0f W eq}
\ee
with
\be
	\cd^G=\cd^K+e^a\,\hat P_a+f^a\,\hat K_a
	=\dd + \o^{ab}\,\hat J_{ab}+b\,\hat D
	+e^a\,\hat P_a+f^a\,\hat K_a\,.
\ee
The Bianchi identity is the consistency 
of the equation \eqref{cov 0f W eq} associated with
\ba
	(\cd^G)^2
	\eq 
	f^{a}\wedge e^{b}\,([\hat K_{a},\hat P_{b}]+\hat J_{ab}-\eta_{ab}\,\hat D)\nn
	&&+\,e^a\wedge e^b\,\hat P_{[a}\,\hat P_{b]}
	+f^a\wedge f^b\,\hat K_{[a}\,\hat K_{b]}
	+\cO(C)\,. 
\ea
Since
the action of $(\cd^G)^2$ on $C$ is at least quadratic in $C$, 
the following should hold.
\be
\begin{split}
	&(\hat P_{[c}\,\hat P_{d]}\,C)^{[\D]a(m),b(n)}=0\,,
	\qquad 
	(\hat K_{[c}\,\hat K_{d]}\,C)^{[\D]a(m),b(n)}=0\,,\\
	&\big(([\hat K_{a},\hat P_{b}]+\hat J_{ab}-\eta_{ab}\,\hat D)\,C\big)^{[\D]a(m),b(n)}=0\,,
	\label{condition}
\end{split}
\ee
and, we find that the consistency of the equation requires the operators $\hat P_a$ and $\hat K_a$
coincide with the actions of translation and special conformal transformation.
Therefore, the linear part of $\cd^G$ can be viewed as a $G=SO(2,d)$-covariant derivative.

The most general form of a $\hat P_a$ action on
the space of tensors with two-row Young symmetry is
simply
\ba
	(\hat P^c\, C)^{[\Delta]a(m),b(n)}\eq 
(\cP^{c}_{1+}\,C^{[\Delta+1]})^{a(m),b(n)}
+(\cP^{c}_{1-}\,C^{[\Delta+1]})^{a(m),b(n)}\nn
&&+\,(\cP^{c}_{2+}\,C^{[\Delta+1]})^{a(m),b(n)}
+(\cP^{c}_{2-}\,C^{[\Delta+1]})^{a(m),b(n)}\,,
\label{P def}
\ea
where the operators $\cP^a_{1\pm}$ and $\cP^a_{2\pm}$
are the operators
which map 
the tensors with the Young symmetry  $(m\mp1,n)$, 
and $(m,n\mp1)$ to the tensor with the Young symmetry $(m,n)$\,.
Since these operators are unique up to overall factors
(see Section \ref{sec: solution} for the details),
 the unknowns are only the proportionality constants
which depend $m,n$ and $\Delta$.
Similarly, the most general form of a $\hat K$ action 
can be written as
\ba
	(\hat K^c\, C)^{[\Delta]a(m),b(n)}\eq 
(\cK^{c}_{1+}\,C^{[\Delta-1]})^{a(m),b(n)}
+(\cK^{c}_{1-}\,C^{[\Delta-1]})^{a(m),b(n)}\nn
&&+\,(\cK^{c}_{2+}\,C^{[\Delta-1]})^{a(m),b(n)}
+(\cK^{c}_{2-}\, C^{[\Delta-1]})^{a(m),b(n)}\,,
\label{K def}
\ea
with similarly defined operators $\cK^a_{1\pm}$ and $\cK^a_{2\pm}$.

\subsection{Linearization around (A)dS background}

Before moving to solve the conditions \eqref{condition},
let us consider the linearization around a (A)dS background where
the zero-forms all vanish:
\be 
	\bar C^{[\D]a(m),b(n)}=0\,,
\ee
and the one-forms satisfy
\be
	 \bar f^a=\Lambda\,\bar e^a\,, \qquad \bar b=0\,,
\ee
and hence,
\be
	\dd \bar \o^{ab}+\bar \o^{ac}\wedge \bar\o_{c}{}^b
	-2\Lambda\,\bar e^a\wedge \bar e^b=0\,.
\ee
Repeating the analysis
for the linear fluctuation,
we find that the background $G$-covariant derivative 
reduces to a background $H$-covariant derivative
\be
	\bar\cd^G=
	\bar\cd^H=\dd + \bar \o^{ab}\,\hat J_{ab}
	+\bar e^a\,(\hat P_a+\L\,\hat K_a)\,,
\ee
where $H$ is the subgroup $SO(1,d)$ for $\Lambda>0$, $SO(2,d-1)$ for $\Lambda<0$
and $ISO(1,d-1)$ for $\Lambda=0$
generated by $\hat J_{ab}$ and $\hat P_a+\Lambda\,\hat K_a$\,.
In this case, the Bianchi identity gives the condition,
\ba
	&(\hat P_{[c}\,\hat P_{d]}\,C)^{[\D]a(m),b(n)}
	+\L^2\,(\hat K_{[c}\,\hat K_{d]}\,C)^{[\D]a(m),b(n)}\nn
	&+\,2\,\L\,\big(([\hat K_{[c},\hat P_{d]}]+\hat J_{cd})\,C\big)^{[\D]a(m),b(n)}=0\,.
	\label{linearization}
\ea
Since the above, being identities, should not impose any relation between fields of different $\D$,
the three terms should separately vanish 
we can recover the three conditions among the ones in \eqref{condition}.
When the cosmological constant vanishes, 
we have only one
consistency condition, $(\hat P_{[c}\,\hat P_{d]}\,C)^{[\D]a(m),b(n)}=0$,
which determines only  the $\hat P_a$ action,
then the  $\hat P_a$ action alone defines
the zero-form field content of linearized conformal geometry around flat spacetime
(that is, the spin-two off-shell Fradkin-Tseyltin system). 
Since the field content of the linearized system should be the same as the field content of  non-linear one,
the $\hat P_a$ action should be enough to determine the zero-form field content of conformal geometry. 
However, when we consider an on-shell reduction of the system,
it is necessary to have the information of the $\hat K_a$ action,
which cannot be obtained from the linearization around flat space.

It would be instructive to rewrite the linearized zero-form equation using
the depth $\delta$ \eqref{depth} instead of the conformal weight $\D$:
\ba
	&&\bar\cd^L C^{\{\delta\}a(m),b(n)}
	+\bar e_c\,(\cP^{c}_{1-}\,C^{\{\delta\}})^{a(m),b(n)}
	+\bar e_c\,(\cP^{c}_{2+}\, C^{\{\delta\}})^{a(m),b(n)}\nn
	&&+\, \L\,\bar e_c\,(\cK^{c}_{1+}\,C^{\{\delta\}})^{a(m),b(n)}
	+\L\,\bar e_c\,(\cK^{c}_{2-}\,C^{\{\delta\}})^{a(m),b(n)}\nn
	&&+\, \bar e_c\,(\cP^{c}_{1+}\,C^{\{\delta+1\}})^{a(m),b(n)}
	+\bar e_c\,(\cP^{c}_{2-}\,C^{\{\delta+1\}})^{a(m),b(n)}\nn
	&&+\,\L\,\bar e_c\,(\cK^{c}_{1-}\,C^{\{\delta-1\}})^{a(m),b(n)}
	+\L\,\bar e_c\,(\cK^{c}_{2+}\,C^{\{\delta-1\}})^{a(m),b(n)}=0\,.\quad 
	\label{0f W eq'}
\ea
Remark that
the half of terms in the $\hat P_a, \hat K_a$ action above
preserve the depth, but not the conformal dimension, of the fields.
If we impose the condition,
\be
	\cP^a_{1+}=\cP^a_{2-}=\cK^a_{1-}=\cK^a_{2+}=0\,,
\ee
then the system can rely on fields of a single depth:
\ba
	&&\bar\cd^L C^{\{\delta\}a(m),b(n)}
	+\bar e_c\,(\cP^{c}_{1-}\,C^{\{\delta\}})^{a(m),b(n)}
	+\bar e_c\,(\cP^{c}_{2+}\, C^{\{\delta\}})^{a(m),b(n)}\nn
	&&+\, \L\,\bar e_c\,(\cK^{c}_{1+}\,C^{\{\delta\}})^{a(m),b(n)}
	+\L\,\bar e_c\,(\cK^{c}_{2-}\,C^{\{\delta\}})^{a(m),b(n)}=0\,.
\ea
The resulting system is nothing but a non-conformal system 
such as massless, partially-massless or even massive spin two.
These systems are studied in \cite{Skvortsov:2006at,Ponomarev:2010st}.
The $\sigma^1_{\pm}$ and $\sigma^2_{\pm}$ operators
used therein correspond in our case to
\be
	\s^1_{-}=\bar e_a\,\cP^a_{1-}\,,
	\qquad
	\s^1_{+}=\L\,\bar e_a\,\cK^a_{1+}\,,
	\qquad
	\s^2_{-}=\L\,\bar e_a\,\cK^a_{2-}\,,
	\qquad 
	\s^2_{+}=\bar e_a\,\cP^a_{2+}\,.
\ee
Furthermore, if we impose  the restriction,
\be
	\cP^a_{2+}=\cK^a_{2-}=0\,,
\ee
then we end up with only two operators $\s_-=\s_-^1=\bar e_a\,\cP^a_{1-}$
and $\s_+=\s_+^1=\L\,\bar e_a\,\cK^a_{1+}$,
which describe the massless spin-$s$ dynamics.
Therefore, the system \eqref{0f W eq'}  encompasses
the dynamics of (partially-)massless and massive
as well as conformal fields---the goal of the current paper---with a consistent
choice of $\cP^a_{r\pm}$ and $\cK^a_{r\pm}$\,.
We will come back to this point 
later in Section \ref{sec: reduction}.

\subsection{Cell operators and Recurrence relations}
\label{sec: solution}

As mentioned earlier, the operators
$\cP^a_{r\pm}$ and $\cK^a_{r\pm}$
 are proportional to the cell operators
 which adds or removes one box with index $a$ 
 to a two-row Young diagram \cite{Skvortsov:2006at,Boulanger:2008up,Boulanger:2008kw,Ponomarev:2010st}.
 These operators are unique up to proportionality 
 and their precise expressions and properties are given in the latter references.
 Here, we use their realizations as differential operators acting on auxiliary variables:
we contract the zero-form fiberwise tensors with two set of auxiliary variables $u_a$ and $v_a$ as
\be
	C^{[\D](m,n)}(u,v)=C^{[\D]a(m),b(n)}\,
	\frac{u_{a_1}\cdots u_{a_m}}{m!}\,\frac{v_{b_1}\cdots v_{b_n}}{n!}\,.
	\label{aux var}
\ee
Then, the irreducibility and traceless conditions of the tensors read
\be
	u\cdot \partial_v\,C^{[\D](m,n)}(u,v)=0\,,
	\qquad 
	\partial_u^2\,C^{[\D](m,n)}(u,v)=0\,,
\ee
and the Lorentz generators act as the differential operator,
\be
	\hat J_{ab}
	=2\,u_{[a}\,\partial_{u^{b]}}+2\,v_{[a}\,\partial_{v^{b]}}\,.
	\label{J gen}
\ee
The one-cell operators, denoted henceforth by $\cY^a_{r\pm}$, can be defined as\footnote{
More explicitly, the one-cell operators, when acted on $(m,n)$ tensors,
have the form,
\be
\cY_{1-}^a = \partial_{u_a}+
\tfrac{1}{m-n+1}\,v\cdot\partial_u\,\partial_{v_a}\,,
\qquad 
\cY_{2-}^a =  \partial_{v_a}\,,
\ee
\ba
\cY^a_{1+} 
\eq u^a -\tfrac{1}{d+2m-2}\,u^2\, \partial_{u_a}
-\tfrac{1}{d+m+n-3}\,u\cdot v\,\partial_{v_a} \nn
&&+\,\tfrac{1}{(d+2m-2)(d+m+n-3)}\,
u^2\,v\cdot \partial_u\,\partial_{v_a}\,,
\ea
\ba
\cY_{2+}^a
\eq {v^a}
-\tfrac{(m-n)}{(d+2n-4)(m-n+1)}\,v^2\,\partial_{v_a}
-\tfrac{1}{(m-n+1)}\, v\cdot\partial_u \,u^a  \nn
&&
-\,\tfrac{m-n-1}{(m-n+1)(d+m+n-3)}\,u\cdot v\, \partial_{u_a} 
+\tfrac{d+2m -4}{(m-n+1) (d+m+n-3)(d+2n-4)} \,u\cdot v\,v\cdot\partial_u\, \partial_{v_a}\\
&&
+\,\tfrac{1}{(m-n+1)(d+m+n-3)} \,u^2\,v\cdot \partial_u\, \partial_{u_a}
-\tfrac{1}{(m-n+1)(d+m+n-3)(d+2n-4)} \,u^2\,(v\cdot \partial_u)^2\,\partial_{v_a}\,.\nonumber
\ea 
} 
\be
	\cY^a_{1+}=\Pi_{\mathbb Y}\,u^a\,, 
	\qquad \cY^a_{1-}=\Pi_{\mathbb Y}\,\partial_{u_a}\,,
	\qquad
	\cY^a_{2+}=\Pi_{\mathbb Y}\,v^a\,,
	\qquad
	\cY^a_{2-}=\Pi_{\mathbb Y}\,\partial_{v_a}\,,
\ee
where $\Pi_{\mathbb Y}$ is the projection operator onto 
the space of traceless tensors of two-row Young diagram symmetry.
Beside the one-cell operators, let us also introduce `two-cell operators' defined by
\ba
	&\cY_{1+}^a{}_{1+}^{b}=\Pi_{\mathbb Y}\,u^a u^b\,, \qquad
	\cY_{1+}^a{}_{2+}^{b}=\Pi_{\mathbb Y}\,u^a v^b\,, \qquad
	&\cY_{2+}^a{}_{2+}^{b}=\Pi_{\mathbb Y}\,v^a v^b\,, \\
	&\cY_{1-}^a{}_{1-}^{b}=\Pi_{\mathbb Y}\,\partial_{u_a} \partial_{u_b}\,, \qquad
	\cY_{1-}^a{}_{2-}^{b}=\Pi_{\mathbb Y}\,\partial_{u_a} \partial_{v_b}\,, \qquad
	&\cY_{2-}^a{}_{2-}^{b}=\Pi_{\mathbb Y}\,\partial_{v_a} \partial_{v_b}\,,
\ea
and
\ba
	&\cY_{1+}^a{}_{2-}^{b}=\Pi_{\mathbb Y}\,u^a \partial_{v_b}\,, \qquad
	&\cY_{2+}^a{}_{1-}^{b}=\Pi_{\mathbb Y}\,v^a \partial_{u_b}\,, \\
	&\cY_{1+}^a{}_{1-}^{b}=\Pi_{\mathbb Y}\,u^a \partial_{u_b}\,, \qquad
	&\cY_{2+}^a{}_{2-}^{b}=\Pi_{\mathbb Y}\,v^a \partial_{v_b}\,.
\ea
We can express the product of two one-cell operators as two-cell operators as
\be
	\cY^a_{1\pm}\,\cY^b_{1\pm}=\cY^b_{1\pm}\,\cY^a_{1\pm}=\cY_{1\pm}^a{}_{1\pm}^{b}\,,
	\qquad
	\cY^a_{2\pm}\,\cY^b_{2\pm}=\cY^b_{2\pm}\,\cY^a_{2\pm}=\cY_{2\pm}^a{}_{2\pm}^{b}\,,
\ee
\ba
	& \cY^b_{2+}\,\cY^a_{1+}=\cY_{1+}^a{}_{2+}^{b}\,,
	\qquad
	& \cY^a_{1+}\,\cY^b_{2+}=\cY^a_{1+}{}^b_{2+}+\tfrac{1}{m-n+1}\,\cY^b_{1+}{}^a_{2+}\,,
	\nn
	&\cY^a_{1-}\,\cY^b_{2-}=\cY_{1-}^a{}_{2-}^{b}\,,
	\qquad 
	&\cY^b_{2-}\,\cY^a_{1-}=\cY^a_{1-}{}^b_{2-}+\tfrac{1}{m-n+1}\,\cY^b_{1-}{}^a_{2-}\,,\nn
	&
	\cY^a_{1+}\,\cY^b_{2-}=\cY_{1+}^a{}_{2-}^{b}\,,
	\qquad
	&\cY^b_{2-}\,\cY^a_{1+}=\cY_{1+}^a{}_{2-}^{b}-\tfrac{1}{d+m+n-3}\,\cY^b_{1+}{}^a_{2-}\,,\\
	&
	\cY^a_{2+}\,\cY^b_{1-}=\cY_{2+}^a{}_{1-}^{b}\,,
	\qquad
	&\cY^b_{1-}\,\cY^a_{2+}=\cY_{2+}^a{}_{1-}^{b}-\tfrac{1}{d+m+n-3}\,\cY^b_{2+}{}^a_{1-}\,,
	\nonumber
\ea
\be
	\cY^a_{1+}\,\cY^b_{1-}=\cY_{1+}^a{}_{1-}^{b}-\tfrac{1}{m-n+1}\,\cY_{2+}^a{}_{2-}^{b}\,,
	\qquad
	\cY^a_{2+}\,\cY^b_{2-}=\cY_{2+}^a{}_{2-}^{b}\,,
\ee
\ba
	\cY^a_{1-}\,\cY^b_{1+}\eq 
	\cY^b_{1+}{}^a_{1-}+\h^{ab}
	- \tfrac{2}{d+2m-2} \,\cY^a_{1+}{}^b_{1-}
	-\tfrac{d+2m}{(d+2m-2)(d+m+n-3)} \,\cY^a_{2+}{}^b_{2-}\,,\\
	\cY^a_{2-}\,\cY^b_{2+}\eq 
	\cY^b_{2+}{}^a_{2-}+
	\tfrac{m-n}{m-n+1}\,\eta^{ab} 
	- \tfrac{2(m-n)(d+m+n-2)+(d+2n-4)}{(m-n+1)(d+2n-4)(d+m+n-3)} \,\cY^a_{2+}{}^b_{2-} \nn
	&&
	-\, \tfrac{m-n-1}{(m-n+1)(d+m+n-3)} \,\cY^a_{1+}{}^b_{1-}
	- \tfrac{1}{m-n+1}\,\cY^b_{1+}{}^a_{1-}\,,
\ea
where $m, n$ are the eigenvalues of $u\cdot\partial_u$ and $v\cdot \partial_v$,
that is, the length of the first and second rows of the Young diagram on which
the operators act.
Since the two-cell operators are independent, we can solve the Bianchi identities
by expressing all the operators appearing there as linear combinations of the two-cell operators. 
In particular, the Lorentz generator can be expressed as
\be
	\hat J^{ab}=\cY^a_{1+}{}^b_{1-}-\cY^b_{1+}{}^a_{1-}+\cY^a_{2+}{}^b_{2-}-\cY^b_{2+}{}^a_{2-}\,.
\ee
The operators $\cP^c_{1\pm}$ and $\cK^c_{r\pm}$ are both proportional to $\cY^c_{r\pm}$\,:
\ba
	&& (\cP^c_{1\pm}\,C^{[\D]})^{a(m\pm1),b(n)}\,\frac{u_{a_1}\cdots u_{a_{m\pm1}}}{(m\pm1)!}\,\frac{v_{b_1}\cdots v_{b_n}}{n!}
	=
	p^{[\D]m,n}_{1\pm}\,\cY^c_{1\pm}\,C^{[\D](m,n)}(u,v)\,,
	\nn
	&& (\cP^c_{2\pm}\,C^{[\D]})^{a(m),b(n\pm1)}\,\frac{u_{a_1}\cdots u_{a_m}}{m!}\,\frac{v_{b_1}\cdots v_{b_{n\pm1}}}{(n\pm1)!}
	=
	p^{[\D]m,n}_{2\pm}\,\cY^c_{2\pm}\,C^{[\D](m,n)}(u,v)\,,
	\nn
	&&(\cK^c_{1\pm}\,C^{[\D]})^{a(m\pm1),b(n)}\,\frac{u_{a_1}\cdots u_{a_{m\pm1}}}{(m\pm1)!}\,\frac{v_{b_1}\cdots v_{b_n}}{n!}=
	k^{[\D]m,n}_{1\pm}\,\cY^c_{1\pm}\,C^{[\D](m,n)}(u,v)\,,
	\nn
	&&(\cK^c_{2\pm}\,C^{[\D]})^{a(m),b(n\pm1)}\,\frac{u_{a_1}\cdots u_{a_m}}{m!}\,\frac{v_{b_1}\cdots v_{b_{n\pm1}}}{(n\pm1)!}=
	k^{[\D]m,n}_{2\pm}\,\cY^c_{2\pm}\,C^{[\D](m,n)}(u,v)\,,
	\label{P K coeff}
\ea
where $p^{[\D]m,n}_{r\pm}$ and $k^{[\D]m,n}_{r\pm}$ are
the proportionality constants.
We will determine these constants  by asking the operators $\hat P^a$ and $\hat K^a$,
defined by \eqref{P def}, \eqref{K def} and \eqref{P K coeff}, satisfy the conditions \eqref{condition} which arose from the Bianchi identities of the zero-form equations
and implies that the operators $\hat P^a$ and $\hat K^a$ form a representation of conformal algebra $\mathfrak{so}(2,d)$
together with $\hat J^{ab}$ and $\hat D$\,.
Note that $\cP^a_{r\pm}$ and $\cK^a_{r\pm}$ can be viewed 
as differential operators acting on $C^{[\D](m,n)}(u,v)$, 
whereas $\hat P^a$ and $\hat K^a$ cannot because
they alter the conformal dimensions $\D$.

Firstly, the condition $[\hat P_a,\hat P_b]=0$ gives
\ba
p^{[\D-1]m,n-1}_{1-}\,p^{[\D]m,n}_{2-} 
- \tfrac{m-n}{m-n+1}\, p^{[\D-1]m-1,n}_{2-}\,p^{[\D]m,n}_{1-}\eq 0\,,\label{PP1}\\
p^{[\D-1]m+1,n}_{2+}\,p^{[\D]m,n}_{1+} 
- \tfrac{m-n}{m-n+1}\, p^{[\D-1]m,n+1}_{1+}\,p^{[\D]m,n}_{2+}\eq 0\,,\label{PP1_2}\\
p^{[\D-1]m-1,n}_{2+}\,p^{[\D]m,n}_{1-}
- \tfrac{d+m+n-2}{d+m+n-3}\,p^{[\D-1]m,n+1}_{1-}\,p^{[\D]m,n}_{2+}\eq0 \,,\label{PP2}\\
p^{[\D-1]m,n-1}_{1+}\,p^{[\D]m,n}_{2-}
- \tfrac{d+m+n-2}{d+m+n-3}\, p^{[\D-1]m+1,n}_{2-}\,p^{[\D]m,n}_{1+}\eq 0 \,,\label{PP3}\\
p^{[\D-1]m-1,n}_{1+}\,p^{[\D]m,n}_{1-}
- \tfrac{d+2m}{d+2m-2}\, p^{[\D-1]m+1,n}_{1-}\,p^{[\D]m,n}_{1+}&&\nn
+\,\tfrac{d+2n-2}{(m-n+1)(d+m+n-3)}\,p^{[\D-1]m,n+1}_{2-}\, p^{[\D]m,n}_{2+}\eq 0\,,
 \label{PP4}\\
p^{[\D-1]m,n-1}_{2+}\,p^{[\D]m,n}_{2-}
-\tfrac{d+2m}{(m-n+1)(d+m+n-3)}\, p^{[\D-1]m+1,n}_{1-}\,p^{[\D]m,n}_{1+} &&\nn
-\tfrac{(m-n)(m-n+2)(d+2n-2)}{(m-n+1)^2(d+2n-4)}\, p^{[\D-1]m,n+1}_{2-}\,p^{[\D]m,n}_{2+}\eq 0\,. \label{PP5}
\ea
Secondly, the condition $[\hat K_a,\hat P_b]=\eta_{ab}\,\hat D-\hat J_{ab}$ gives
two kinds of equations: homogeneous ones and inhomogeneous ones.
The homogeneous equations are
\ba 
k^{[\D-1]m\pm1,n}_{1\pm}\,p^{[\D]m,n}_{1\pm} -p^{[\D+1]m\pm1,n}_{1\pm}\,k^{[\D]m,n}_{1\pm}=0\,, 
\label{KP0 first} \\
k^{[\D-1]m,n\pm1}_{2\pm}\,p^{[\D]m,n}_{2\pm} - p^{[\D+1]m,n\pm1}_{2\pm}\,k^{[\D]m,n}_{2\pm}=0\,, 
\ea
and
\ba
k^{[\D-1]m,n-1}_{1-}\,p^{[\D]m,n}_{2-} 
- p^{[\D+1],m-1,n}_{2-}\,k^{[\D]m,n}_{1-}
+\tfrac{1}{m-n+1}\,k^{[\D-1]m-1,n}_{2-}\,p^{[\D],m,n}_{1-} \eq 0 \,,\\
\tfrac{m-n}{m-n+1}\,k^{[\D-1]m,n+1}_{1+}\,p^{[\D]m,n}_{2+}
- \tfrac{1}{m-n+2}\,k^{[\D-1]m+1,n}_{2+}\,p^{[\D],m,n}_{1+} && \nn
-\tfrac{m-n+1}{m-n+2}\,p^{[\D+1]m+1,n}_{2+}\,k^{[\D]m,n}_{1+} \eq 0\,, \\
p^{[\D+1]m,n+1}_{1-}\,k^{[\D]m,n}_{2+} - k^{[\D-1]m-1,n}_{2+}\,p^{[\D]m,n}_{1-}
+\tfrac1{d+m+n-3}\,k^{[\D-1]m,n+1}_{1-}\,p^{[\D]m,n}_{2+}\eq 0\,. \\
k^{[\D-1]m+1,n}_{2-}\,p^{[\D]m,n}_{1+} - p^{[\D+1]m,n-1}_{1+}\,k^{[\D]m,n}_{2-}
+\tfrac1{d+m+n-3}\,p^{[\D+1]m+1,n}_{2-}\,k^{[\D]m,n}_{1+}\eq 0\,. \\ 
(k^{[\D-1]}\,p^{[\D]}\ \leftrightarrow\ p^{[\D+1]}\,k^{[\D]})\,, \label{KP0 last}\hspace{80pt}&&
\ea 
The inhomogeneous equations are
\ba
	k^{[\D-1]m-1,n}_{1+}p^{[\D]m,n}_{1-} -  \tfrac{2}{d+2m-2}\,k^{[\D-1]m+1,n}_{1-}\,p^{[\D]m,n}_{1+} &&\nn
	 -\,\tfrac{m-n-1}{(m-n+1)(d+m+n-3)}\,k^{[\D-1]m,n+1}_{2-}\,p^{[\D]m,n}_{2+} 
	-p^{[\D+1]m+1,n}_{1-}\,k^{[\D]m,n}_{1+} && \nn
	+\,\tfrac{1}{m-n+1}\,p^{[\D+1]m,n+1}_{2-}\,k^{[\D]m,n}_{2+} = -1\,,&&
	\label{KP1}
\ea
\ba
	k^{[\D-1]m,n-1}_{2+}\,p^{[\D]m,n}_{2-} 
	-\tfrac{2(m-n)(m-n+2)}{(m-n+1)^2(d+2n-4)}\,k^{[\D-1]m,n+1}_{2-}\,p^{[\D]m,n}_{2+} &&\nn
	 -\tfrac{m-n+3}{(m-n+1)(d+m+n-3)} \,k^{[\D-1]m+1,n}_{1-}\,p^{[\D]m,n}_{1+}  
	 -\tfrac{(m-n)(m-n+2)}{(m-n+1)^2}\,p^{[\D+1]m,n+1}_{2-}k^{[\D]m,n}_{2+}&& \nn
	 - \tfrac1{m-n+1}\,p^{[\D+1]m+1,n}_{1-}\,k^{[\D]m,n}_{1+}= -\tfrac{m-n+2}{m-n+1}\,,&& \label{KP2}
\ea
\be
	(k^{[\D-1]}\,p^{[\D]}\ \leftrightarrow\ p^{[\D+1]}\,k^{[\D]})\,,\label{KP3}
\ee
and 
\ba
	 &&k^{[\D-1]m+1,n}_{1-}p^{[\D]m,n}_{1+} - p^{[\D+1]m+1,n}_{1-}k^{[\D]m,n}_{1+}  \nn
	&&+ \tfrac{m-n}{m-n+1}\,(k^{[\D-1]m,n+1}_{2-}p^{[\D]m,n}_{2+} - p^{[\D+1]m,n+1}_{2-}k^{[\D]m,n}_{2+}) = -\D\,, 
	\label{KP4}
\ea
Finally, $[\hat K_a,\hat K_b]=0$ gives
\ba
k^{[\D+1]m,n-1}_{1-}\,k^{[\D]m,n}_{2-} 
- \tfrac{m-n}{m-n+1}\, k^{[\D+1]m-1,n}_{2-}\,k^{[\D]m,n}_{1-}\eq 0\,,\label{KK1}\\
k^{[\D+1]m+1,n}_{2+}\,k^{[\D]m,n}_{1+} 
- \tfrac{m-n}{m-n+1}\, k^{[\D+1]m,n+1}_{1+}\,k^{[\D]m,n}_{2+}\eq 0\,,\label{KK1_2}\\
k^{[\D+1]m-1,n}_{2+}\,k^{[\D]m,n}_{1-}
- \tfrac{d+m+n-2}{d+m+n-3}\,k^{[\D+1]m,n+1}_{1-}\,k^{[\D]m,n}_{2+}\eq0 \,,\label{KK2}\\
k^{[\D+1]m,n-1}_{1+}\,k^{[\D]m,n}_{2-}
- \tfrac{d+m+n-2}{d+m+n-3}\, k^{[\D+1]m+1,n}_{2-}\,k^{[\D]m,n}_{1+}\eq 0 \,,\label{KK3}\\
k^{[\D+1]m-1,n}_{1+}\,k^{[\D]m,n}_{1-}
- \tfrac{d+2m}{d+2m-2}\, k^{[\D+1]m+1,n}_{1-}\,k^{[\D]m,n}_{1+}&&\nn
+\,\tfrac{d+2n-2}{(m-n+1)(d+m+n-3)}\,k^{[\D+1]m,n+1}_{2-}\, k^{[\D]m,n}_{2+}\eq 0\,,
\label{KK4}\\
k^{[\D+1]m,n-1}_{2+}\,k^{[\D]m,n}_{2-}
-\tfrac{d+2m}{(m-n+1)(d+m+n-3)}\, k^{[\D+1]m+1,n}_{1-}\,k^{[\D]m,n}_{1+} &&\nn
-\tfrac{(m-n)(m-n+2)(d+2n-2)}{(m-n+1)^2(d+2n-4)}\, k^{[\D+1]m,n+1}_{2-}\,k^{[\D]m,n}_{2+}\eq 0\,. \label{KK5}
\ea
When identifying consistent sets of equations for zero-form fields $C^{[\D]m,n}$,
one needs to take into account the ambiguities (or redundancies) of field redefinitions,
\be
	C^{[\D](m,n)}\quad \longrightarrow \quad \rho^{[\D]m,n}\,C^{[\D](m,n)}\,.
\ee
This would affect the action of $\hat P_a$ and $\hat K_a$ as
\be
	(p^{[\D]m,n}_{1\pm}, p^{[\D]m,n}_{2\pm})
	\quad \longrightarrow\quad
	\left(\frac{\rho^{[\D]m,n}}{\rho^{[\D-1]m\pm1,n}}\,p^{[\D]m,n}_{1\pm},
	\frac{\rho^{[\D]m,n}}{\rho^{[\D-1]m,n\pm1}}\,p^{[\D]m,n}_{2\pm}\right),
\ee
and 
\be
	(k^{[\D]m,n}_{1\pm},k^{[\D]m,n}_{2\pm})
	\quad \longrightarrow\quad
	\left(\frac{\rho^{[\D]m,n}}{\rho^{[\D+1]m\pm1,n}}\,k^{[\D]m,n}_{1\pm},
	\frac{\rho^{[\D]m,n}}{\rho^{[\D+1]m,n\pm1}}\,k^{[\D]m,n}_{2\pm}\right).
\ee

\subsection{Solutions}

We can first consider the equations \eqref{PP1}--\eqref{PP5} from $[\hat P_a,\hat P_b]=0$\,,
which are homogeneous equations for the $p^{[\D]m,n}_{r\pm}$ coefficients.
One can first observe that the `seed' zero-form $C^{[2](2,2)}$
generate other zero-forms with integer $\D\ge 2$
and their Lorentz label $(m,n)$ are restricted since they
should be related to $(2,2)$ by actions of $\D-2$ one-cell operators.
A simple reasoning reveals the following. For even and odd $\D$, the admissible Lorentz labels 
of zero-forms are depicted as the black bullets in the $(m,n)$ lattice of Figure \ref{fig: even 0f content}
and Figure \ref{fig: odd 0f content}, respectively.
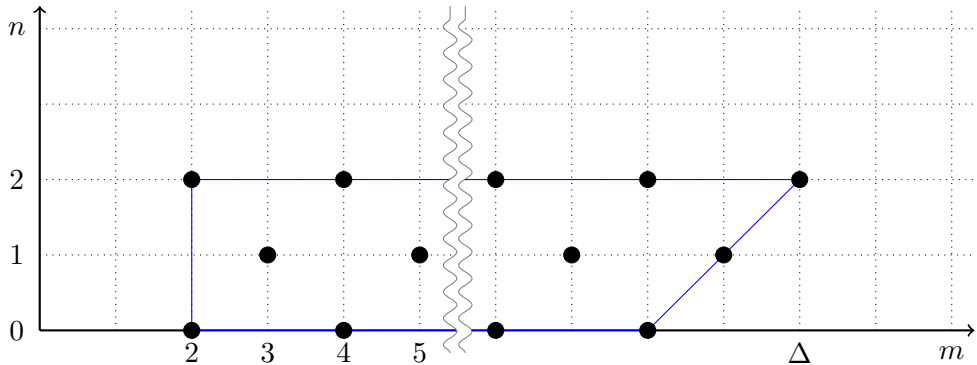
\begin{figure}[H]
	\centering
	\begin{tikzpicture}
	\draw[thick](0,0)--(5.48,0);\draw[thick,->](5.68,0)--(12.3,0);
	\draw[thick,->](0,0)--(0,4.3);
	\draw[dotted](0,1)--(5.46,1);\draw[dotted](5.66,1)--(12.3,1);
	\draw[dotted](0,2)--(5.4,2);\draw[dotted](5.6,2)--(12.3,2);
	\draw[dotted](0,3)--(5.4,3);\draw[dotted](5.6,3)--(12.3,3);
	\draw[dotted](0,4)--(5.4,4);\draw[dotted](5.6,4)--(12.3,4);
	\draw[dotted](1,0)--(1,4.3);\draw[dotted](2,0)--(2,4.3);\draw[dotted](3,0)--(3,4.3);
		\draw[dotted](4,0)--(4,4.3);\draw[dotted](5,0)--(5,4.3);\draw[dotted](6,0)--(6,4.3);
	\draw[dotted](7,0)--(7,4.3);\draw[dotted](8,0)--(8,4.3);\draw[dotted](9,0)--(9,4.3);
	\draw[dotted](10,0)--(10,4.3);\draw[dotted](11,0)--(11,4.3);\draw[dotted](12,0)--(12,4.3);
	 \path [draw=gray,snake it] (5.4,-0.3) -- (5.4,4.3);
	\path [draw=gray,snake it] (5.6,-0.3) -- (5.6,4.3);
		\draw[blue] (5.59,2)--(10,2)--(8,0)--(5.68,0);
		\draw[blue](5.39,2)--(2,2)--(2,0)--(5.48,0);
	\filldraw[black] (2,0) circle (3pt);\filldraw[black] (4,0) circle (3pt);\filldraw[black] (6,0) circle (3pt);
	\filldraw[black] (8,0) circle (3pt);
	\filldraw[black] (3,1) circle (3pt);\filldraw[black] (5,1) circle (3pt);\filldraw[black] (7,1) circle (3pt);\filldraw[black] (9,1) circle (3pt);
	\filldraw[black] (2,2) circle (3pt);\filldraw[black] (4,2) circle (3pt);\filldraw[black] (6,2) circle (3pt);
	\filldraw[black] (8,2) circle (3pt);\filldraw[black] (10,2) circle (3pt);
	\node at (-0.3,0) {0};\node at (-0.3,1) {1};\node at (-0.3,2) {2};\node at (-0.3,4) {$n$};
		\node at (2,-0.3) {2};\node at (3,-0.3) {3};\node at (4,-0.3) {4};\node at (5,-0.3) {5};
		\node at (10,-0.3) {$\D$};\node at (12,-0.3) {$m$};	
	\end{tikzpicture}
	\caption{Admissible Lorentz labels of zero-forms for even $\D$}\label{fig: even 0f content}
\end{figure}
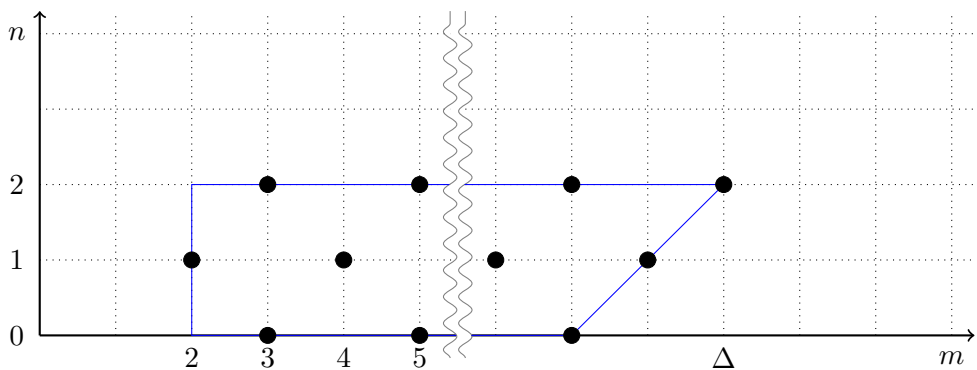
\begin{figure}[H]
	\centering
	\begin{tikzpicture}	
	\draw[thick](0,0)--(5.48,0);\draw[thick,->](5.68,0)--(12.3,0);
	\draw[thick,->](0,0)--(0,4.3);
	\draw[dotted](0,1)--(5.46,1);\draw[dotted](5.66,1)--(12.3,1);
	\draw[dotted](0,2)--(5.4,2);\draw[dotted](5.6,2)--(12.3,2);
	\draw[dotted](0,3)--(5.4,3);\draw[dotted](5.6,3)--(12.3,3);
	\draw[dotted](0,4)--(5.4,4);\draw[dotted](5.6,4)--(12.3,4);
	\draw[dotted](1,0)--(1,4.3);\draw[dotted](2,0)--(2,4.3);\draw[dotted](3,0)--(3,4.3);
		\draw[dotted](4,0)--(4,4.3);\draw[dotted](5,0)--(5,4.3);\draw[dotted](6,0)--(6,4.3);
	\draw[dotted](7,0)--(7,4.3);\draw[dotted](8,0)--(8,4.3);\draw[dotted](9,0)--(9,4.3);
	\draw[dotted](10,0)--(10,4.3);\draw[dotted](11,0)--(11,4.3);\draw[dotted](12,0)--(12,4.3);
	 \path [draw=gray,snake it] (5.4,-0.3) -- (5.4,4.3);
	\path [draw=gray,snake it] (5.6,-0.3) -- (5.6,4.3);
	\draw[blue] (5.59,2)--(9,2)--(7,0)--(5.68,0);
		\draw[blue](5.39,2)--(2,2)--(2,0)--(5.48,0);	
	\filldraw[black] (3,0) circle (3pt);\filldraw[black] (5,0) circle (3pt);\filldraw[black] (7,0) circle (3pt);
	\filldraw[black] (2,1) circle (3pt);\filldraw[black] (4,1) circle (3pt);\filldraw[black] (6,1) circle (3pt);\filldraw[black] (8,1) circle (3pt);
	\filldraw[black] (3,2) circle (3pt);\filldraw[black] (5,2) circle (3pt);\filldraw[black] (7,2) circle (3pt);
	\filldraw[black] (9,2) circle (3pt);
		\node at (-0.3,0) {0};\node at (-0.3,1) {1};\node at (-0.3,2) {2};\node at (-0.3,4) {$n$};
		\node at (2,-0.3) {2};\node at (3,-0.3) {3};\node at (4,-0.3) {4};\node at (5,-0.3) {5};
		\node at (9,-0.3) {$\D$};\node at (12,-0.3) {$m$};	
		\end{tikzpicture}
	\caption{Admissible Lorentz labels of zero-forms for odd $\D$}\label{fig: odd 0f content}
\end{figure} 
\noindent In other words, the conformal dimension $\D$ of the zero-forms with Lorentz label $(m,n)$ is
restricted to the ones with non-negative integer depth \eqref{depth}:
\be
	\D=m-n+2\,\delta\,, \qquad \d=1,2,\ldots.
\ee 
In the following we shall label the conformal dimension in terms of the depth $\delta$.
The above content of zero-forms coincides
with the basis tensors of Weyl invariants
used in  \cite{Boulanger:2004eh}.

When we build the $\hat P_a$ and $\hat K_a$ action, we assumed that there exist one zero-form
for each $\{\d\}(m,n)$ label, which is now restricted to admissible ones.
 This `multiplicity-one' assumption can be verified by demonstrating that the recurrence relations arising from $[\hat P_a,\hat P_b]=0$
 uniquely determines  $p^{\{\delta\}m,n}_{r\pm}$ after fixing all field redefinition ambiguities.  
As mentioned before, the content of zero-form is determined solely by the $\hat P_a$ action
because  if we linearize the system around the flat background, we only obtain the condition $[\hat P_a,\hat P_b]=0$\,.
The other consistency conditions will be served to determine the $\hat K_a$ action together with the value $\D$. 

Let us demonstrate how the recurrence relations \eqref{PP1}--\eqref{PP5} can be solved.
First, using the field redefinitions $\rho^{\{\d\}m,n}$ with $m\ge 3$, we fix all $p^{\{\d\}m,n}_{1-}$
	with $m\ge 3$ to 1. Note that this leaves out the field redefinitions
	$\rho^{\{\d\}2,n}$ as the corresponding coefficients $p^{\{\d\}2,n}_{1-}$ already vanish.
Then, the equations \eqref{PP1} and \eqref{PP2} determine $p^{\{\d\}m,n}_{2\pm}$ 
in terms of $p^{\{\d\}2,n}_{2\pm}$ as
\ba
	 &p^{\{\d\}m,n}_{2+} =\tfrac{d+n}{d+m+n-2} \,p^{\{\d\}2,n}_{2+}\qquad &[n=0,1]\,, \nn
	&p^{\{\d\}m,n}_{2-} = \tfrac{3-n}{m-n+1}\,p^{\{\d\}2,n}_{2-} \qquad &[n=1,2]\,. \label{ppsol:2+,2-}
 \ea
The remaining equations \eqref{PP3}--\eqref{PP5} determine also $p^{\{\d\}m,n}_{1+}$ in terms of $p^{\{\d\}2,n}_{2\pm}$ as
\ba
&& p^{\{\d\}m,2}_{1+} = \tfrac{d+1}{d+2m}\,p^{\{\d\}2,2}_{2-}\,p^{\{\d-1\}2,1}_{2+} \,,\nn
&& p^{\{\d\}m,n}_{1+}= \tfrac{(m-1)(d+m)(d+n)(d+2n-2)}{(d+2m)(m-n+1)(d+n-1)(d+m+n-2)}\,
p^{\{\d\}2,n+1}_{2-}\,p^{\{\d\}2,n}_{2+}\qquad [n=0,1]\,, \nn 
&& \label{ppsol: 1+}
\ea
and impose the constraint on $p^{\{\d\}2,n}_{2\pm}$,
\be
p^{\{\d\}2,2}_{2-}\,p^{\{\d\}2,1}_{2+}
	=\tfrac{d\,(d-2)}{(d-1)(d+1)}\,
	p^{\{\d\}2,1}_{2-}\,p^{\{\d-1\}2,0}_{2+}
	\qquad [\delta\ge2]\,.
	\label{pp const}
\ee
In the end, we are left with $p^{\{\d\}2,2}_{2-},\ p^{\{\d\}2,1}_{2+},\ p^{\{\d\}2,1}_{2-},\ p^{\{\d\}2,0}_{2+}$ subject to the above equations.
For $\delta=1$, we have only two non-zero coefficients $p^{\{1\}2,1}_{2+}$ and $p^{\{1\}2,0}_{2+}$ 
and they can be fixed by the field redefinitions $\rho^{\{1\}2,1}$ and $\rho^{\{1\}2,0}$, respectively.
If all the coefficients and field redefinitions, except for $\rho^{\{1\}2,2}$, are fixed up to an order $\delta-1$ with $\delta\ge1$,
we can determine the four coefficients 
$p^{\{\d\}2,2}_{2-},\ p^{\{\d\}2,1}_{2+},\ p^{\{\d\}2,1}_{2-},\ p^{\{\d\}2,0}_{2+}$
by using the equation \eqref{pp const} and the three field redefinitions $\rho^{\{\d\}2,2},\  \rho^{\{\d\}2,1},\ \rho^{\{\delta\}2,0}$.
By induction, we can determine all the coefficients $p^{\{\d\}2,n}_{2\pm}$
and hence the $\hat P_a$ action.
In doing this, we used all the field redefinitions except for $\rho^{\{1\}2,2}$.
The latter field redefinition corresponds to the overall rescaling of the homogeneous equation,
and hence it is trivial.
To recapitulate, we have shown that the recurrence relations \eqref{PP1}--\eqref{PP5}
determine uniquely the $p^{[\D]m,n}_{r\pm}$ coefficients, namely the $\hat P_a$ action,
by making use of the freedom of field redefinition.
This proves the correctness of the `multiplicity-one' assumption. 
Let us restate the zero-form field content of the conformal geometry:
\ba
	&& \bigoplus_{\delta=1}^\infty \bigoplus_{m=2}^{\infty} 
	C^{\{\d\}(m,2)}\oplus C^{\{\d\}(m,1)}\oplus C^{\{\d\}(m,0)}\nn
	&&=\bigoplus_{\delta=1}^\infty \bigoplus_{m=2}^{\infty} 
	C^{[m-2+2\d](m,2)}\oplus C^{[m-1+2\delta](m,1)}\oplus C^{[m+2\delta](m,0)}\,.
	\label{0content}
\ea

If we do not aim for conformal geometry, then there are more possibilities:
instead of determining $p^{\{\d\}2,2}_{2-},\ p^{\{\d\}2,1}_{2+},\ p^{\{\d\}2,1}_{2-},\ p^{\{\d\}2,0}_{2+}$ with field redefinitions,
we can set some of them to zero and solve the conditions \eqref{pp const} as $0=0$.
In this case, the fields whose redefinition could fix such $p^{\{\d\}2,n}_{2\pm}$ coefficients
simply decouple from the theory, and hence we can remove them.
Since we have less zero-form fields compared to the off-shell system, the resulting system will be an on-shell theory.
See Appendix \ref{sec: other PP} for more details.

Next we can move to the inhomogeneous equations \eqref{KP1}--\eqref{KP4}
 arising from $[\hat K_a,\hat P_b]=\eta_{ab}\,\hat D-\hat J_{ab}$\,.
They uniquely determine the following quadratics:
for $n=0,1,2$ we obtain
\ba 
	&& p^{[\D+1]m+1,n}_{1-}\,k^{[\D]m,n}_{1+} = \tfrac{(m-1)(d+m)(m-n+2+\D)(d+m+n-2+\D)}{2(m-n+1)(d+2m)(d+m+n-2)}\,, \nn
	&& k^{[\D-1]m+1,n}_{1-}\,p^{[\D]m,n}_{1+} = \tfrac{(m-1)(d+m)(m-n+2-\D)(d+m+n-2-\D)}{2(m-n+1)(d+2m)(d+m+n-2)}\,, \label{1m1p}
\ea
and for $n=0,1$ we obtain
\ba
	&& p^{[\D+1]m,n+1}_{2-}k^{[\D]m,n}_{2+} = \tfrac{(d-1)(n-m+\D)(d+m+n-2+\D)}{(n+1)(m-n)(d+n-2)(d+m+n-2)}\,, \nn
	&& k^{[\D-1]m,n+1}_{2-}p^{[\D]m,n}_{2+} = \tfrac{(d-1)(n-m-\D)(d+m+n-2-\D)}{(n+1)(m-n)(d+n-2)(d+m+n-2)}\,. \label{2m2p}
\ea
Since the $p^{[\D]m,n}_{r\pm}$ are already determined to be non-vanishing numbers, the above
equations alone determine all the $k^{[\D]m,n}_{r\pm}$ coefficients.
One can check that such  $k^{[\D]m,n}_{r\pm}$ 
readily satisfy the remaining equations  \eqref{KP0 first}--\eqref{KP0 last} 
and \eqref{KK1}--\eqref{KK5}
arising respectively from 
$[\hat K_a,\hat P_b]=\eta_{ab}\,\hat D-\hat J_{ab}$
and $[\hat K_a,\hat K_b]=0$.
Therefore, both $\hat P_a$ and $\hat K_a$ action is entirely determined
and hence the linear part of the unfolded equation for conformal geometry.

\begin{figure}[H]
	\centering
	\begin{tikzpicture}	
	\draw[thick](0,0)--(1.47,0);\draw[thick](1.67,0)--(6.47,0);\draw[thick,->](6.67,0)--(12.3,0);
	\draw[thick,->](0,0)--(0,4.3);
	\draw[dotted](0,1)--(1.46,1);\draw[dotted](1.66,1)--(6.46,1);\draw[dotted](6.66,1)--(12.3,1);
	\draw[dotted](0,2)--(1.43,2);\draw[dotted](1.63,2)--(6.42,2);\draw[dotted](6.64,2)--(12.3,2);
	\draw[dotted](0,3)--(1.35,3);\draw[dotted](1.55,3)--(6.35,3);\draw[dotted](6.55,3)--(12.3,3);
	\draw[dotted](0,4)--(1.35,4);\draw[dotted](1.55,4)--(6.35,4);\draw[dotted](6.55,4)--(12.3,4);
	\draw[dotted](1,0)--(1,4.3);\draw[dotted](2,0)--(2,4.3);\draw[dotted](3,0)--(3,4.3);
		\draw[dotted](4,0)--(4,4.3);\draw[dotted](5,0)--(5,4.3);\draw[dotted](6,0)--(6,4.3);
	\draw[dotted](7,0)--(7,4.3);\draw[dotted](8,0)--(8,4.3);\draw[dotted](9,0)--(9,4.3);
	\draw[dotted](10,0)--(10,4.3);\draw[dotted](11,0)--(11,4.3);\draw[dotted](12,0)--(12,4.3);
	 \path [draw=gray,snake it] (1.4,-0.3) -- (1.4,4.3);
	\path [draw=gray,snake it] (1.6,-0.3) -- (1.6,4.3);
	 \path [draw=gray,snake it] (6.4,-0.3) -- (6.4,4.3);
	\path [draw=gray,snake it] (6.6,-0.3) -- (6.6,4.3);
	\draw[blue] (1.6,2)--(6.4,2);\draw[blue] (1.67,0)--(6.47,0);
	\draw[blue] (6.6,2)--(10,2)--(8,0)--(6.67,0);	
	\draw[orange,very thick,<-](3.12,2)--(4,2);
	\draw[orange,very thick,->](4,2)--(4,1.12);
	\draw[orange,very thick,<-](4.12,1)--(5,1);
	\draw[orange,very thick,->](5,1)--(5,0.12);
	\draw[orange,very thick,<-](5.12,0)--(6,0);
	\draw[green,very thick,<-](2.88,2)--(2,2);
	\draw[green,very thick,<-](3,1.88)--(3,1.1);
	\draw[green,very thick,<-](3.88,1)--(3,1);
	\draw[green,very thick,<-](4,0.88)--(4,0);
	\draw[green,very thick,<-](4.88,0)--(4,0);
	
	\draw[red,thick] (3.47,1.9)--(3.67,2.1);\draw[red,thick] (3.67,1.9)--(3.47,2.1);
	\draw[red,thick] (3.9,1.47)--(4.1,1.67);\draw[red,thick] (4.1,1.47)--(3.9,1.67);
	\draw[red,thick] (4.47,0.9)--(4.67,1.1);\draw[red,thick] (4.67,0.9)--(4.47,1.1);
	\draw[red,thick] (4.9,0.47)--(5.1,0.67);\draw[red,thick] (5.1,0.47)--(4.9,0.67);
	\draw[red,thick] (5.47,-0.1)--(5.67,0.1);\draw[red,thick] (5.67,-0.1)--(5.47,0.1);
	\filldraw[black] (3,0) circle (3pt);\filldraw[black] (5,0) circle (3pt);\filldraw[black] (8,0) circle (3pt);
	\filldraw[black] (2,1) circle (3pt);\filldraw[black] (4,1) circle (3pt);\filldraw[black] (6,1) circle (3pt);
	\filldraw[black] (7,1) circle (3pt);\filldraw[black] (9,1) circle (3pt);
	\filldraw[black] (3,2) circle (3pt);\filldraw[black] (5,2) circle (3pt);\filldraw[black] (8,2) circle (3pt);
	\filldraw[black] (10,2) circle (3pt);
	\filldraw[gray] (2,0) circle (3pt);\filldraw[gray] (4,0) circle (3pt);\filldraw[gray] (6,0) circle (3pt);\filldraw[gray] (7,0) circle (3pt);
	\filldraw[gray] (3,1) circle (3pt);\filldraw[gray] (5,1) circle (3pt);\filldraw[gray] (8,1) circle (3pt);
	\filldraw[gray] (2,2) circle (3pt);\filldraw[gray] (4,2) circle (3pt);\filldraw[gray] (6,2) circle (3pt);
	\filldraw[gray] (7,2) circle (3pt);\filldraw[gray] (9,2) circle (3pt);
		\node at (-0.3,0) {0};\node at (-0.3,1) {1};\node at (-0.3,2) {2};\node at (-0.3,4) {$n$};
		\node at (5,-0.3) {$\ell\!+\!2$};\node at (10,-0.3) {$\D\!=\!d\!+\!\ell$};	\node at (12,-0.3) {$m$};	
		\end{tikzpicture}
	\caption{Vanishing $k$ cofficients}\label{fig: vanishing coeff}
\end{figure}
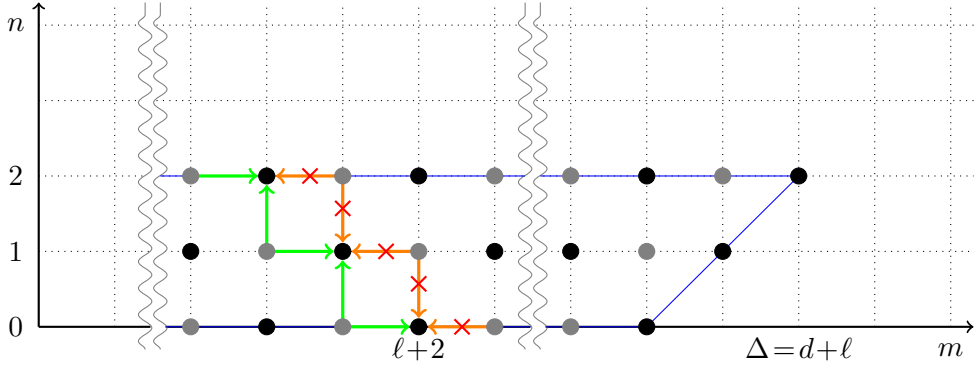

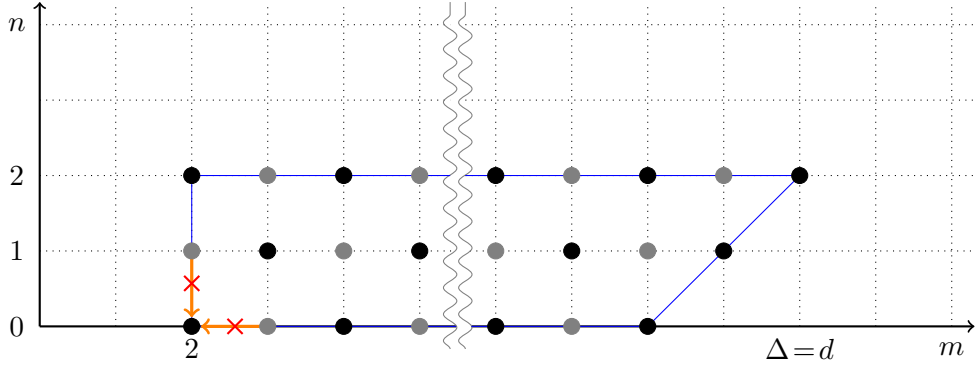
\begin{figure}[H]
	\centering
	\begin{tikzpicture}
	\draw[thick](0,0)--(5.48,0);\draw[thick,->](5.68,0)--(12.3,0);
	\draw[thick,->](0,0)--(0,4.3);
	\draw[dotted](0,1)--(5.46,1);\draw[dotted](5.66,1)--(12.3,1);
	\draw[dotted](0,2)--(5.4,2);\draw[dotted](5.6,2)--(12.3,2);
	\draw[dotted](0,3)--(5.4,3);\draw[dotted](5.6,3)--(12.3,3);
	\draw[dotted](0,4)--(5.4,4);\draw[dotted](5.6,4)--(12.3,4);
	\draw[dotted](1,0)--(1,4.3);\draw[dotted](2,0)--(2,4.3);\draw[dotted](3,0)--(3,4.3);
		\draw[dotted](4,0)--(4,4.3);\draw[dotted](5,0)--(5,4.3);\draw[dotted](6,0)--(6,4.3);
	\draw[dotted](7,0)--(7,4.3);\draw[dotted](8,0)--(8,4.3);\draw[dotted](9,0)--(9,4.3);
	\draw[dotted](10,0)--(10,4.3);\draw[dotted](11,0)--(11,4.3);\draw[dotted](12,0)--(12,4.3);
	 \path [draw=gray,snake it] (5.4,-0.3) -- (5.4,4.3);
	\path [draw=gray,snake it] (5.6,-0.3) -- (5.6,4.3);
		\draw[blue] (5.6,2)--(10,2)--(8,0)--(5.68,0);
		\draw[blue](5.4,2)--(2,2)--(2,0)--(5.48,0);
	\draw[orange,very thick,->](2,1)--(2,0.12); 
	\draw[orange,very thick,<-](2.12,0)--(3,0);
	\draw[red,thick] (1.9,0.47)--(2.1,0.67);\draw[red,thick] (2.1,0.47)--(1.9,0.67);
	\draw[red,thick] (2.47,-0.1)--(2.67,0.1);\draw[red,thick] (2.67,-0.1)--(2.47,0.1);
	\filldraw[black] (2,0) circle (3pt);\filldraw[black] (4,0) circle (3pt);\filldraw[black] (6,0) circle (3pt);
	\filldraw[black] (8,0) circle (3pt);
	\filldraw[black] (3,1) circle (3pt);\filldraw[black] (5,1) circle (3pt);\filldraw[black] (7,1) circle (3pt);\filldraw[black] (9,1) circle (3pt);
	\filldraw[black] (2,2) circle (3pt);\filldraw[black] (4,2) circle (3pt);\filldraw[black] (6,2) circle (3pt);
	\filldraw[black] (8,2) circle (3pt);\filldraw[black] (10,2) circle (3pt);
	\filldraw[gray] (3,0) circle (3pt);\filldraw[gray] (5,0) circle (3pt);\filldraw[gray] (7,0) circle (3pt);
	\filldraw[gray] (2,1) circle (3pt);\filldraw[gray] (4,1) circle (3pt);\filldraw[gray] (6,1) circle (3pt);\filldraw[gray] (8,1) circle (3pt);
	\filldraw[gray] (3,2) circle (3pt);\filldraw[gray] (5,2) circle (3pt);\filldraw[gray] (7,2) circle (3pt);
	\filldraw[gray] (9,2) circle (3pt);	\node at (-0.3,0) {0};\node at (-0.3,1) {1};\node at (-0.3,2) {2};\node at (-0.3,4) {$n$};
		\node at (2,-0.3) {2};
		\node at (10,-0.3) {$\D\!=\!d$};\node at (12,-0.3) {$m$};	
	\end{tikzpicture}
	\caption{Decoupling of $C^{[d](2,0)}$}\label{fig: decoupling}
\end{figure}

At this point, one can observe 
from \eqref{1m1p} and \eqref{2m2p}
that $p^{[\D+1]m+1,n}_{1-}\,k^{[\D]m,n}_{1+}$ and $p^{[\D+1]m,n+1}_{2-}k^{[\D]m,n}_{2+}$
never vanish but
$k^{[\D-1]m+1,n}_{1-}\,p^{[\D]m,n}_{1+}$ and $k^{[\D-1]m,n+1}_{2-}\,p^{[\D]m,n}_{2+}$ can vanish
for some $\D, m, n$.
Since $p^{[\D]m,n}_{r\pm}$ are not zero,
we find that the coefficients 
$k^{[\D-1]m,n}_{r-}$ (depicted as orange arrows in Figure \ref{fig: vanishing coeff}) should vanish.
\noindent For a generic $\ell$, the black bullets, which have vanishing $k^{[\D-1]m,n}_{r-}$ coefficients,
still keeps  non-vanishing $k^{[\D-1]m,n}_{r+}$ coefficients (depicted as green arrows in Figure \ref{fig: vanishing coeff}),
and hence the corresponding zero-forms $C^{[d+\ell](m,n)}$ can be still obtained from $C^{[d+\ell-1](m-1,n)}$
or  $C^{[d+\ell-1](m,n-1)}$ by the $\hat K_a$ action.
On the contrary, for $\ell=0$, the zero-forms $C^{[d](2,0)}$ 
can never be obtained from $C^{[d-1](m,n)}$ by the $\hat K_a$ action: See Figure \ref{fig: decoupling}.
\noindent Therefore, the field $C^{[d](2,0)}$ cannot be obtained from a $\hat K$ action.
In fact, the tensor $C^{[d](2,0)}$ corresponds to the Bach tensor,
and the decoupling of $C^{[d](2,0)}$ under $\hat K_a$ action assures 
that one can consistently  impose  the Bach-flat condition $C^{[d](2,0)}=0$.
As mentioned earlier, the zero-form fields  carry certain dual representations of highest weight representations 
of $\mathfrak{so}(2,d)$.
The decoupling of $C^{[d](2,0)}$ implies that its dual state is a lowest $\hat D$ state
and generates an invariant sub-representation, which can be quotiented out
to get an ``on-shell'' representation.
In order to understand better the relation between the zero-form module and the highest weight representation
of $\mathfrak{so}(2,d)$,
we review a few results of the representation theory in the following section.
\begin{figure}[h]
\includegraphics[width=\textwidth]{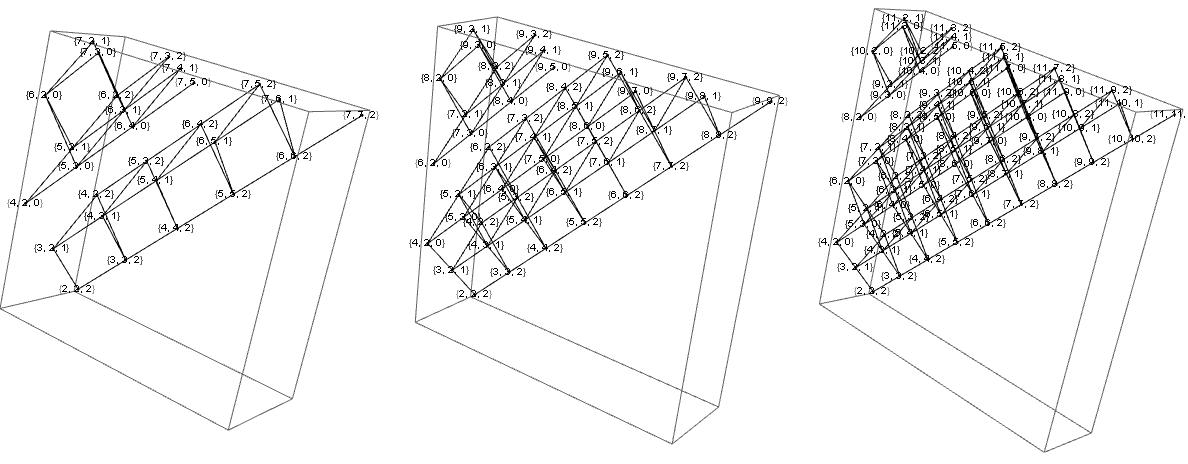}
\caption{$\hat K$ actions on zero-form fields in $4, 6, 8$ dimensions are depicted as the edges of graph.}
\label{fig:Kactions_on_zeroforms}
\end{figure}

\section{Representation}
\label{sec: representation}

\subsection{Off-shell Fradkin-Tseytlin module}
\label{sec: FT module}

The actions of $\hat P_a$ and $\hat K_a$ identified in the previous section,
together with those of $\hat J_{ab}$ and $\hat D$,
define a $\mathfrak{so}(2,d)$-representation realized on the space of 
zero-forms $C^{[\D]a(m),b(n)}$ with field content \eqref{0content}.
In this section, we demonstrate
how one can recover the same field contents
from an analysis of $\mathfrak{so}(2,d)$ representations. 
For that, let us review a few basics:
the (generalized) Verma module $\cV(\D,\mathbb Y)$
of $\mathfrak{so}(2,d)$ is given by
\be
	\cV(\D,\mathbb Y)= \bigoplus_{n,m=0}^\infty 
	[\D+n+2m,\mathbb Y\otimes (n)]
\ee
where $[\Delta,\mathbb Y]$ is the lowest-weight representation (or primary state),
\be
	\hat K_a\,[\Delta,\mathbb Y]=0\,,
\ee
which carries a finite-dimensional 
irrep of $\mathfrak{so}(2)\oplus \mathfrak{so}(d)$. 
We denote the Young diagram $\mathbb Y$ as a row vector,
\be
	(n,m,1^k)=(n,m,\underbrace{1,\ldots,1}_k)\,.
\ee
 The Bernstein-Gelfand-Gelfand resolution provides
various (non-unitary) representations of $\mathfrak{so}(2,d)$ 
as a successive quotient of Verma modules $\cV(\D,\mathbb Y)$
\cite{Shaynkman:2004vu,Beccaria:2014jxa} (see also \cite{Basile:2018eac}).
The spin-$s$ Fradkin-Tseytlin (FT) module,
\be
	\cD(2,(s,s))=\cS(2-s,(s))\ominus \cD(s+d-2,(s))\,,
\ee
with
\ba
	&\cS(2-s,(s))= \cV(2-s,(s))\ominus \cV(1-s,(s-1))\oplus\cD(1-s,(s-1))\,,
	\nn 
	&\cD(s+d-2,(s))=\cV(s+d-2,(s))\ominus\cV(s+d-1,(s-1))\,,
\ea
is the one related to the on-shell conformal spin-$s$ field,
but what we need is the module related to the off-shell conformal spin $s$,
in particular the off-shell conformal spin two,
namely  conformal geometry.
Above, $\cD(1-s,(s-1))$ is the module of the spin-$s$ conformal Killing tensors.
Note that the module associated with an on-shell system is the quotient of the 
module associated with its off-shell system
by the module associated with the equation of motion. 
For the conformal geometry with $s=2$, the equation of motion is the $d$-derivative Bach equation.
Therefore, conformal geometry must be associated with
the module $\cS(0,(2))$ 
which  does not involve any $d$ dependent quotient.
The logic extends to other spins,
and $\cS(2-s,(s))$ is the off-shell FT module,
also sometimes referred to as the shadow module.
Let us decompose $\cS(2-s,(s))$ into 
$\mathfrak{so}(2)\oplus \mathfrak{so}(d)$ modules $[\D,\mathbb Y]$ for $s=2$.
The spin-two conformal Killing tensors are nothing
but the $\mathfrak{so}(2,d)$ adjoint representation,
\be
	\cD(-1,(1))=[-1,(1)]\oplus [0,(1,1)]\oplus [0,(0)]\oplus [1,(1)]\,,
\ee
and hence we find 
\ba
	&&\cS(0,(2))= \cV(0,(2))\ominus\cV(-1,(1))\oplus \cD(-1,(1))
		\label{S decomp}\\
	&&=\, 
	\bigoplus_{m,n=0}^\infty
	[2+n+2m,(n+2,2)]\oplus [3+n+2m,(n+2,1)]
	\oplus [4+n+2m,(n+2)]\,.\nonumber
\ea
The same result can be also obtained from
\be
	\cS(0,(2))=\bigoplus_{k=0} (-1)^k\,\cV(2+k,(2,2,1^k))\,,
	\label{FT module}
\ee
where the successive quotients represent the implementation of the Bianchi identity
and the Bianchi identity of the Bianchi identity etc.
The decomposition of the above into $\mathfrak{so}(2)\oplus \mathfrak{so}(d)$ modules
are shown, in Appendix \ref{sec: FT},  to reproduce the result \eqref{S decomp}.

In \eqref{S decomp}, the lowest conformal weight state is $(2,(2,2))$, which is the representation of the Weyl tensor,
and the space coincides with the zero-form field content \eqref{0content}.
Remark that  in the unfolded equation, the zero-form fields carry finite-dimensional representations
of $K=SO(1,1)\times SO(1,d-1)$, which is different from 
$SO(2)\times SO(d)$.
However, they are related by the ``Wick rotation'' on the $0$-th and $d$-th components,
which maps the generators $\hat D=\hat M_{0'd}$ to ${\rm i}\,\hat D$  and $(\hat J_{ij},\hat J_{0i})$ to $(\hat J_{ij},{\rm i}\,\hat J_{0i})$.
To repeat, the Wick rotation maps the Lie algebras
$\mathfrak{so}(1,1)={\rm Span}_{\mathbb R}\{{\rm i}\,\hat D\}$
and $\mathfrak{so}(1,d-1)={\rm Span}_{\mathbb R}\{{\rm i}\,\hat J_{ab}\}$
 into the Lie algebras $\mathfrak{so}(2)={\rm Span}_{\mathbb R}\{\hat D\}$
and $\mathfrak{so}(d)={\rm Span}_{\mathbb R}\{{\rm i}\,\hat J_{ij}, \hat J_{0i} \}$, respectively.
It is important to note here that the Wick rotation does not alter the vector space on which the generators act,
and the same vector space carries a representation of $\mathfrak{so}(1,1)\oplus \mathfrak{so}(1,d-1)$
as well as a representation of $\mathfrak{so}(2)\oplus \mathfrak{so}(d)$.
In a vector space carrying an irreducible unitary representation of $\mathfrak{so}(2)\oplus\mathfrak{so}(d)$,
$\hat D$ has a real eigenvalue $\D$,
and ${\rm i}\,\hat J_{ij}$ and $\hat J_{0i}$ are represented as finite-dimensional Hermitian matrices.
The same vector space carries a finite but non-unitary representation of $\mathfrak{so}(1,1)\oplus \mathfrak{so}(1,d-1)$,
which was used to label the zero-form fields.
Let us note also that
the decomposition of a $G=SO(2,d)$ representation into (infinitely many) non-unitary finite-dimensional representations
of $K$ does not imply
that the  $G$-representation is a non-unitary one.
In fact, a state in the $G$-representation is not 
a finite but an infinite linear combination of finite-dimensional $K$-representations.
If we adopt the standard norm of finite-dimensional representations (which is not positive definite
since $K$ is non-compact),
the norm of a state in the  $G$-representation may diverge.
Therefore, we need to introduce a new scalar product for the $G$-representation, which will be singular for a finite-dimensional $K$-representation.
This new scalar product can be positive definite, though the Fradkin-Tseytlin module is not of this type.
What we discussed just above is analogous to
what happens in the Taylor expansion of $L^2$ space: each Taylor coefficient has a divergent $L^2$ norm,
whereas a $L^2$ state may have a divergent norm with respect to a well-defined scalar product in the space of Taylor coefficients.

Before moving to the next section, let us make a remark
on the relation between $\cS(0,(2))$ and $\cD(d,(2))$,
namely the relation between the off-shell spin-2 FT module (or the spin-2 shadow module) 
and the on-shell massless spin-2 module.
In terms of nonlinear system, such a relation is about the interplay
between the $d$-dimensional conformal geometry and 
the $(d+1)$-dimensional Einstein gravity with negative cosmological constant.
The massless spin-2 module $\cD(d,(2))$
can be branched into $\mathfrak{so}(2)\oplus \mathfrak{so}(d)$ as
\ba
	&&\cD(d,(2))= \cV(d,(2))\ominus\cV(d+1,(1))
		\label{D decomp}\\
	&&=\, 
	\bigoplus_{m,n=0}^\infty
	[d+n+2m,(n+2)]
	\oplus [d+1+n+2m,(n+2,1)]
	\oplus[d+2+n+2m,(n+2,2)]\,.\nonumber
\ea
Comparing \eqref{D decomp} with \eqref{S decomp},
one can see that the two vector spaces are isomorphic as $\mathfrak{so}(d)$ representations,
but have different $\mathfrak{so}(2)$ eigenvalues.
One can obtain the same result \eqref{D decomp} via the $\mathfrak{so}(1,d)$ decomposition of  $\cD(d,(2))$,
\be
	\cD(d,(2))=\bigoplus_{k=0}^\infty 
	(k+2,2)_{\mathfrak{so}(1,d)}\,,
	\label{massless unfold}
\ee 
which is the zero-form content  of massless spin-2 field in its unfolded formulation.
Here, one need to remind that 
the expansion in $k$ should be considered as a Taylor expansion of a function
even though it is not normalizable in Taylor basis
but in a harmonic basis.
Each of the $(k+2,2)_{\mathfrak{so}(1,d)}$ module can be further branched into $so(d)$ irreps as
\be
	(k+2,2)_{\mathfrak{so}(1,d)}=
	\bigoplus_{m=0}^k 
	(k-m+2,2)_{\mathfrak{so}(d)}\oplus 
	(k-m+2,1)_{\mathfrak{so}(d)}\oplus (k-m+2)_{\mathfrak{so}(d)}\,.
	\label{further decomp}
\ee
 The expansion \eqref{massless unfold} with \eqref{further decomp}
 reproduces the decomposition \eqref{D decomp}
 with $m=k-n$, but without the $\mathfrak{so}(2)$ labels.

\subsection{Zero-form module}
\label{sec: rep}

In this section, we provide a more detailed explanation about
the connection between the representations of zero-form fields 
and the off-shell FT module.  
All zero-form fields of conformal geometry can be packed into
\be
	C=\sum_{\D,m,n} C^{[\D]a(m),b(n)}\,\tilde \Psi^{[-\D]}_{a(m),b(n)}\,,
\ee
by contracting them with the basis vectors, denoted by $\tilde\Psi^{[-\D]}_{a(m),b(m)}$, of the underlying representation,
denoted by $\tilde \cS(0,(2))$.
Then, all zero-form equations can be obtained from
\be
	\dd C+\hat A\, C=\cO(C^2)\,,
\ee 
and  the gauge symmetry simply reads 
\be
	\delta C=-(\epsilon^a\,\hat P_a+\l^{ab}\,\hat J_{ab}+\k^a\,\hat K_a+\s\,\hat D)\,C
	+\cO(C^2)\,.
\ee
with
\ba
	\hat D\, \tilde \Psi^{[-\D]}_{a(m),b(n)} \eq -\D\,\tilde \Psi^{[-\D]}_{a(m),b(n)}\,,\nn
	 \hat P_c\,\tilde \Psi^{[-\D]}_{a(m),b(n)} \eq 
	(m+1)\,p^{[\D]m,n}_{1+}\,\tilde \Psi^{[-\D+1]}_{a(m)c,b(n)} \nn 
	&& +\,p^{[\D]m,n}_{2+}
	\,\big(\tfrac{(m-n)(n+1)}{m-n+1}\,\tilde \Psi^{[-\D+1]}_{a(m),b(n)c}  -\tfrac{m\,(n+1)}{m-n+1}\,\tilde{\Psi}^{[-\D+1]}_{a(m-1)c,ab(n)}\big) \nn 
	&&+p^{[\D]m,n}_{1-}\,\eta_{ca}\,\tilde \Psi^{[-\D+1]}_{a(m-1),b(n)} +p^{[\D]m,n}_{2-}\,\eta_{cb}\,\tilde \Psi^{[-\D+1]}_{a(m),b(n-1)}\,,\nn
	 \hat K_c\,\tilde \Psi^{[-\D]}_{a(m),b(n)} \eq 
	(m+1)\,k^{[\D]m,n}_{1+}\,\tilde \Psi^{[-\D-1]}_{a(m)c,b(n)} \nn 
	&& +\,k^{[\D]m,n}_{2+}
	\,\big(\tfrac{(m-n)(n+1)}{m-n+1}\,\tilde \Psi^{[-\D-1]}_{a(m),b(n)c}  -\tfrac{m(n+1)}{m-n+1}\,\tilde{\Psi}^{[-\D-1]}_{a(m-1)c,ab(n)}\big) \nn 
	&&+k^{[\D]m,n}_{1-}\,\eta_{ca}\,\tilde \Psi^{[-\D-1]}_{a(m-1),b(n)} +k^{[\D]m,n}_{2-}\,\eta_{cb}\,\tilde \Psi^{[-\D-1]}_{a(m),b(n-1)}\,.
\ea
The basis vector $\tilde \Psi^{[-\D]}_{a(m),b(m)}$ can be realized various ways,
for instance as we have introduced earlier in the paper,
as a function of auxiliary variables:
\be
	\tilde\Psi^{[-\Delta]a(m),b(n)}
	=\frac{t^{\delta}}{\delta!}\,\Pi_{\mathbb Y}\,\frac{u^{a_1}\cdots u^{a_m}}{m!}\,\frac{v^{b_1}\cdots v^{b_n}}{n!}\,,
\ee
with $\D=m-n+2\delta$\,.
Note here that we have introduce also a Lorentz scalar auxiliary variable $t$
to generate fields of different conformal dimensions. 
The above realization is compatible with the grading,
\be
	\hat D\,u^a=-\,u^a\,,
	\qquad 
	\hat D\,v^a=+\,v^a\,,
	\qquad
	\hat D\,t=-2\,t\,,
\ee
In terms of these, all the generators of $\mathfrak{so}(2,d)$ will be
realized as differential operators acting on the auxiliary variables:
\ba
	&&\hat J_{ab}
	=2\,u_{[a}\,\partial_{u^{b]}}+2\,v_{[a}\,\partial_{v^{b]}}\,,
	\nn
	&& \hat D=u\cdot \partial_u-v\cdot\partial_v+2\,t\,\partial_t\,,\nn
	&& \hat P^a=\cP^a_{1+}+\cP^a_{2-}
	+(t\,\partial_t)^{-1} t\,(\cP^a_{2+}+\cP^a_{1-})\,,\nn
	&& \hat K^a=\cK^a_{1+}+\cK^a_{2-}
	+(\cK^a_{2+}+\cK^a_{1-})\,\partial_t\,,
\ea
where $\cP^a_{r\pm}$ and $\cK^a_{r\pm}$ are defined in \eqref{P K coeff}.
Therefore, we constructed an oscillator representation which
is realized on the space of functions\,,
\be
	\tilde\cS(0,(2))=\{ t\,u^a\,u^b\,v^c\,v^d\,f_{ab,cd}(t,u)\}\cup
	\{ t^2\,u^a\,u^b\,v^c\,g_{ab,c}(t,u)\}\cup\{t^3\,u^a\,u^b\,h_{ab}(t,u)\}\,,
\ee
subject to the conditions $u\cdot\partial_v\,\tilde\cS(0,(2))=0$ and $\partial_u^2\,\tilde\cS(0,(2))=0$\,.
It will be interesting to relate the above representation to the one that can be obtained from the reductive
dual pair correspondence of $(O(2,d), Sp(4,\mathbb R))$ 
(see \cite{Basile:2020gqi} for a review of the correspondence).

Note that $\tilde\cS(0,(2))$ has vectors with negative conformal weights,
which are bounded from above ($-\Delta\le -2$).
The representation $\tilde\cS(0,(2))$ is in fact 
isomorphic to the dual representation (aka contragredient representation) of the off-shell FT model $\cS(0,(2))$.
 The isomorphism is given by the anti-involution,\footnote{The anti-involution $\rho$ flips only the sign of $\mathfrak{k}$ generators.
 If we take $\mathfrak{so}(2)\oplus \mathfrak{so}(d)$ at the place of $\mathfrak{k}$,
 then the anti-involution would correspond to minus the Cartan anti-involution.
 Note that $\rho$ is different from the Chevallay anti-involution $\tau$, 
 \be
	  \tau(\hat P_a)=\hat K_a\,,\qquad \tau(\hat K_a)=\hat P_a\,,
	\qquad \tau(\hat J_{ab})=-\hat J_{ab}\,, \qquad 
	\tau(\hat D)=\hat D\,.
\ee
used in \cite{Shaynkman:2004vu}. They are related by $\rho=\tau \circ \i$  through the involution $\i$ given by
\be
	 \i(\hat P_a)=\hat K_a\,,\qquad \i(\hat K_a)=\hat P_a\,,
	\qquad \i(\hat J_{ab})=\hat J_{ab}\,, \qquad 
	\i(\hat D)=-\hat D\,.
\ee 
 }
\be 
	\rho(\hat P_a)=\hat P_a\,,\qquad \rho(\hat K_a)=\hat K_a\,,
	\qquad \rho(\hat J_{ab})=-\hat J_{ab}\,, \qquad 
	\rho(\hat D)=-\hat D\,.
\ee
More precisely, we can define a non-degenerate bi-linear form $\la\cdot|\cdot\ra$ for
$\cS(0,(2)) \times \tilde \cS(0,(2))$ as
\be
	\la\,\Psi^{[\D]}_{a(m),b(n)}\,|\,\tilde \Psi^{[-\D']}_{c(m'),d(n')}\,\ra
	=\delta_{\D,\D'}\,\delta_{m,m'}\,\delta_{n,n'}\,
	\Pi_{a(m),b(n)|c(m),d(n)}\,,
\ee
where $\Pi_{a(m),b(n)|c(m),d(n)}$ is the projection operator onto the space $(m,n)$,
and  $\Psi^{[\D]}_{a(m),b(n)}$ 
and $\tilde\Psi^{[-\D']}_{c(m'),d(n')}$
are the basis vectors of  $[\D,(m,n)]\subset\cS(0,(2))$ and  $[-\D',(m',n')]\subset\tilde\cS(0,(2))$, respectively.
Then, for any $v\in \cS(0,(2))$ and $w\in \tilde \cS(0,(2))$\,,
the actions of an element $\hat X\in \mathfrak{so}(2,d)$ on each spaces satisfy
\be
	\la \,v\,|\,\hat X\,w\,\ra= \la\,\rho(\hat X)\,v\,|\,w\,\ra\,.
\ee
The duality between $\cS(0,(2))$ and $\tilde\cS(0,(2))$ implies that  $[\D,(m,n)] \subset \cS(0,(2))$ is the subspace of (generalized) highest weight states annihilated by $\hat K_a$\,, 
if and only if    there is no $w\in  \tilde \cS(0,(2))$ such that $\hat K_a\,w\in [-\D,(m,n)]\subset\tilde\cS(0,(2))$\,:
\be
	\hat K_a\,[\D,(m,n)]=0\quad {\rm in}\quad \cS(0,(2))
	\qquad \Longleftrightarrow \qquad 
	\hat K_a\,\tilde\cS(0,(2))\not\subset [-\D,(m,n)]\,.
\ee
This shows that the field $C^{[d](2,0)}$ is dual to an invariant highest-weight representation 
generated by $[d,(2,0)]\in \cS(0,(2))$
as it cannot be obtained from $C^{[d-1](m,n)}$ by $\hat K_a$ actions.

So far, we have considered only the passive transformation of $\mathfrak{so}(2,d)$,
which acts only on the basis vectors $\tilde\Psi^{[-\D]}_{a(m),b(n)}$
but not on the coefficients $C^{[\D]a(m),b(n)}$.
We can also consider the active transformation  of $\mathfrak{so}(2,d)$,
which acts only on $C^{[\D]a(m),b(n)}$ but not on $\tilde\Psi^{[-\D]}_{a(m),b(n)}$, 
as the linear part of the gauge transformation with constant parameters, namely the global transformation:
the active action of an element $\hat X$ in $\mathfrak{so}(2,d)$  can be defined as 
\be
	\hat X\, C^{[\D]a(m),b(n)}:=\delta_{\rho(\hat X)}\,C^{[\D]a(m),b(n)}\,\big|_{\cO(C^2)=0}\,.
\ee 
Here, $C^{[\D]a(m),b(b)}$ should not be viewed as the coefficients, but as
the dual basis vectors.
Indeed, the active representation of the zero-form fields is dual to the passive representation, $\tilde\cS(0,(2))$.
Therefore, the active representation of the zero-form fields is  simply 
the off-shell FT module $\cS(0,(2))$.
What is noteworthy in the active representation is
that it is related to the gauge transformation $\delta\,C^{[\D]a(m),b(n)}$
which is in fact nonlinear in $C$.
We will come back to this point in Section \ref{sec: reduction}.

\section{Higher order unfolding of conformal geometry}
\label{sec: nonlinear}

\subsection{General structure}

Let us discuss the structure of the nonlinear unfolded equations for
the zero-form fields. For that, let us label the zero-form fields 
by a collective index $I=[\D]a(m),b(n)$ as
\be
	C^I=C^{[\D]a(m),b(n)}\,.
\ee
In this notation, the nonlinear unfolded equations read simply\footnote{One may think that it could be more useful to express the equation as
\be
	\cd^G C^I= e_a\,\cE^{I,a}_{n\ge2}(C)+f_a\,\cF^{I,a}_{n\ge2}(C)\,,
	\label{G cov eq}
\ee
then, act $\cd^G$ on it to get the Bianchi identity. 
However, such a computation could lead to an inconsistency:
the LHS of the Bianchi identity is quadratic in $C$ since $(\cd^G)^2=\cO(C)$,
whereas the RHS is cubic in $C$. The mistake is due to the incorrect assumption 
of the $G$-covariance of the multi-linear forms hidden in the RHS of \eqref{G cov eq}.
More formally, the cocyle of the corresponding Chevalley-Eilenberg complex 
is covariant under $K$ but not under $G$.
    }
\be
	\cd^K C^I= e_a\,\cE^{I,a}(C)+f_a\,\cF^{I,a}(C)\,,
	\label{nonlin eq}
\ee
where the conformal weight of $C^I, \cE^I$ and $\cF^I$ are
$\D_I, \D_I+1$ and $\D_I-1$\,, respectively.
Using the relations,
\ba
&& \cd^Ke^a=0\,,\qquad \cd^K f^a=\frac12\,e_b\wedge e_c\,C^{[3]ab,c}\,,\nn
&&
(\cd^K)^2=\frac14\,e_a\wedge e_b\,C^{[2]ac,bd}\,\hat J_{cd}+e^a\wedge f^b\,(\hat J_{ab}-\eta_{ab}\,\hat D)\,.
\ea
we can simplify the action of $\cd^K$ on \eqref{nonlin eq} as
\ba 
0 \eq e_a\wedge e_b \left( \frac14\,C^{[2]ac,bd}(\hat J_{cd}\,C)^I 
-\frac12 C^{[3]ca,b}\cF^I{}_c(C) + \cE^{J,b}(C)\frac{\partial \cE^{I,a}}{\partial C^J} \right) \nn
&& +e_a\wedge f_b \left( (\hat J^{ab}\,C)^I + \D_I\,\h^{ab}\,C^I+\cF^{J,b}(C)\frac{\partial \cE^{I,a}}{\partial C^J} -\cE^{J,a}(C)\frac{\partial \cF^{I,b}}{\partial C^J} \right) \nn
&& +f_a\wedge f_b \,\cF^{J,b}(C) \frac{\partial \cF^{I,a}}{\partial C^J}\,.
\label{nonlin BI}
\ea 
We require the above equation to hold identically --- it becomes the Bianchi identity for the 
zero-form fields. 
Moreover, each line should vanish separately since the equation should hold independently of 
the choice of $e^a$ and $f^a$. 
This requirement, together with the Bianchi identities \eqref{f BI} of the one-form fields,
 determines the functions $\cE^{I,a}(C)$ and $\cF^{I,a}(C)$\,,
 up to (nonlinear) field redefinitions of $C$.

Since the identity should hold for any value of $C^I$, we can expand
the functions in the powers of $C$ as
\be
	\cE^{I,a}(C)=\sum_{n=1}^\infty \cE^{I,a}_n(C)\,,
	\qquad 
	\cF^{I,a}(C)=\sum_{n=1}^\infty \cF^{I,a}_n(C)\,,
\ee
and the equation \eqref{nonlin BI} should hold identically order by order in 
 powers of $C$. 
The first order part of the equation,
\ba
	&&e_a\wedge e_b\,\cE^{J,b}_1(C)\frac{\partial\cE^{I,a}_1(C)}{\partial C^J}
	+ f_a\wedge f_b\,\cF^{J,b}_1(C)\frac{\partial\cF^{I,a}_1(C)}{\partial C^J}
	\nn
	&&+\,
	e_a\wedge f_b
	\left[(\hat J^{ab}\,C)^I-\D_I\,\eta^{ab}\,C^I 
	+\cF^{J,b}_1(C)\,\frac{\partial\cE^{I,a}_1(C)}{\partial C^J}
	-\cE^{J,a}_1(C) \,\frac{\partial\cF^{I,b}_1(C)}{\partial C}\right]=0\,,
\ea
is what we have worked out in the previous section,
and $\cE^{I,a}_1$ and $\cF^{I,a}_1$ are determined by the actions of $\hat P_a$ and $\hat K_a$\,.
Moving to the second order, we find
\ba
	&&e_a\wedge e_b
	\left[\frac14\,C^{[2]ac,bd}\,(\hat J_{cd}\,C)^I
	-\frac12\,C^{[3]}{}_c{}^{a,b}\,\cF^{I,c}_{1}(C)+
	\cE^{J,b}_1(C)\frac{\partial\cE^{I,a}_2(C)}{\partial C^J}
	+\cE^{J,b}_2(C)\frac{\partial\cE^{I,a}_1(C)}{\partial C^J}\right]
	 \nn
	&&+\,e_a\wedge f_b
	\left[
	\cF^{J,b}_1(C)\frac{\partial\cE^{I,a}_2(C)}{\partial C^J}
	+\cF^{J,b}_2(C)\frac{\partial\cE^{I,a}_1(C)}{\partial C^J}
	-\cE^{J,a}_1(C)\frac{\partial\cF^{I,b}_2(C)}{\partial C^J}
	-\cE^{J,a}_2(C)\frac{\partial\cF^{I,b}_1(C)}{\partial C^J}
\right]\nn
&&+\,f_a\wedge f_b
	\left[
	\cF^{J,b}_1(C)\frac{\partial\cF^{I,a}_2(C)}{\partial C^J}
	+\cF^{J,b}_2(C)\frac{\partial\cF^{I,a}_1(C)}{\partial C^J}\right]=0\,.
	\label{2 BI}
\ea
Since $\cE^{I,a}_1$ and $\cF^{I,a}_1$ are already determined,
the above defines inhomogeneous linear equations for $\cE^{I,a}_2$ and $\cF^{I,a}_2$\,.
Finally, at the order $n\ge 3$, we have
\ba
	&&e_a\wedge e_b
	\left(\frac12\,
	C^{[3]}{}_c{}^{a,b}\,\cF^{I,c}_{n-1}(C)+\sum_{m=1}^n	\cE^{J,b}_m(C)\frac{\partial\cE^{I,a}_{n+1-m}(C)}{\partial C^J}
	\right)
	\nn
	&&+\,e_a\wedge f_b
	\sum_{m=1}^n
	\left(\cF^{J,b}_m(C)\frac{\partial\cE^{I,a}_{n+1-m}(C)}{\partial C^J}
	-\cE^{J,a}_m(C)\frac{\partial\cF^{I,b}_{n+1-m}(C)}{\partial C^J}\right)\nn
	&&+\,f_a\wedge f_b
	\sum_{m=1}^n	\cF^{J,b}_m(C)\frac{\partial\cF^{I,a}_{n+1-m}(C)}{\partial C^J}
	=0\,,
	\label{n BI}
\ea
which define again 
inhomogeneous linear equations for $\cE^{I,a}_n$ and $\cF^{I,a}_n$ 
where the coefficients are given by $(\cE^{I,a}_m,\cF^{I,a}_m)$ with $m=1,\ldots, n-1$\,.
In this way, one can iteratively determine 
all power series coefficients $\cE^{I,a}_n$ and $\cF^{I,a}_n$
by solving the linear equations.
The index $I=[\D]a(p),b(q)$ takes infinitely many values,
but at a fixed order $n$ only finitely many $\D$ can appear because 
the conformal dimensions $\D$ is additive and bounded from below ($\D\ge 2$)\,:
\be
	\cE^{[\D],a}(C)=\sum_{k=1}^{[\frac{\D+1}2]} \cE^{[\D],a}_k(C)\,,
	\qquad
	\cF^{[\D],a}(C)=\sum_{k=1}^{[\frac{\D-1}2]} \cF^{[\D],a}_k(C)\,.
\ee
Moreover, for a finite $\D$, finitely many $(m,n)$ tensors appear.
Therefore, the procedure of determining the functions $\cE^{I,a}(C)$ and $\cF^{I,a}(C)$
--- that is, the unfolded equations for the zero-forms ---
can be decomposed into finite-dimensional linear equations. 

We have seen that the linear part of the unfolded equation for conformal geometry 
defines the  representation $\cS(0,(2))$ of conformal group $SO(2,d)$, but the system is not consistent 
without higher order completion.
In general the unfolded equations can be viewed as a Lie algebroid with an infinite-dimensional base manifold
corresponding to the zero-forms.
 In the conformal geometry case, the structure constant of the Lie algebroid
is at most linear in $C$, whereas the anchor, given by $\cE^{[\D],a}(C)$ and $\cF^{[\D],a}(C)$,
are higher order  polynomials in $C$.

\subsection{A few lowest $\D$ }
\label{low D}

Let us show more explicitly, but still schematically, how the nonlinear part of the 
consistency equations \eqref{2 BI} and \eqref{n BI} can be organized as
a series of finite-dimensional linear equations by 
arranging them in the total conformal dimensions $\D_{\rm tot}=\D_e+\D_f+\D_C$\,.

\begin{itemize}
\item 
$\underline{\D_{\rm tot}=4}$\,: We have only one non-trivial condition from $e^a\wedge e^b$,
\be
	\frac14\,C^{[2]ac,bd}\,(\hat J_{cd}\,C)^{[2]}
	+\bm{\cE_2^{[3],[a}(C^{[2]},C^{[2]})}\frac{\partial\cE^{[2],b]}_1(C^{[3]})}{\partial C^{[3]}}=0\,,
\ee
which is an equation for $\cE_2^{[3],a}(C^{[2]},C^{[2]})$\,.
Here, we suppressed the Lorentz label $a(m),b(n)$ in the tensors $C^{[\D]a(m),b(n)}$ for simplicity of the expressions,
and one should note that there are more than one tensors for a given $\D\ge 3$.
Since $\cE^{[3],a}(C)$ is at most quadratic in $C$, by solving the above equations,
we can determine $\cE^{[3],a}(C)$ completely.

\item

$\underline{\D_{\rm tot}=5}$\,:
We still have only one equation from $e^a\wedge e^b$, which reads,
\ba
	&&\frac14\,C^{[2]ac,bd}\,(\hat J_{cd}\,C)^{[3]}
	-\frac12\,C^{[3]}{}_c{}^{[a,b]}\,\cF^{[3],c}_{1}(C^{[2]})\nn
	&&+\,\cE^{[2],[a}_1(C^{[3]})\frac{\partial\cE^{[3],b]}_2(C^{[2]},C^{[2]})}{\partial C^{[2]}}
	+\bm{\cE^{[4],[a}_{2}(C^{[2]},C^{[3]})}\frac{\partial\cE^{[3],b]}_1(C^{[4]})}{\partial C^{[4]}}=0\,,
\ea
and it determines $\cE^{[4],a}_{2}(C^{[2]},C^{[3]})$. Since $\cE^{[4],a}(C)$ is at most quadratic in $C$,
it is also determined. 
\item

$\underline{\D_{\rm tot}=6}$\,:
In this case, 
we find two non-trivial conditions from  $e^a\wedge e^b$ and $e^a\wedge f^b$\,.
\begin{itemize}
\item 
First, from $e^a\wedge e^b$, 
we obtain two quadratic and one cubic conditions.
The first quadratic condition is proportional to $C^{[2]}\,C^{[4]}$,
\ba
	&&\frac14\,C^{[2]ac,bd}\,(\hat J_{cd}\,C)^{[4]}
	+\cE^{[3],[a}_1(C^{[4]})\frac{\partial\cE^{[4],b]}_2(C^{[2]},C^{[3]})}{\partial C^{[3]}}\nn
	&&+\bm{\cE^{[5],[a}_2(C^{[2]},C^{[4]})}\frac{\partial\cE^{[4],b]}_1(C^{[5]})}{\partial C^{[5]}}=0\,,
\ea
and it determines $\cE^{[5],a}_2(C^{[2]},C^{[4]})$\,.
The second quadratic condition is proportional to $C^{[3]}\,C^{[3]}$,
\ba
	&&-\frac12\,C^{[3]}{}_c{}^{[a,b]}\,\cF^{[4],c}_{1}(C^{[3]})
	+\cE^{[2],[a}_1(C^{[3]})\frac{\partial\cE^{[4],b]}_2(C^{[2]},C^{[3]})}{\partial C^{[2]}}\nn
	&&+\bm{\cE^{[5],[a}_2(C^{[3]},C^{[3]})}\frac{\partial\cE^{[4],b]}_1(C^{[5]})}{\partial C^{[5]}}=0\,,
\ea
and it determines $\cE^{[5],a}_2(C^{[3]},C^{[3]})$\,.
The cubic condition is 
\be
	\cE^{[3],[a}_2(C^{[2]},C^{[2]})\frac{\partial\cE^{[4],b]}_{2}(C^{[2]},C^{[3]})}{\partial C^{[3]}}
	+\bm{\cE^{[5],[a}_3(C^{[2]},C^{[2]},C^{[2]})}\frac{\partial\cE^{[4],b]}_{1}(C^{[5]})}{\partial C^{[5]}}
	=0\,,
\ee
and it determines $\cE^{[5],[a}_3(C^{[2]},C^{[2]},C^{[2]})$\,.
With these, $\cE^{[5],a}(C)$ is completely determined as it is at most cubic in $C$.

\item From
$e^a\wedge f^b$, we find one condition,
\ba
	&&\cF^{[3],b}_1(C^{[2]})\frac{\partial\cE^{[4],a}_2(C^{[2]},C^{[3]})}{\partial C^{[3]}}
	-\cE^{[3],b}_2(C^{[2]},C^{[2]})\frac{\partial\cF^{[4],a}_1(C^{[3]})}{\partial C^{[3]}}\nn
	&&+\,\bm{\cF^{[5],b}_2(C^{[2]},C^{[2]})}\frac{\partial\cE^{[4],a}_1(C^{[5]})}{\partial C^{[5]}}
	=0\,,
\ea
which determines $\cF^{[5],a}_2(C^{[2]},C^{[2]})$,
and hence $\cF^{[5],a}(C)$ as it is at most quadratic. 

\end{itemize}

\end{itemize}

\noindent Even though that the above sets of equations are finite-dimensional linear equations,
they are tedious to solve and hence it would be more desirable to use a computer program code,
which is beyond the scope of the current paper.

\section{Weyl invariant densities}
\label{sec: Weyl}

In this section we discuss how Weyl invariants can
be classified within the unfolded formulation of conformal geometry.

\subsection{Gauge symmetry}
\label{sec: sym}

Let us consider our system within
the general framework of the unfolded formulation.
Our system has two kinds of differential forms, the one-forms
taking values in the adjoint representation of $\mathfrak{so}(2,d)$ and  
the zero-forms taking (infinitely many) values in the off-shell FT module $\cS(0,(2))$ of $\mathfrak{so}(2,d)$,
\be
	W^A\,: \qquad e^a\,,\quad \o^{ab},\quad  b\,, \quad  f^a\,,\quad  C^I\,.
\ee
Their equations have the general structure,
\be
	\dd\,W^A+G^A(W)=0\,,
\ee
with the (generalized) Jacobi identity,
\be
	G^{B}(W) \frac{\partial G^A(W)}{\partial W^B}=0\,.
\ee
The system is invariant under the gauge transformations,
\be
	\delta W^A=\dd w^A+w^B\,\frac{\partial G^A(W)}{\partial W^B}\,,
\ee
where $w^A$ are the $(p-1)$-form gauge parameters
associated with the $p$-form field $W^A$\,.
Applying this to our system, we find the ``standard'' gauge transformations
for $e^a$ and $b$ as
\be
	\delta\,e^a=\cd^K \e^a+\l^{ab}\,e_b-\s\,e_b\,,
	\qquad 
	\delta\,b=\cd^K \s +\e^a\,f_a+\k^a\,e_a\,,
\ee
and ``deformed'' gauge transformations for $\o^{ab}$ and $f^a$ as
\ba
	\delta\,\o^{ab}\eq \cd^K \l^{ab}+
	\k^{[a}\,e^{b]}+\e_{[c}\,e_{d]} \left(\eta^{c[a}\,f^{b],d}+C^{[2]ac,bd}\right)\,,\nn
	 \delta\,f^a\eq \cd^K \k^a +\l^{ab}\,f_b+\s\,f_b
	+\e_{[b}\,e_{c]}\,C^{[3]ab,c}\,.
\ea
The modification terms are proportional to the curvatures and 
corresponds to the  ``non-geometrical"  terms considered in \cite{Kaku:1977pa}.

The zero-forms  transform as
\be
	\delta\,C^I
	= \l^{ab}\,(\hat J_{ab}\,C)^I - \D_I\,\s\,C^I
	+\e_a\,\cE^{I,a}(C)+\k_a\,\cF^{I,a}(C)\,,	
\ee
involving nonlinear functions $\cE^{I,a}(C)$ and $\cF^{I,a}(C)$\,.

As we discussed earlier, the gauge transformations by $\l^{ab}$ and $\k^a$ can be used
to reduce the system to the metric form,
whereas the transformations generated by $\ve^a$ and $\sigma$ become
the diffeomorphisms and Weyl transformations.

\subsection{Constructing Weyl invariants \`a la unfolding}

Let us revisit the classification of Weyl invariants,
which consist of the type-B Weyl anomalies,
within the unfolded formulation of conformal geometry.
Weyl invariants are the scalar densities made by curvatures,
which are strictly invariant under Weyl rescaling,
without relying on a total derivative term. 

The Weyl invariants should correspond to a gauge invariant $d$-form within the unfolded formulation.
The invariance under $\hat D$ and $\hat J_{ab}$ requires the 
basis for the $d$-form to be made by $e^a$ and $f^a$ only.
The $\hat P_a$ gauge symmetry is the diffeomorphism, so its invariance can be achieved only up to a total derivative term.
On the contrary, the $\hat K_a$ gauge symmetry is related to a Weyl rescaling, 
so we require the strict invariance without relying on an integration by part.
From this, we can rule out the dependency of $f^a$  as it transforms with derivatives,
which can never be compensated if integrations by part are not allowed.
In the end, the ansatz for the Weyl invariants is 
\be
	I_d=\e_{a_1\cdots a_d}\,e^{a_1}\wedge \cdots \wedge e^{a_d} \,\cI_d(C)\,.
\ee
The invariance under the Lorentz and dilatation is guaranteed
by considering $\cI_d(C)$ 
where all the Lorentz indices of $C$ are fully contracted without using any external tensors
and the total conformal dimension is $\D_{\rm tot}=d$\,.
The gauge variation under translation and special conformal transformations give
\ba
	\delta I_d
	\eq d\,\e_{a_1\cdots a_d}\,\cd^K\ve^{a_1}\wedge e^{a_2}\wedge\cdots \wedge e^{a_d} \,\cI_d(C)\nn
	&& +\,
	\e_{a_1\cdots a_d}\,e^{a_1}\wedge\cdots \wedge e^{a_d} 
	\left[\ve_c\,\cE^{I,c}(C)+\k_c\,\cF^{I,c}(C)\right]\frac{\partial \cI_d(C)}{\partial C^I}\,.
\ea
With a total derivative term, the above can be expressed as  
\ba
	\delta I_d
	\eq 
	\dd\left[d\,\e_{a_1\cdots a_d}\,\ve^{a_1}\,e^{a_2}\wedge\cdots \wedge e^{a_d} \,\cI_d(C)\right]
	\nn
	&&+\,\e_{a_1\cdots a_d}\left[
	d\,\ve^{a_1}\,e^{a_2}\wedge\cdots \wedge  e^{a_d}\wedge e_c+
	\ve_c\,\,e^{a_1}\wedge\cdots \wedge e^{a_d}\right] 
	\cE^{I,c}(C)\,\frac{\partial \cI_d(C)}{\partial C^I}
	\nn
	&& 
	+\,
	\e_{a_1\cdots a_d}
	\left(\ve^{a_1}\,e^{a_2}\wedge\cdots \wedge  e^{a_d}
	\wedge f_c
	+\k_c\,e^{a_1}\wedge\cdots \wedge  e^{a_d}\right)
	\cF^{I,c}(C)\,\frac{\partial \cI_d(C)}{\partial C^I}\,.
\ea
The first line is a total derivative term proportional to the translation gauge parameter $\ve^a$\,.
The second line vanishes identically:
\be
	\e_{a_1\cdots a_d}\left[
	d\,\ve^{a_1}\,e^{a_2}\wedge\cdots \wedge  e^{a_d}\wedge e_c+
	\ve_c\,\,e^{a_1}\wedge\cdots \wedge e^{a_d}\right]=0\,,
\ee
due to the properties of antisymmetrizations.
Only the third line poses as a non-trivial condition,
\be
	\cF^{I,a}(C)\,\frac{\partial \cI_d(C)}{\partial C^I}=0\,,
\ee
which ensures the gauge invariance under translation and special conformal transformation at the same time.
In the end, it is sufficient to ask the special conformal invariance
of the density $\cI_d(C)$\,:
\be
	\delta_{\k}\,\cI_d(C)=\k^a\,\cF^{I,a}(C)\,\frac{\partial \cI_d(C)}{\partial C^I}=0\,.
	\label{Weyl inv eq}
\ee
To find an explicit form of Weyl invariants, one needs to begin with a general ansatz
\be
	\cI_d(C)=\cI_d(C^{[2]},C^{[3]},\ldots, C^{[d-2]})\,,
\ee
with a certain finite number of undetermined coefficients $c_i$\,.
For instance, for $d=4$ we have only one term,
\be
	\cI_4(C)
	=c_1\,\frac{C^{[2]a(2),b(2)}\,C^{[2]}{}_{a(2),b(2)}}{4}\,,
\ee
whereas for $d=6$, we have three terms,
\ba
	\cI_6(C)
	\eq c_1\,\frac{C^{[4]a(2),b(2)}\,C^{[2]}{}_{a(2),b(2)}}4
	+c_2\,\frac{C^{[3]a(3),b(2)}\,C^{[3]}{}_{a(3),b(2)}}{12}
	+c_3\,\frac{C^{[3]a(2),b}\,C^{[3]}{}_{a(2),b}}2\nn
	\eq c_1\,\la C^{[4](2,2)}|C^{[2](2,2)}\ra
	+c_2\,\la C^{[3](3,2)}|C^{[3](3,2)}\ra
	+c_3\,\la C^{[3](2,1)}|C^{[3](2,1)}\ra\,.
\ea
In the last line, we used the convention,
\be
	\la f|g\ra
	=\exp\left(\partial_{u_1}\cdot\partial_{u_2}+\partial_{v_1}\cdot\partial_{v_2}\right)
	f(u_1,v_2)\,g(u_2,v_2)\,\big|_{\substack{u_1=u_2=0\\v_1=v_2=0}}\,.
\ee
By checking the special conformal transformations of $\cI_4$ and $\cI_6$, we find
$\cI_4$ is already conformally invariant:
\be
	\delta_{\k}\,\cI_4(C)=0\,,
\ee
whereas the conformal invariance of $\cI_6$ requires 
\ba
	\delta_{\k}\,\cI_6(C)
	\eq 
	\k^c\,(c_1\,k^{[3]3,2}_{1-}+2\,c_2\,k^{[2]2,2}_{1+})\,
	\langle \partial_{u^c}\,C^{[3](3,2)}| C^{[2](2,2)} \rangle \nn 
	&& +\,\k^c\,(c_1\,k^{[3] 2,1}_{2+} + 2\,c_3\,k^{[2]2,2}_{2-})\,
	\langle u_c\,C^{[3](2,1)} | C^{[2](2,2)} \rangle 
	=0\,.
\ea
The above defines the system of linear equations for $c_1, c_2, c_3$,
\be 
\begin{pmatrix}
	k^{[3]3,2}_{1-} & 2k^{[2]2,2}_{1+} & 0 \\ 
	k^{[3]2,1}_{2+} &  0& 2k^{[2]2,2}_{2-} 
\end{pmatrix}
\begin{pmatrix}
	c_1 \\ c_2 \\ c_3 
\end{pmatrix} =
\begin{pmatrix}
	0 \\ 0 
\end{pmatrix}. \label{dim 6 sc to vec}
\ee 
In any $d$, the above has one-dimensional solution space,
\be
	(c_1,c_2,c_3)=c_3 \left(
	4\,\frac{d-2}{d+4} \,, 
	\frac{(d-2)^2}{(d+1)(d+4)} \,, 1 \right),
\label{dim6 weylinv quad:gen}
\ee
where we fixed $p^{[\D]2,2}_{2-}=p^{[\D]2,1}_{2+}=p^{[\D]2,1}_{2-}=1$ using field redefinition freedoms
(we will keep this choice in the following analysis).
Likewise, once the form of $\cF^{I,a}(C)$ is determined,
the problem of finding Weyl invariants becomes a pure algebraic exercise.

\subsection{Quadratic Weyl invariants}

As we have seen in $6d$ case, the ansatz for quadratic Weyl invariants are easy to handle
so it allowed to determine the invariant using the explicit expression of $\hat K_a$ action.
Let us extend this analysis to $8d$ and $10d$.
In these cases, the quadratic Lorentz scalars that are invariant under the linear part of $\hat K_a$ action should
be complemented by cubic (and also quartic for $d=10$) terms
to compensate the nonlinear  part of  $\hat K_a$ action.

In the following, we determine the quadratic part of Weyl invariants,
that is, the quadratic Lorentz scalars invariant under the linear part of $\hat K_a$ action.
In $6d$, there are 3 quadratic Lorentz scalars with $\D=6$, and 2 quadratic Lorentz vectors with $\D=5$,
so a generic action of $\hat K_a$ would have left one dimensional solution space for quadratic Weyl invariant.
This is to be contrasted with the situations in higher dimensions:
as shown in Appendix \ref{sec: Char},
there are 7  quadratic Lorentz scalars with $\D=8$,
and 8 quadratic Lorentz vectors with $\D=7$; 
and there are 12  quadratic Lorentz scalars with $\D=10$,
and 19 quadratic Lorentz vectors with $\D=9$.
See Table \ref{tab: number} for the summary.
\begin{center}
\begin{tabular}{ c | c c c}
$\D$ & 6 & 8 & 10 \\
\hline
Number of quadratic scalars with $\D$ & 3 & 7 & 12\\ 
 Number of quadratic vectors with $\D-1$ & 2 & 8 & 19\\     
\end{tabular}
\label{tab: number}
\end{center}
Because of the ``inversion of numbers''
in $8d$ and $10d$, we need explicit computations to identify the kernel of $\hat K_a$ action
in these cases.

\paragraph{Eight dimensions}

As mentioned above, there are 7 quadratic Lorentz scalar with $\D=8$
leading to the ansatz,
\ba
	\cI_{8}(C) \eq	
	c_1\, \la C^{[6](2,2)}\,|\,C^{[2](2,2)}\ra
	+c_2\, \la C^{[5](3,2)}\,|\,C^{[3](3,2)} \ra
	+ c_3\,\la C^{[5](2,1)}\,|\,C^{[3](2,1)}\ra\nn
	&&
	+\,c_4\,\la C^{[4](4,2)}\,|\,C^{[4](4,2)}\ra
	+\,c_5\,\la C^{[4](3,1)}\,|\,C^{[4](3,1)}\ra\nn
	&&
	+\,c_6\,\la C^{[4](2,2)}\,|\,C^{[4](2,2)}\ra 
	+c_7\,\la C^{[4](2,0)}\,|\,C^{[4](2,0)}\ra
	+\cO(C^3)\,.
	\label{I8: quad}
\ea
The invariance of $\cI_8(C)$ under $\hat K_a$ transformation requires
\be
\begin{pmatrix}
	k^{[5]3,2}_{1-} & k^{[2]2,2}_{1+} & 0 & 0& 0& 0& 0\\
	k^{[5]2,1}_{2+} &0 & k^{[2]2,2}_{2-} &0 &0 & 0& 0\\ 
	0& k^{[4]4,2}_{1-} & 0&2k^{[3]3,2}_{1+} & 0& 0& 0\\ 
	0& k^{[4]3,1}_{2+} & 0&0 &2k^{[3]3,2}_{2-} &0 &0 \\
	0& k^{[4]2,2}_{1+} & 0&0 & 0&2k^{[3]3,2}_{1-} & 0\\
	0& 0& k^{[4]3,1}_{1-} & 0&2k^{[3]2,1}_{1+} & 0& 0\\
	0& 0& k^{[4]2,2}_{2-} &0 &0 &2k^{[3]2,1}_{2+} &0 \\ 
	0& 0& k^{[4]2,0}_{2+} & 0&0 & 0&2k^{[3]2,1}_{2-} \\
\end{pmatrix}
\begin{pmatrix}
	c_1 \\ c_2 \\c_3 \\ c_4 \\ c_5 \\ c_6 \\ c_7
\end{pmatrix} 
=\begin{pmatrix}
	0 \\ 0 \\ 0 \\ 0 \\ 0 \\ 0 \\ 0 \\ 0
\end{pmatrix}\,.
\ee
With explicit values of $k^{[\D]m,n}_{r\pm}$, we find 
the solution space is one-dimensional:
\ba
	&&(c_1,c_2,c_3,c_4,c_5,c_6,c_7)\nn
	&&=
	c_7\left(
	\tfrac{24\,d\, (d-4) (d-2)^2 }{(d+4) (d+6) \left(d^2-1\right)} ,\, 
	\tfrac{24\,d\, (d-4)^2 (d-2)^2 }{(d-1) (d+1)^2 (d+4) (d+6)} ,\, 
	\tfrac{24\,d\, (d-4) (d-2) }{(d+4) \left(d^2-1\right)} ,\right. \nn
	&& \qquad\quad
	\left.\tfrac{4\,d\, (d-4)^2 (d-2)^3 }{(d-1) (d+1)^3 (d+4) (d+6)} ,\, 
	 \tfrac{8\,d\, (d-4)^2 (d-2)}{(d+2) (d+4) \left(d^2-1\right)} ,\,
	\tfrac{36\,d\, (d-4)^2 (d-2) }{(d+4)^2 \left(d^2-1\right)} ,\, 
	  1 \right).
\ea

\paragraph{Ten dimensions}

We have 12 dimensional ansatz,
\ba
	&&\cI_{10}(C)
	=
	c_1\,\la C^{[8](2,2)}\,|\,C^{[2](2,2)}\ra
	+c_2 \, \la C^{[7](3,2)}\,|\,C^{[3](3,2)}\ra
	+c_3 \, \la C^{[7](2,1)}\,|\,C^{[3](2,1)}\ra
	 \nn
	&&\qquad 
	+\,c_4\,\la C^{[6](4,2)}\,|\,C^{[4](4,2)}\ra
	+c_5 \,\la C^{[6](3,1)}\,|\,C^{[4](3,1)}\ra   
	+c_6\,\la C^{[6](2,2)}\,|\,C^{[4](2,2)}\ra 
	 \nn
	&&\qquad 
	+\,c_7\,\la C^{[6](2,0)}\,|\,C^{[4](2,0)}\ra
	+c_8\,\la C^{[5](5,2)}\,|\,C^{[5](5,2)}\ra  
	+c_9\,\la C^{[5](4,1)}\,|\,C^{[5](4,1)}\ra   
	  \\
	&&\qquad
	+\,c_{10} \,\la C^{[5](3,2)}\,|\,C^{[5](3,2)}\ra 
	+c_{11}\,\la C^{[5](3,0)}\,|\,C^{[5](3,0)}\ra
	+c_{12}\,\la C^{[5](2,1)}\,|\,C^{[5](2,1)}\ra+\cO(C^3)\,.
	\nonumber
	 \label{I10: quad}
\ea
The $\hat K_a$ invariance of $\cI_{10}(C)$ gives
\be
 \tiny
\left( \begin{smallmatrix}
	k^{[7]3,2}_{1-} & k^{[2]2,2}_{1+} & 0&0 &0 &0 & 0& 0& 0& 0& 0& 0\\
	k^{[7]2,1}_{2+} &0 & k^{[2]2,2}_{2-} & 0&0 &0 &0 &0 &0 &0 &0 &0 \\ 
	0& k^{[6]4,2}_{1-} & 0& k^{[3]3,2}_{1+} & 0& 0& 0& 0& 0& 0& 0& 0\\
	0& k^{[6]3,1}_{2+} &0 &0 & k^{[3]3,2}_{2-} & 0& 0& 0&0 &0 &0 &0 \\
	0& k^{[6]2,2}_{1+} & 0& 0&0 & k^{[3]3,2}_{1-} &0 &0 &0 &0 &0 &0 \\
	0& 0& k^{[6]3,1}_{1-} &0 & k^{[3]2,1}_{1+} &0 &0 &0 &0 &0 &0 &0 \\
	0&0 & k^{[6]2,2}_{2-} & 0& 0& k^{[3]2,1}_{2+} &0 &0 &0 &0 &0 & 0\\
	0&0 & k^{[6]2,0}_{2+} & 0& 0& 0& k^{[3]2,1}_{2-} & 0& 0& 0& 0&0 \\ 
	0& 0&0 & k^{[5]5,2}_{1-} & 0& 0& 0&2k^{[4]4,2}_{1+} & 0&0 &0 &0 \\
	0& 0&0 & k^{[5]4,1}_{2+} & 0& 0&0 &0 &2k^{[4]4,2}_{2-} &0 &0 & 0\\
	0& 0& 0& k^{[5]3,2}_{1+} & 0&0 & 0& 0& 0&2k^{[4]4,2}_{1-} & 0& 0\\
	0& 0& 0&0 & k^{[5]4,1}_{1-} &0 &0 &0 &2k^{[4]3,1}_{1+} & 0& 0& 0\\
	0&0 &0 &0 & k^{[5]3,2}_{2-} & 0&0 &0 &0 &2k^{[4]3,1}_{2+} &0 &0 \\
	0& 0& 0& 0& k^{[5]3,0}_{2+} & 0&0 &0 &0 &0 &2k^{[4]3,1}_{2-} &0 \\
	0&0 &0 &0 & k^{[5]2,1}_{1+} &0 &0 &0 &0 &0 &0 &2k^{[4]3,1}_{1-} \\
	0& 0& 0&0 &0 & k^{[5]3,2}_{1-} &0 &0 &0 &2k^{[4]2,2}_{1+} & 0&0 \\
	0& 0&0 &0 &0 & k^{[5]2,1}_{2+} & 0& 0&0 &0 &0 &2k^{[4]2,2}_{2-} \\
	0&0 &0 &0 &0 &0 & k^{[5]3,0}_{1-} &0 &0 &0 &2k^{[4]2,0}_{1+} &0 \\
	0& 0&0 &0 &0 &0 & k^{[5]2,1}_{2-} &0 &0 &0 &0 &2k^{[4]2,0}_{2+} \\
\end{smallmatrix}\right)
{\footnotesize
 \begin{pmatrix}
	c_1 \\ c_2 \\ c_3 \\ c_4 \\ c_5 \\ c_6 \\ c_7 \\ c_8 \\ c_9 \\ c_{10} \\ c_{11} \\ c_{12}
\end{pmatrix}} =
{\footnotesize
\begin{pmatrix}
0\\0\\0\\0\\
0\\0\\0\\0\\
0\\0\\0\\0\\
0\\0\\0\\0\\
0\\0\\0
\end{pmatrix}},
\label{V10}
\ee 
which has one-dimensional solution space,
\ba 
	&&(c_1,c_2,c_3,c_4,c_5,c_6,c_7,c_8,c_9,c_{10},c_{11},c_{12})\nn
	&&=c_{12}\left(
	\tfrac{(d-4) (d-2) (d+4)}{2 (d-6) (d+6) (d+8)} ,\, 
	\tfrac{3 (d-4) (d-2) (d+4)}{4 (d+1) (d+6) (d+8)} ,\, 
	\tfrac{3 (d-4)(d+4)}{4 (d-6) (d+6)} ,\, 
	\tfrac{(d-4)^2 (d-2) (d+4)}{2 (d+1)^2 (d+6) (d+8)} ,\right. \nn
	&&\qquad \quad
	\tfrac{(d-4) (d+4)}{(d+2) (d+6)} ,\, \tfrac{3 (d-4)}{d+6} ,\,
	\tfrac{(d-1) (d+1) (d+4)}{2 (d-6) (d-2) d} ,\,
	\tfrac{(d-4)^2 (d-2)^2 (d+4)}{16 (d+1)^3 (d+6) (d+8)} ,\,\nn
	&& \qquad\quad \left.
	\tfrac{3 (d-4)^2 (d+4)}{16 (d+1) (d+3) (d+6)} ,\,
	\tfrac{(d-4)^2 (d+4)}{(d+1) (d+6)^2} ,\,
	\tfrac{3 (d-1) (d+4)}{16 (d-2) (d+2)} ,\, 
	1 \right).
\ea

Let us note that the densities $\cI_8$ and $\cI_{10}$ are invariant under $\hat K_a$ action in
any dimensions.
Multiplied by the volume form, they become Weyl invariants in 8 and 10 dimensions.

\subsection{Higher order Weyl invariants}

In the previous section, we identified the quadratic part of Weyl invariants in $d=8$ and $10$,
and this analysis can be straightforwardly extended to any even dimensions as we know
the linear part of the $\hat K_a$ action explicitly.
In principle, to identify higher order parts, we would need explicit expressions
of $\cF^{I,a}(C)$ up to $\D_I=d-2$, which requires many steps even for $d=8$ as we have seen in Section \ref{low D}.
However, even without the nonlinear part of $\hat K_a$ action being identified,
one can still guess the number of Weyl invariants.
Let us explain this point in the following.
The $\hat K_a$ gauge transformation is a linear differential operator on the space of $C$
and hence can be expanded as powers of $C$ as
\be
	\delta=\delta^{\sst (1)}+\delta^{\sst (2)}+\delta^{\sst (3)}+\cdots\,,
\ee
where $\delta^{\sst (n)}\sim C^n\,\partial_C$\,.
The commutativity of $\hat K_a$ transformation implies
\be
	\delta_{[a}\,\delta_{b]}=0
	\quad \Longrightarrow \quad 
	\sum_{m=1}^{n-1}\,\delta^{\sst (m)}_{[a}\,\delta_{b]}^{\sst (n-m)}=0\,.
\ee
In $8d$ example, the Weyl invariants have at most cubic terms:
\be
	\cI_8=\cI_8^{\sst (2)}+\cI_8^{\sst (3)}\,,
\ee
and the $\hat K_a$ invariance of the above is equivalent to
\be
	\delta^{\sst (1)}\cI_8^{\sst (2)}=0\,,
	\qquad
	\delta^{\sst (1)}\cI_8^{\sst (3)}+\delta^{\sst (2)}\cI_8^{\sst (2)}=0\,.
\ee
The first equation is what we have solved in the previous section, and we already identified $\cI_8^{\sst (2)}$.
Turning to the second equation, we need to solve again linear equations for $\cI_8^{\sst (3)}$
with inhomogeneous term $\delta^{\sst (2)}\cI_8^{\sst (2)}$,
which obscures the existence of the solution  $\cI_8^{\sst (3)}$.
By taking the antisymmetrized $\delta^{\sst (1)}$ variation of the second equation, we find
\be
	\delta^{\sst (1)}_{[a}(\delta^{\sst (2)}_{b]}\cI_8^{\sst (2)})=-\delta^{\sst (2)}_{[a}(\delta^{\sst (1)}_{b]}\cI_8^{\sst (2)})=0\,.
\ee
This tells that the inhomogeneous term $\delta^{\sst (2)}_a\cI_8^{\sst (2)}$ is $\delta^{\sst (1)}$ closed.
If $\delta^{\sst (2)}_a\cI_8^{\sst (2)}$ is not in the $\delta^{\sst (1)}$ cohomology of the space of $C^3$ vectors with $\D=7$,
then we know the solution will exist, and it will be sufficient
to identify the kernel of $\delta^{\sst (1)}$ in the space of $C^3$ scalars with $\D=8$.
We know from \cite{Boulanger:2004zf} this is indeed the case and the dimension of the kernel is four:
the number of $C^3$ scalars and vectors are 11 and 7, and hence $\delta^{\sst (1)}$ is surjective.

Let us consider also the $10d$ example, where the Weyl invariants have at most quartic terms:
\be
	\cI_{10}=\cI_{10}^{\sst (2)}+\cI_{10}^{\sst (3)}+\cI_{10}^{\sst (4)}\,.
\ee
The $\hat K_a$ invariance gives
\be
	\delta^{\sst (1)}\cI_{10}^{\sst (2)}=0\,,
	\qquad
	\delta^{\sst (1)}\,\cI_{10}^{\sst (3)}+\delta^{\sst (2)}\cI_{10}^{\sst (2)}=0\,,
	\qquad
	\delta^{\sst (1)}\,\cI_{10}^{\sst (4)}+\delta^{\sst (2)}\,\cI_{10}^{\sst (3)}+\delta^{\sst (3)}\cI_{10}^{\sst (2)}=0\,,
\ee
where the inhomogeneous terms in the second and third equations are $\delta^{\sst (1)}$ closed:
\ba
	&\delta^{\sst (1)}_{[a}(\delta^{\sst (2)}_{b]}\,\cI_{10}^{\sst (2)})
	=-\delta^{\sst (2)}_{[a}(\delta^{\sst (1)}_{b]}\,\cI_{10}^{\sst (2)})=0\,, \nn
	&\delta^{\sst (1)}_{[a}(\delta^{\sst (2)}_{b]}\,\cI_{10}^{\sst (3)}+\delta^{\sst (3)}_{b]}\cI_{10}^{\sst (2)})
	=-\delta^{\sst (2)}_{[a}(\delta^{\sst (1)}_{b]}\,\cI_{10}^{\sst (3)}+\delta^{\sst (2)}_{b]}\cI_{10}^{\sst (2)})
	-\delta^{\sst (3)}_{[a}\,\delta^{\sst (1)}_{b]}\cI_{10}^{\sst (2)}=0\,.
\ea
Again, if the inhomogeneous terms are not in the $\delta^{\sst (1)}$ cohomology of the space of $C^3$ and $C^4$ vectors with $\D=9$,
then the solution will exist for $\cI_{10}^{\sst (3)}$ and $\cI_{10}^{\sst (4)}$, respectively.
Therefore, in such case, one can just identify the kernel of $\delta^{\sst (1)}$ in the space of $C^3$ and $C^4$ scalars with $\D=10$.

Let us discuss more about the underlying algebraic structure.
At a fixed order $C^m$, the variation $\delta^{\sst (1)}=\delta^{\sst (1)}_{\hat K}$ defines 
 the cochain complex $(A^{m,\bullet}, \delta^{\sst (1)}_{\hat K})$
where $A^{m,n}\, :\, \mathfrak{n}^{\wedge n}\, \rightarrow \, \cS(0,(2))^{\odot m}$
 and $\mathfrak{n}$ is the subalgebra of $\mathfrak{so}(2,d)$ generated by $\hat K_a$.
 By introducing the co-differential  $\delta^{\sst (1)}_{\hat P}$ from the $\hat P_a$ action,
 the associated homotopy operator $H$ is given by
 \be
 	H=\{\delta^{\sst (1)}_{\hat K},\delta^{\sst (1)}_{\hat P}\}=\hat P^a\,\hat K_a+n(\D-d+n)\,.
\ee
We find that the $\delta^{\sst (1)}_{\hat K}$ cohomology is trivial if $\D\neq d-n$
as it lies in the kernel of $H$.
Weyl invariants  precisely concern the other cases with $\D=d-n$ and we need more detalied analysis. 
For explicit computations of the $\delta^{\sst (1)}_{\hat K}$ cohomology, 
it will be useful to recast all higher-order forms of $C$ as differential operators 
in auxiliary variables.
If we combine different $C^m$, we have 
$A^n=\oplus_{m=0}^{\frac d2} A^{m,n}\,:\, \mathfrak{n}^{\wedge n}\otimes P_{\frac d2}(\cS(0,(2))\, \rightarrow \, \mathbb R$
and the differential and co-differential $\delta^{\sst (1)}_{\hat K}$ and $\delta^{\sst (1)}_{\hat P}$
are deformed to the full gauge variations $\delta_{\hat K}$ and $\delta_{\hat P}$
which are nonlinear in $C$.
This  can be viewed as a deformation of Lie algebra cohomology to a Lie algebroid one.

\section{Reduction by constraints}
\label{sec: reduction}

So far, we have considered the unfolded equation for conformal geometry,
that is, the off-shell system for conformal gravity.
In this section, we discuss how an off-shell system can be reduced to various on-shell systems. 
The reduction can be achieved by imposing certain algebraic constraints $\Phi^I$ on the fields,
\be
	\Phi^I(e,\o,f,b,C)=0\,.
\ee 
A constraint generates infinitely many consequent algebraic constraints through (successive) gauge variations, 
\be
	\delta \cdots \delta\,\Phi^I(e,\o,f,b,C)=0\,,
\ee
where each gauge variation $\delta$ makes use of different gauge parameters,
and they are not nilpotent operators.
If the constraints are Lorentz tensors, Lorentz transformation will not generate any consequent constraint.
If the constraints can be decomposed into homogeneous ones under dilatation, 
dilatation will also leave the constraints invariant. 
On the contrary, translation always generates additional algebraic constraints at the level of the unfolded system,
but they are related to the derivatives of original constraints.  
Therefore, we can disregard the constraints which can be obtained by a $\hat P_a$ transformation.
We will refer  to this class of constraints as descendant constraint.
In this way, we are left with an assessment of the effect of special conformal transformation on the constraints.
Below, we consider two cases of non-descendant constraints.

\begin{itemize}
\item Primary constraint:\footnote{This is not to be confused with the primary constraint of Hamiltonian system. 
We use this term because this class of constraints generalizes the concept of primary fields.} the constraints which are left invariant under dilatation and special conformal transformation.
If we further restrict to the constraints which contain linear term in fields, we find only two possibilities
because only $C^{[2](2,2)}$ and $C^{[d](2,0)}$ satisfy the primary field condition 
$\hat K_a\,C^{[2](2,2)}=0$ and $\hat K_a\,C^{[d](2,0)}=0$, respectively. 
The former case corresponds to the conformally flat geometry,
\be
	\Phi^{[2](2,2)}(C)=C^{[2](2,2)}=0\,.
\ee
The latter case corresponds to the Bach flat geometry,
\be
	\Phi^{[d](2,0)}(C)=C^{[d](2,0)}+\cO(C^2)=0\,,
	\label{Bach nonlin}
\ee
where we have included non-linear terms $\cO(C^2)$,
which is necessary in compensating $\delta_{\kappa} C^{[d](2,0)}=\cO(C^2)$.
This nonlinear term in \eqref{Bach nonlin} can be 
removed by a nonlinear field redefinition of $C^{[d](2,0)}$\,.
If we consider primary constraints which are at least quadratic in fields, then the Weyl invariant densities 
belong to this class. They are all scalars, but one could equally consider tensor analogues of these nonlinear constraints.
These constraints generalize the concept of primary fields to a nonlinear level because
the $\hat K_a$ transformation acts nonlinearly on the field $C$
and the constraints are in general nonlinear functions of $C$.

\item  Section constraint:
the constraints which are neither primary nor generate new constraints under $\hat K_a$ action.
In this case, the variation necessarily constrains the gauge parameters and breaks a part of conformal symmetry.  
A simple example is the Einstein equation,
\be
	\Phi^a(f,e)=f^a-\L\,e^a\,,
\ee
whose gauge variation, dropping the local Lorentz part, is
\be
	\delta\Phi^a(f,e)
	=\cd^L(\kappa^a-\L\,\varepsilon^a)-(\kappa^a+\L\,\varepsilon^a)\,b+\sigma(f^a+\L\,e^a)
	+\varepsilon_{[b}\,e_{c]}\,C^{[3]ab,c}=0\,.
\ee
The above results in
the symmetry breaking,
\be
	\kappa^a=\L\,\varepsilon^a\,,\qquad \sigma=0\,,
\ee
and $\mathfrak{so}(2,d)$ will be reduced to
either $\mathfrak{so}(1,d), \mathfrak{so}(2,d-1)$ or $\mathfrak{iso}(1,d-1)$
depending on the parameter $\L$.
Another example is
\be
	\Phi^a(f,e,C)
	=f^a-\mu^{ab}(C)\,e_b\,,
\ee
where $\mu^{ab}(C)$ is a rank-two tensor function of $C$.
The simplest non-trivial case is $\mu^{ab}(C)=\frac1{\ell^2}\,C^{[4]ab}$\,,
and it leads to a four-derivative theory. 
Therefore, in the unfolded system of conformal geometry,
the dynamical equations of various on-shell gravitational systems
can be treated as  algebraic constraints. 

\end{itemize}

In order to obtain conformal geometry, we have used the constraints \eqref{F constr}.
These constraints can be also regarded as primary constraints imposed on an even larger system,
where none of two form curvatures are constrained:
\ba
	&F_{\hat P}^a = e_b\wedge e_c \,B^{[1]a;bc}\,,\qquad
	&F_{\hat K}^a = e_b\wedge e_c\,B^{[3]a;bc}\,,\nn
	&F_{\hat J}^{ab} = e_c\wedge e_d\,B^{[2]ab;cd}\,,\qquad  
	&F_{\hat D} = e_a\wedge e_b\,B^{[2]ab}\,.
\ea
Here, the zero-form fields $B^{[\D]a_1\cdots a_m;b_1\cdots b_n}$ are 
reducible Lorentz tensors (no Young symmetry condition is imposed on them)\footnote{For more concrete analysis, 
it would be better to decompose $B^{[\D]a_1\cdots a_m;b_1\cdots b_n}$ into irreducible tensors,
but for the current discussion it is enough to deal with reducible tensors.}
 with two groups of fully antisymmetric indices, $a_1\cdots a_m$ and $b_1\cdots b_n$.
They are subject to the Bianchi identities,
\ba 
	&& \cd^K_{[a}\,B^{[1]d;}{}_{bc]}-\d_{[a}^d\,B^{[2]}{}_{bc]} - B^{[2]d}{}_{[a;bc]} = \cO(B^2)\,,
	\label{BB1} \\
	&&\cd^K_{[a}\,B^{[2]de;}{}_{bc]}-2\,\delta^{[d}_{[a}\,B^{[3]e];}{}_{bc]}
	-2\,f^{[d}_{[a}\,B^{[1]e];}{}_{bc]}
	=\cO(B^2)\,, 
	\label{BB2} \\
	&& \cd ^K_{[a} B^{[2]}{}_{bc]}-B^{[3]}{}_{[a;bc]}
	+f_{[a}^{d}\,B^{[1]}{}_{d;|bc]}=\cO(B^2)\,,
	\label{BB3} \\
	&& \cd^K_{[a}\,B^{[3]d;}{}_{bc]}-f_{[a}^{e}\,B^{[2]d}{}_{e;|bc]}
	+f^d_{[a}\,B^{[2]}{}_{bc]}=\cO(B^2)\,.
	\label{BB4}
\ea 
Here, the nonlinearities are due to non-vanishing torsion $B^{[1]a;bc}$.
In order to obtain  the gauge transformation $\delta B^{[\D]a_1\cdots a_m;b_1\cdots b_n}$, we need  to determine
 their unfolded equations.
But since they are equivalent to these Bianchi identities, we can  read off the relevant information from 
\eqref{BB1}--\eqref{BB4} directly. First, the field $B^{[1]a;bc}$, being the lowest $\D$ field, does not have any $f$ term in 
its equation \eqref{BB1}.
This means that its $\hat K_a$ transformation vanishes and hence it is primary.
From the antisymmetrization $[abc]$ in \eqref{BB1},  we find that the descendants of $B^{[1]a;bc}$ do not contain the 
traceful $\tiny\yng(2,2)$ projection $B^{[2]ab;cd}$.
The projection of \eqref{BB2} to  such components of $B^{[2]ab;cd}$
contains $f$ term, and hence these fields are 
neither descendant nor primary. 
We may call these constraints ``secondary'' since their $\hat K_a$ variations vanish upon imposing primary constraints.
Finally, all the components of $B^{[3]a;bc}$ are descendants of the primary or the secondary constraints.
Therefore, imposing all the secondary constraints would trivialize the system
and imposing only the traceful $\tiny\yng(2)$ part of the secondary constraints gives conformal geometry.
To recapitulate,  in this enlarged unfolded system, which is nothing but the off-shell $\mathfrak{so}(2,d)$ gauge theory
with invertible $e^a_\mu$, 
conformal geometry is obtained by imposing all the primary and a part of secondary constraints.

The unfolded system asks us to work with the space of functions in the zero-form fields.
The vector space that the zero-form fields take value in carry an infinite-dimensional representation,
which can be viewed as the Hilbert space of the corresponding quantum system.
Then, it is natural to interpret the space of functions in the zero-form fields 
as the Fock space of the corresponding quantum field theory.
In this regard, it would be tempting to understand whether and how 
 a suitable quantization of the unfolded system can actually associate the space of functions in
 the zero-forms with the Fock space.
After such a quantization, the Fock space will be endowed with nonlinear actions of $SO(2,d)$.
It will be equally interesting to understand the physical meaning of this nonlinear actions and 
the associated representations in the context of conformal field theory.

\acknowledgments

We thank Thomas Basile and Nicolas Boulanger for useful discussions and 
helpful comments on our first draft.
This work was supported by National Research Foundation (Korea) through the grant NRF-2019R1F1A1044065.

\appendix

\section{Unfolding free fields}
\label{sec: other PP}

As shown in Section \ref{sec: linear},
the $\hat P_a$ action is determined by 
the coefficients $p^{\{\d\}2,2}_{2-}$, $p^{\{\d\}2,1}_{2+}$, $p^{\{\d\}2,1}_{2-}$ and $p^{\{\d\}2,0}_{2+}$
subject to the condition \eqref{pp const}.
We also showed that all these coefficients can be determined by field redefinitions
 leading to the off-shell system (conformal geometry)
 or some of them can be set to zero and solve the condition  \eqref{pp const}
as $0=0$. In the latter cases, the corresponding fields decouple from the system.
In the following, we provide a few solutions of this kind.

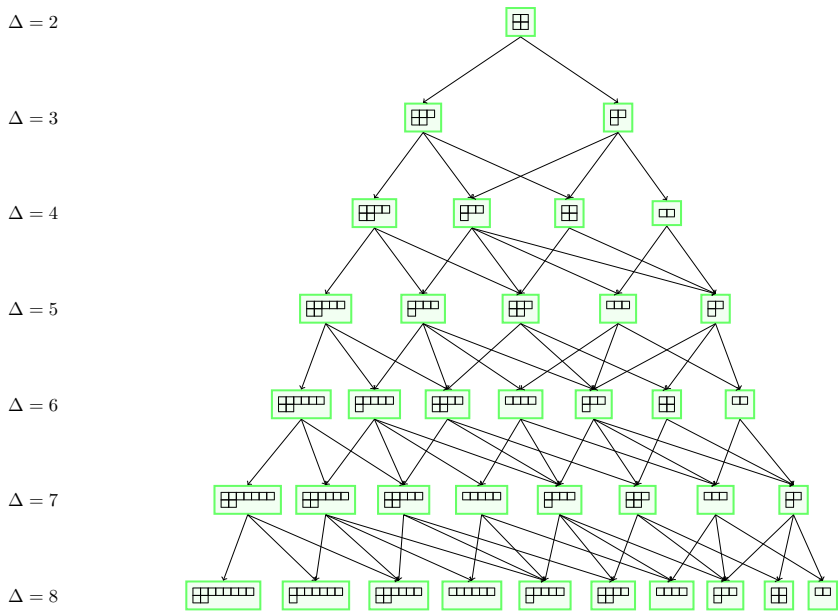
\begin{figure}[H]
	\centering
	\resizebox{110mm}{80mm}{%
		\begin{tikzpicture}[
			squarednode/.style={rectangle, draw=green!60, fill=green!5, very thick, minimum size=5mm},
			]
			\ytableausetup{boxsize = 4pt}
			\node(D2) at (-10,0){$\D=2$};
			\node(D3) at (-10,-2){$\D=3$};
			\node(D4) at (-10,-4){$\D=4$};
			\node(D5) at (-10,-6){$\D=5$};
			\node(D6) at (-10,-8){$\D=6$};
			\node(D7) at (-10,-10){$\D=7$};
			\node(D8) at (-10,-12){$\D=8$};
			\node[squarednode](222) at (0,0){$\ydiagram{2,2}$};
			\node[squarednode](332) at (-2,-2){$\ydiagram{3,2}$};
			\node[squarednode](321) at (2,-2){$\ydiagram{2,1}$};
			\node[squarednode](442) at (-3,-4){$\ydiagram{4,2}$};
			\node[squarednode](431) at (-1,-4){$\ydiagram{3,1}$};
			\node[squarednode](422) at (1,-4){$\ydiagram{2,2}$};
			\node[squarednode](420) at (3,-4){$\ydiagram{2}$};
			\node[squarednode](552) at (-4,-6){$\ydiagram{5,2}$};
			\node[squarednode](541) at (-2,-6){$\ydiagram{4,1}$};
			\node[squarednode](532) at (0,-6){$\ydiagram{3,2}$};
			\node[squarednode](530) at (2,-6){$\ydiagram{3,0}$};
			\node[squarednode](521) at (4,-6){$\ydiagram{2,1}$};
			\node[squarednode](662) at (-4.5,-8){$\ydiagram{6,2}$};
			\node[squarednode](651) at (-3,-8){$\ydiagram{5,1}$};
			\node[squarednode](642) at (-1.5,-8){$\ydiagram{4,2}$};
			\node[squarednode](640) at (0,-8){$\ydiagram{4,0}$};
			\node[squarednode](631) at (1.5,-8){$\ydiagram{3,1}$};
			\node[squarednode](622) at (3,-8){$\ydiagram{2,2}$};
			\node[squarednode](620) at (4.5,-8){$\ydiagram{2,0}$};
			\node[squarednode](772) at (-5.6,-10){$\ydiagram{7,2}$};
			\node[squarednode](761) at (-4.0,-10){$\ydiagram{6,2}$};
			\node[squarednode](752) at (-2.4,-10){$\ydiagram{5,2}$};
			\node[squarednode](750) at (-0.8,-10){$\ydiagram{5,0}$};
			\node[squarednode](741) at (0.8,-10){$\ydiagram{4,1}$};
			\node[squarednode](732) at (2.4,-10){$\ydiagram{3,2}$};
			\node[squarednode](730) at (4.0,-10){$\ydiagram{3,0}$};
			\node[squarednode](721) at (5.6,-10){$\ydiagram{2,1}$};
			\node[squarednode](882) at (-6.1,-12){$\ydiagram{8,2}$};
			\node[squarednode](871) at (-4.2,-12){$\ydiagram{7,1}$};
			\node[squarednode](862) at (-2.5,-12){$\ydiagram{6,2}$};
			\node[squarednode](860) at (-1.0,-12){$\ydiagram{6,0}$};
			\node[squarednode](851) at (0.5,-12){$\ydiagram{5,1}$};
			\node[squarednode](842) at (1.9,-12){$\ydiagram{4,2}$};
			\node[squarednode](840) at (3.1,-12){$\ydiagram{4,0}$};
			\node[squarednode](831) at (4.2,-12){$\ydiagram{3,1}$};
			\node[squarednode](822) at (5.3,-12){$\ydiagram{2,2}$};
			\node[squarednode](820) at (6.2,-12){$\ydiagram{2,0}$};
			
			\draw[->] (222.south) -- (332.north);
			\draw[->] (222.south) -- (321.north);
			\draw[->] (332.south) -- (442.north);
			\draw[->] (332.south) -- (431.north);
			\draw[->] (332.south) -- (422.north);
			\draw[->] (321.south) -- (431.north);
			\draw[->] (321.south) -- (422.north);
			\draw[->] (321.south) -- (420.north);
			\draw[->] (442.south) -- (552.north);
			\draw[->] (442.south) -- (541.north);
			\draw[->] (442.south) -- (532.north);
			\draw[->] (431.south) -- (541.north);
			\draw[->] (431.south) -- (532.north);
			\draw[->] (431.south) -- (530.north);
			\draw[->] (431.south) -- (521.north);
			\draw[->] (422.south) -- (532.north);
			\draw[->] (422.south) -- (521.north);
			\draw[->] (420.south) -- (530.north);
			\draw[->] (420.south) -- (521.north);
			\draw[->] (552.south) -- (662.north);
			\draw[->] (552.south) -- (651.north);
			\draw[->] (552.south) -- (642.north);
			\draw[->] (541.south) -- (651.north);
			\draw[->] (541.south) -- (642.north);
			\draw[->] (541.south) -- (640.north);
			\draw[->] (541.south) -- (631.north);
			\draw[->] (532.south) -- (642.north);
			\draw[->] (532.south) -- (631.north);
			\draw[->] (532.south) -- (622.north);
			\draw[->] (530.south) -- (640.north);
			\draw[->] (530.south) -- (631.north);
			\draw[->] (530.south) -- (620.north);
			\draw[->] (521.south) -- (631.north);
			\draw[->] (521.south) -- (622.north);
			\draw[->] (521.south) -- (620.north);
			\draw[->] (662.south) -- (772.north);
			\draw[->] (662.south) -- (761.north);
			\draw[->] (662.south) -- (752.north);
			\draw[->] (651.south) -- (761.north);
			\draw[->] (651.south) -- (752.north);
			\draw[->] (651.south) -- (750.north);
			\draw[->] (651.south) -- (741.north);
			\draw[->] (642.south) -- (752.north);
			\draw[->] (642.south) -- (741.north);
			\draw[->] (642.south) -- (732.north);
			\draw[->] (640.south) -- (750.north);
			\draw[->] (640.south) -- (741.north);
			\draw[->] (640.south) -- (730.north);
			\draw[->] (631.south) -- (741.north);
			\draw[->] (631.south) -- (732.north);
			\draw[->] (631.south) -- (730.north);
			\draw[->] (631.south) -- (721.north);
			\draw[->] (622.south) -- (732.north);
			\draw[->] (622.south) -- (721.north);
			\draw[->] (620.south) -- (730.north);
			\draw[->] (620.south) -- (721.north);
			\draw[->] (772.south) -- (882.north);
			\draw[->] (772.south) -- (871.north);
			\draw[->] (772.south) -- (862.north);
			\draw[->] (761.south) -- (871.north);
			\draw[->] (761.south) -- (862.north);
			\draw[->] (761.south) -- (860.north);
			\draw[->] (761.south) -- (851.north);
			\draw[->] (752.south) -- (862.north);
			\draw[->] (752.south) -- (851.north);
			\draw[->] (752.south) -- (842.north);
			\draw[->] (750.south) -- (860.north);
			\draw[->] (750.south) -- (851.north);
			\draw[->] (750.south) -- (840.north);
			\draw[->] (741.south) -- (851.north);
			\draw[->] (741.south) -- (842.north);
			\draw[->] (741.south) -- (840.north);
			\draw[->] (741.south) -- (831.north);
			\draw[->] (732.south) -- (842.north);
			\draw[->] (732.south) -- (831.north);
			\draw[->] (732.south) -- (822.north);
			\draw[->] (730.south) -- (840.north);
			\draw[->] (730.south) -- (831.north);
			\draw[->] (730.south) -- (820.north);
			\draw[->] (721.south) -- (831.north);
			\draw[->] (721.south) -- (822.north);
			\draw[->] (721.south) -- (820.north);	
	\end{tikzpicture}}
	\caption{Off-shell system}\label{fig: off-shell}
\end{figure}

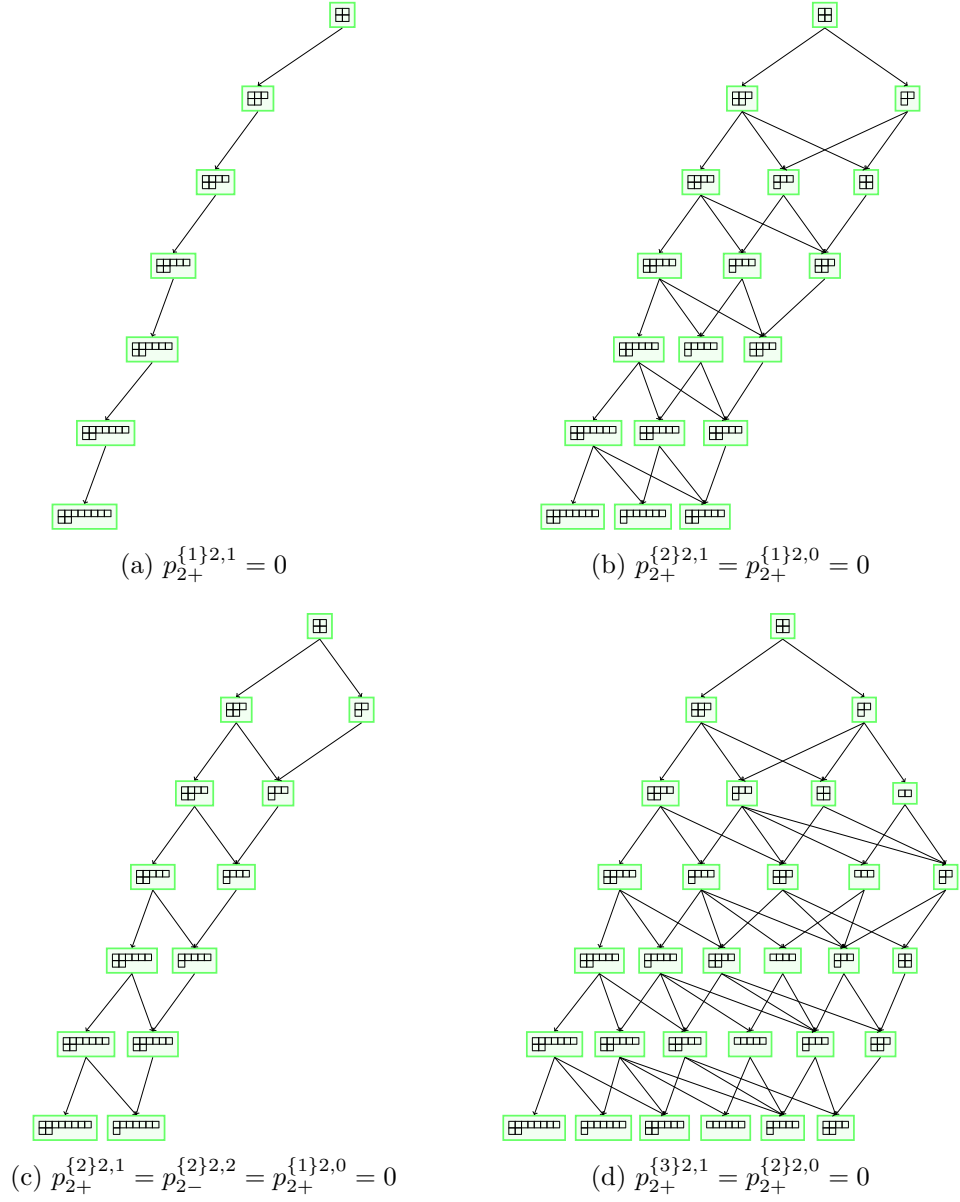
\begin{figure}[H]
	\centering
	\begin{subfigure}[b]{0.45\textwidth}
		\centering
		\resizebox{40mm}{70mm}{%
			\begin{tikzpicture}[
				squarednode/.style={rectangle, draw=green!60, fill=green!5, very thick, minimum size=5mm},
				]
				\ytableausetup{boxsize = 4pt}
				\node[squarednode](222) at (0,0){$\ydiagram{2,2}$};
				\node[squarednode](332) at (-2,-2){$\ydiagram{3,2}$};
				\node[squarednode](442) at (-3,-4){$\ydiagram{4,2}$};
				\node[squarednode](552) at (-4,-6){$\ydiagram{5,2}$};
				\node[squarednode](662) at (-4.5,-8){$\ydiagram{6,2}$};
				\node[squarednode](772) at (-5.6,-10){$\ydiagram{7,2}$};
				\node[squarednode](882) at (-6.1,-12){$\ydiagram{8,2}$};
				
				\draw[->] (222.south) -- (332.north);
				\draw[->] (332.south) -- (442.north);
				\draw[->] (442.south) -- (552.north);
				\draw[->] (552.south) -- (662.north);
				\draw[->] (662.south) -- (772.north);
				\draw[->] (772.south) -- (882.north);
		\end{tikzpicture}}
		\caption{$p^{\{1\}2,1}_{2+}=0$}\label{fig:d222zero}
	\end{subfigure}
	\begin{subfigure}[b]{0.45\textwidth}
		\centering
		\resizebox{50mm}{70mm}{%
			\begin{tikzpicture}[
				squarednode/.style={rectangle, draw=green!60, fill=green!5, very thick, minimum size=5mm},
				]
				\ytableausetup{boxsize = 4pt}
				\node[squarednode](222) at (0,0){$\ydiagram{2,2}$};
				\node[squarednode](332) at (-2,-2){$\ydiagram{3,2}$};
				\node[squarednode](321) at (2,-2){$\ydiagram{2,1}$};
				\node[squarednode](442) at (-3,-4){$\ydiagram{4,2}$};
				\node[squarednode](431) at (-1,-4){$\ydiagram{3,1}$};
				\node[squarednode](422) at (1,-4){$\ydiagram{2,2}$};
				\node[squarednode](552) at (-4,-6){$\ydiagram{5,2}$};
				\node[squarednode](541) at (-2,-6){$\ydiagram{4,1}$};
				\node[squarednode](532) at (0,-6){$\ydiagram{3,2}$};
				\node[squarednode](662) at (-4.5,-8){$\ydiagram{6,2}$};
				\node[squarednode](651) at (-3,-8){$\ydiagram{5,1}$};
				\node[squarednode](642) at (-1.5,-8){$\ydiagram{4,2}$};
				\node[squarednode](772) at (-5.6,-10){$\ydiagram{7,2}$};
				\node[squarednode](761) at (-4.0,-10){$\ydiagram{6,2}$};
				\node[squarednode](752) at (-2.4,-10){$\ydiagram{5,2}$};
				\node[squarednode](882) at (-6.1,-12){$\ydiagram{8,2}$};
				\node[squarednode](871) at (-4.4,-12){$\ydiagram{7,1}$};
				\node[squarednode](862) at (-2.9,-12){$\ydiagram{6,2}$};
				
				\draw[->] (222.south) -- (332.north);
				\draw[->] (222.south) -- (321.north);
				\draw[->] (332.south) -- (442.north);
				\draw[->] (332.south) -- (431.north);
				\draw[->] (332.south) -- (422.north);
				\draw[->] (321.south) -- (431.north);
				\draw[->] (321.south) -- (422.north);
				\draw[->] (442.south) -- (552.north);
				\draw[->] (442.south) -- (541.north);
				\draw[->] (442.south) -- (532.north);
				\draw[->] (431.south) -- (541.north);
				\draw[->] (431.south) -- (532.north);
				\draw[->] (422.south) -- (532.north);
				\draw[->] (552.south) -- (662.north);
				\draw[->] (552.south) -- (651.north);
				\draw[->] (552.south) -- (642.north);
				\draw[->] (541.south) -- (651.north);
				\draw[->] (541.south) -- (642.north);
				\draw[->] (532.south) -- (642.north);
				\draw[->] (662.south) -- (772.north);
				\draw[->] (662.south) -- (761.north);
				\draw[->] (662.south) -- (752.north);
				\draw[->] (651.south) -- (761.north);
				\draw[->] (651.south) -- (752.north);
				\draw[->] (642.south) -- (752.north);
				\draw[->] (772.south) -- (882.north);
				\draw[->] (772.south) -- (871.north);
				\draw[->] (772.south) -- (862.north);
				\draw[->] (761.south) -- (871.north);
				\draw[->] (761.south) -- (862.north);
				\draw[->] (752.south) -- (862.north);
		\end{tikzpicture}}
		\caption{$p^{\{2\}2,1}_{2+}=p^{\{1\}2,0}_{2+}=0$}\label{fig:d422zero}
	\end{subfigure}
	
\vspace{10pt}

\begin{subfigure}[b]{0.45\textwidth}
	\centering
	\resizebox{45mm}{70mm}{%
		\begin{tikzpicture}[
			squarednode/.style={rectangle, draw=green!60, fill=green!5, very thick, minimum size=5mm},
			]
			\ytableausetup{boxsize = 4pt}
			\node[squarednode](222) at (0,0){$\ydiagram{2,2}$};
			\node[squarednode](332) at (-2,-2){$\ydiagram{3,2}$};
			\node[squarednode](321) at (1,-2){$\ydiagram{2,1}$};
			\node[squarednode](442) at (-3,-4){$\ydiagram{4,2}$};
			\node[squarednode](431) at (-1,-4){$\ydiagram{3,1}$};
			\node[squarednode](552) at (-4,-6){$\ydiagram{5,2}$};
			\node[squarednode](541) at (-2,-6){$\ydiagram{4,1}$};
			\node[squarednode](662) at (-4.5,-8){$\ydiagram{6,2}$};
			\node[squarednode](651) at (-3,-8){$\ydiagram{5,1}$};
			\node[squarednode](772) at (-5.6,-10){$\ydiagram{7,2}$};
			\node[squarednode](761) at (-4.0,-10){$\ydiagram{6,2}$};
			\node[squarednode](882) at (-6.1,-12){$\ydiagram{8,2}$};
			\node[squarednode](871) at (-4.4,-12){$\ydiagram{7,1}$};
			
			\draw[->] (222.south) -- (332.north);
			\draw[->] (222.south) -- (321.north);
			\draw[->] (332.south) -- (442.north);
			\draw[->] (332.south) -- (431.north);
			\draw[->] (321.south) -- (431.north);
			\draw[->] (442.south) -- (552.north);
			\draw[->] (442.south) -- (541.north);
			\draw[->] (431.south) -- (541.north);
			\draw[->] (552.south) -- (662.north);
			\draw[->] (552.south) -- (651.north);
			\draw[->] (541.south) -- (651.north);
			\draw[->] (662.south) -- (772.north);
			\draw[->] (662.south) -- (761.north);
			\draw[->] (651.south) -- (761.north);
			\draw[->] (772.south) -- (882.north);
			\draw[->] (772.south) -- (871.north);
			\draw[->] (761.south) -- (871.north);
	\end{tikzpicture}}
	\caption{$p^{\{2\}2,1}_{2+}=p^{\{2\}2,2}_{2-}=p^{\{1\}2,0}_{2+}=0$}\label{fig: pm}
\end{subfigure}	
\begin{subfigure}[b]{0.45\textwidth}
		\centering
		\resizebox{60mm}{70mm}{%
			\begin{tikzpicture}[
				squarednode/.style={rectangle, draw=green!60, fill=green!5, very thick, minimum size=5mm},
				]
				\ytableausetup{boxsize = 4pt}
				\node[squarednode](222) at (0,0){$\ydiagram{2,2}$};
				\node[squarednode](332) at (-2,-2){$\ydiagram{3,2}$};
				\node[squarednode](321) at (2,-2){$\ydiagram{2,1}$};
				\node[squarednode](442) at (-3,-4){$\ydiagram{4,2}$};
				\node[squarednode](431) at (-1,-4){$\ydiagram{3,1}$};
				\node[squarednode](422) at (1,-4){$\ydiagram{2,2}$};
				\node[squarednode](420) at (3,-4){$\ydiagram{2}$};
				\node[squarednode](552) at (-4,-6){$\ydiagram{5,2}$};
				\node[squarednode](541) at (-2,-6){$\ydiagram{4,1}$};
				\node[squarednode](532) at (0,-6){$\ydiagram{3,2}$};
				\node[squarednode](530) at (2,-6){$\ydiagram{3,0}$};
				\node[squarednode](521) at (4,-6){$\ydiagram{2,1}$};
				\node[squarednode](662) at (-4.5,-8){$\ydiagram{6,2}$};
				\node[squarednode](651) at (-3,-8){$\ydiagram{5,1}$};
				\node[squarednode](642) at (-1.5,-8){$\ydiagram{4,2}$};
				\node[squarednode](640) at (0,-8){$\ydiagram{4,0}$};
				\node[squarednode](631) at (1.5,-8){$\ydiagram{3,1}$};
				\node[squarednode](622) at (3,-8){$\ydiagram{2,2}$};
				\node[squarednode](772) at (-5.6,-10){$\ydiagram{7,2}$};
				\node[squarednode](761) at (-4.0,-10){$\ydiagram{6,2}$};
				\node[squarednode](752) at (-2.4,-10){$\ydiagram{5,2}$};
				\node[squarednode](750) at (-0.8,-10){$\ydiagram{5,0}$};
				\node[squarednode](741) at (0.8,-10){$\ydiagram{4,1}$};
				\node[squarednode](732) at (2.4,-10){$\ydiagram{3,2}$};
				\node[squarednode](882) at (-6.1,-12){$\ydiagram{8,2}$};
				\node[squarednode](871) at (-4.4,-12){$\ydiagram{7,1}$};
				\node[squarednode](862) at (-2.9,-12){$\ydiagram{6,2}$};
				\node[squarednode](860) at (-1.4,-12){$\ydiagram{6,0}$};
				\node[squarednode](851) at (0,-12){$\ydiagram{5,1}$};
				\node[squarednode](842) at (1.3,-12){$\ydiagram{4,2}$};
				
				\draw[->] (222.south) -- (332.north);
				\draw[->] (222.south) -- (321.north);
				\draw[->] (332.south) -- (442.north);
				\draw[->] (332.south) -- (431.north);
				\draw[->] (332.south) -- (422.north);
				\draw[->] (321.south) -- (431.north);
				\draw[->] (321.south) -- (422.north);
				\draw[->] (321.south) -- (420.north);
				\draw[->] (442.south) -- (552.north);
				\draw[->] (442.south) -- (541.north);
				\draw[->] (442.south) -- (532.north);
				\draw[->] (431.south) -- (541.north);
				\draw[->] (431.south) -- (532.north);
				\draw[->] (431.south) -- (530.north);
				\draw[->] (431.south) -- (521.north);
				\draw[->] (422.south) -- (532.north);
				\draw[->] (422.south) -- (521.north);
				\draw[->] (420.south) -- (530.north);
				\draw[->] (420.south) -- (521.north);
				\draw[->] (552.south) -- (662.north);
				\draw[->] (552.south) -- (651.north);
				\draw[->] (552.south) -- (642.north);
				\draw[->] (541.south) -- (651.north);
				\draw[->] (541.south) -- (642.north);
				\draw[->] (541.south) -- (640.north);
				\draw[->] (541.south) -- (631.north);
				\draw[->] (532.south) -- (642.north);
				\draw[->] (532.south) -- (631.north);
				\draw[->] (532.south) -- (622.north);
				\draw[->] (530.south) -- (640.north);
				\draw[->] (530.south) -- (631.north);
				\draw[->] (521.south) -- (631.north);
				\draw[->] (521.south) -- (622.north);
				\draw[->] (662.south) -- (772.north);
				\draw[->] (662.south) -- (761.north);
				\draw[->] (662.south) -- (752.north);
				\draw[->] (651.south) -- (761.north);
				\draw[->] (651.south) -- (752.north);
				\draw[->] (651.south) -- (750.north);
				\draw[->] (651.south) -- (741.north);
				\draw[->] (642.south) -- (752.north);
				\draw[->] (642.south) -- (741.north);
				\draw[->] (642.south) -- (732.north);
				\draw[->] (640.south) -- (750.north);
				\draw[->] (640.south) -- (741.north);
				\draw[->] (631.south) -- (741.north);
				\draw[->] (631.south) -- (732.north);
				\draw[->] (622.south) -- (732.north);
				\draw[->] (772.south) -- (882.north);
				\draw[->] (772.south) -- (871.north);
				\draw[->] (772.south) -- (862.north);
				\draw[->] (761.south) -- (871.north);
				\draw[->] (761.south) -- (862.north);
				\draw[->] (761.south) -- (860.north);
				\draw[->] (761.south) -- (851.north);
				\draw[->] (752.south) -- (862.north);
				\draw[->] (752.south) -- (851.north);
				\draw[->] (752.south) -- (842.north);
				\draw[->] (750.south) -- (860.north);
				\draw[->] (750.south) -- (851.north);
				\draw[->] (741.south) -- (851.north);
				\draw[->] (741.south) -- (842.north);
				\draw[->] (732.south) -- (842.north);
		\end{tikzpicture}}
		\caption{$p^{\{3\}2,1}_{2+}=p^{\{2\}2,0}_{2+}=0$}\label{fig:d622zero}
	\end{subfigure}
	\caption{Various on-shell systems}
	\label{fig: hd FT}
\end{figure}

Let us begin with the off-shell system where none of $p^{\{\d\}2,2}_{2-}$, $p^{\{\d\}2,1}_{2+}$, $p^{\{\d\}2,1}_{2-}$ and $p^{\{\d\}2,0}_{2+}$
vanish: in Figure \ref{fig: off-shell}, 
the  coefficients $p^{[\D]m,n}_{1\pm}$ and $p^{[\D]m,n}_{2\pm}$ are depicted as
the lines which start from the Young diagrams $[\D,(m,n)]$ and ends at $[\D-1,(m\pm1,n)]$  and $[\D-1,(m,n\pm1)]$, respectively.
Let us examine the consequences of setting some of
$p^{\{\d\}2,2}_{2-}$, $p^{\{\d\}2,1}_{2+}$, $p^{\{\d\}2,1}_{2-}$ and $p^{\{\d\}2,0}_{2+}$
to zero.
At the lowest depth, we find that the possibility $p^{\{1\}2,1}_{2+}=0$ which gives massless system since the zero-form content depicted in Figure \ref{fig:d222zero} matches that of massless system.
Moving to the next depth $\d=2$, we can consider $p^{\{2\}2,1}_{2+}=0$
where the condition \eqref{pp const} reduces to $p^{\{2\}2,1}_{2-}\,p^{\{1\}2,0}_{2+}=0$\,.
If we take the possibility $p^{\{1\}2,0}_{2+}=0$, then 
the zero-form content in Figure \ref{fig:d422zero} matches 
that of the on-shell Fradkin-Tseyltin system (on-shell conformal spin two).
If we take the other possibility $p^{\{2\}2,1}_{2-}=0$, then we find yet another system with a scalar degrees of freedom.
We can impose also $p^{\{2\}2,2}_{2-}=0$ together with $p^{\{2\}2,1}_{2+}=p^{\{1\}2,0}_{2+}=0$,
and get a system  (see Figure \ref{fig: pm}) with one helicity-two and one helicity-one degrees of freedom, which match the degrees of freedom of partially-massless spin-two field.
Moving to the third depth, we find many possibilities, among which
the condition $p^{\{3\}2,1}_{2+}=p^{\{2\}2,0}_{2+}=0$ gives the spectra (see Figure \ref{fig:d622zero})
which match
the system given by  $\Box B_{\m\n}=0$, where $B_{\m\n}$ is the four-derivative linearized Bach tensor.

\section{Decomposition of the off-shell spin-2 Fradkin-Tseytlin module}
\label{sec: FT}

Let us decompose each Verma modules appearing in 
the off-shell FT module \eqref{FT module} into $\mathfrak{so}(2)\oplus \mathfrak{so}(d)$ irreps,
\be
	\cV(2+k,(2,2,1^k))
	=\bigoplus_{n,m=0}^\infty 
	[2+k+n+2m,(2,2,1^k)\otimes (n)]\,.
\ee
The tensor product of $\mathfrak{so}(d)$ irreps are given by
\be
	(2,2,1^k)\otimes (n)=V_{n,k}\oplus V_{n-1,k+1}\qquad [k\ge 0]\,,
\ee
where $V_{n,0}$ has three irreducible pieces,
\be
	V_{n,0}= 
	(n+2,2)\oplus \delta_{n\ge1}\, (n+1,1)\oplus \delta_{n\ge2}\,	(n)\,,
\ee
whereas $V_{n,k}$ with $k\ge 1$ is
\ba
	 V_{n,k}\eq 
	(n+2,2,1^k)\oplus (n+1,2,2,1^{k-1})\oplus 
	(n+1,2,1^{k-1})\oplus (n,2,2,1^{k-2})\nn
	&& \oplus \delta_{n\ge1}\,(n+1,1^{k+1})\oplus (n,2,1^k)\oplus
	\delta_{n\ge 2}\,(n,1^k)\oplus (n-1,2,1^{k-1}) \,.
\ea
Using these results, the Verma module can be expressed as  
\be
	\cV(2+k,(2,2,1^k))
	=\bigoplus_{n,m=0}^\infty 
	[2+k+n+2m,V_{n,k}]\oplus [2+k+n+2m,V_{n-1,k+1}]\,.
\ee
Finally, using the above result in \eqref{FT module}
and redefining the summation variable, we find
\ba
	&& \cD_{\rm off}(2,(2,2))= \nn
	&&= \bigoplus_{m=0}^\infty
	\bigoplus_{n=0}^\infty \bigoplus_{k=0}^{\infty} (-1)^k\,[2+k+n+2m,V_{n,k}]\oplus [2+k+n+2m,V_{n-1,k+1}]\nn
	&&= \bigoplus_{m=0}^\infty\left(\bigoplus_{n=0}^\infty \bigoplus_{k=0}^{\infty} (-1)^k\,[2+k+n+2m,V_{n,k}]
	\ominus \bigoplus_{n=0}^\infty \bigoplus_{k=1}^{\infty} (-1)^k\, [2+k+n+2m,V_{n,k}] \right)\nn
	&& =\bigoplus_{m=0}^\infty	\bigoplus_{n=0}^\infty [2+n+2m,V_{n,0}]
\ea
which can be simplified as 
\be
	\cD_{\rm off}(2,(2,2))=
	\bigoplus_{m,n=0}^\infty	
	[2+n+2m,(n+2,2)]\oplus [3+n+2m,(n+2,1)]
	\oplus [4+n+2m,(n+2)]\,.
\ee

\section{Number of ansatz for Weyl  invariants}
\label{sec: Char}

Since each zero-form fields correspond to a $\mathfrak{so}(2)\oplus \mathfrak{so}(d)$ modules
inside of the spin 2 FT module $\cS(0,(2))$,
the number of possible contractions of the zero-forms
can be obtained from multiple symmetrized tensor products of $\cS(0,(2))$.  
The latter can be conveniently handled in terms of Lie algebra character,
and all symmetrized tensor products are generated by
the plethystic exponential,
\be
	 PE\left[\chi\right](q,\bm x)\nn
	= \exp\left(\sum_{n=1}^\infty\frac1n\,\chi(q^n, \bm x^n)\right),
	\label{pl ex}
\ee
where $\bm x=(x_1,\ldots, x_{d/2})$ and $\bm x^n=(x_1^n,\ldots,x_{d/2}^n)$\,.
The above can be expanded as a series of $\mathfrak{k}$ characters,
and the number of all possible contractions of zero-forms,
which is a  Lorentz tensor $\mathbb Y$ of dimension $\D$,
is equal to the coefficient of the $\mathfrak{k}$ character corresponding
to the  irrep $[\D,\mathbb Y]$. 
Therefore,
the numbers $N$ and $M$ are the coefficients of the character 
$\chi^{\mathfrak{k}}_{[d,(0)]}(q,\bm x)=q^d$
and $\chi^{\mathfrak{k}}_{[d-1,(1)]}(q,\bm x)=q^{d-1}\,\chi^{\mathfrak{so}(d)}_{(1)}(\bm x)$
in the expansion of $PE[\chi_{\cS(0,(2)}](q,\bm x)$  \eqref{pl ex}.
Using the expression of the $\mathfrak{so}(d)$ character,
\be
	\chi^{\mathfrak{so}(d)}_{(\ell_1,\ldots, \ell_{d/2})}(\bm x)
	=\frac{\det\left(x_i^{\ell_j+\frac d2-j}+x_i^{-(\ell_j+\frac d2-j)}\right)+
	\det\left(x_i^{\ell_j+\frac d2-j}-x_i^{-(\ell_j+\frac d2-j)}\right)}
	{2\,\D(x_1+x_1^{-1},\ldots, x_{d/2}+x_{d/2}^{-1})}\,,
\ee
where $\D(\bm x)=\prod_{i<j} (x_i-x_j)$ is the Vandermonde determinant,
one can extract the number, namely the multiplicity, of the module  $[\D,(\ell_1,\ldots, \ell_{d/2})]$ as
the multiple integral,
\be
	\oint \frac{\dd q}{2\pi\,i\,q^{[\D+1]}}
	\left[\prod_{n=1}^{d/2} \oint \frac{\dd x_n}{2\pi\,i\,x_n^{\ell_n+\frac d2-n+1}}\right]
	(2-\delta_{\ell_{d/2},0})\,\D(x_1+x_1^{-1},\ldots, x_{d/2}+x_{d/2}^{-1})\,
	PE[\chi_{\cS(0,(2))}](q,\bm x)\,.
\ee
It is also useful to expand $PE[\chi](q,\bm x)$ as 
\be
	PE[\chi](g)=\sum_{n=0}^\infty \chi^{\odot n}(g)\,,
	\qquad
	\chi^{\odot n}(g)=\sum_{j_1+2\,j_2+\cdots +n\,j_n=n} \prod_{k=1}^n\frac{\chi(g^k)^{j_k}}{k^{j_k}\,j_k!}\,,
\ee
where we used the compact notation $g=(q,\bm x)$
and $g^n=(q^n,\bm x^n)$\,.
The $n$-th order part $\chi^{\odot n}(g)$ corresponds to symmetrized tensor product of 
the $n$ copies of the relevant representation. First few $\chi^{\odot n}$ reads
\ba
	&& \chi^{\odot 1}(g)=\chi(g)\,,\nn
	&& \chi^{\odot 2}(g)= \frac12\,\chi(g)^2+\frac12\,\chi(g^2)\,,\nn
	&&\chi^{\odot 3}(g)=\frac16\,\chi(g)^3+ \frac12\,\chi(g)\,\chi(g^2)+\frac13\,\chi(g^3)\,,\\
	&&\chi^{\odot 4}(g)=\frac1{24}\,\chi(g)^4 +\frac14\,\chi(g)^2\,\chi(g^2)
	  +\frac1{8}\,\chi(g^2)^2+\frac13\,\chi(g)\,\chi(g^3)
	  +\frac1{4}\,\chi(g^4)\,.\nonumber
	  \label{Z_n}
\ea
For our purpose --- computing the numbers of the dimension $d$ scalars
and the dimension $d-1$ vectors ---
it  is sufficient to consider $n=2,\ldots, d/2$ since $C$ has minimum conformal dimension $2$.
The last piece $\chi^{\odot  \frac d2}_{\cS(0,(2))}$ corresponds to the contractions of $d/2$ copies of $C^{[2]a(2),b(2)}$,
which are trivially Weyl invariant.
Therefore, we consider only $\chi^{\odot  2}_{\cS(0,(2))},\ldots, \chi^{\odot (\frac d2-1)}_{\cS(0,(2))}$\,,
and the expressions \eqref{Z_n} are sufficient up to $d=10$. 

\paragraph{Eight dimensions}
In $d=8$, the relevant $\chi^{\odot n}_{\cS(0,(2))}$'s are the $\chi^{\odot 2}_{\cS(0,(2))}$ and $\chi^{\odot 3}_{\cS(0,(2))}$.
From 
\ba
\chi^{\odot 2}_{\cS(0,(2))}\eq q^7\,\Big[\chi_{(2,2)}\left(\chi_{(3,2)}+\chi_{(2,1)}\right)+
\chi_{(3,2)}(\chi_{(4,2)}+\chi_{(3,1)}+\chi_{(2,2)})\nn
&&
\qquad +\,\chi_{(2,1)}\left(\chi_{(3,1)}+\chi_{(2,0)}+\chi_{(2,2)} \right)\Big]\nn
&&+\,q^8\,\Big[\chi_{(2,2)} ^2+\chi_{(3,2)} ^2+\chi_{(2,1)} ^2 +\,\chi_{(4,2)} ^{\odot 2}+\chi_{(3,1)} ^{\odot 2}+\chi_{(2,0)} ^{\odot 2}+\chi_{(2,2)} ^{\odot 2} \Big]\nn
&&+\,(\textrm{irrelevant terms})\,,
\ea
we find $M_2=8$ and $N_2=7$, which simply coincide to the numbers of terms. From 
\ba
\chi^{\odot 3}_{\cS(0,(2))} \eq q^7\bigg[
\left(\chi_{(3,2)} +\chi_{(2,1)} \right)\chi_{(2,2)} ^{\odot 2}\bigg]\nn
&&+q^8\bigg[
\left(\chi_{(4,2)} +\chi_{(3,1)} 
+\chi_{(2,0)} +\chi_{(2,2)} \right)\chi_{(2,2)} ^{\odot 2}\nn 
&&\qquad\quad +\,\chi_{(2,2)} \left( \c_{(3,2)}\c_{(2,1)} + \c_{(3,2)}^{\odot 2} + \c_{(2,1)}^{\odot 2} \right)
\bigg]\nn
&&+\,(\textrm{irrelevant terms})\,,
\ea
we find $M_3=7$ and $N_3=11$ from the integral.

\paragraph{Ten dimensions}

In $d=10$, the relevant $\chi^{\odot n}_{\cS(0,(2))}$'s are the $\chi^{\odot 2}_{\cS(0,(2))}$, $\chi^{\odot 3}_{\cS(0,(2))}$ and $\chi^{\odot 4}_{\cS(0,(2))}$.
From 
\ba
\chi^{\odot 2}_{\cS(0,(2))}\eq q^9\,\Big[ \c_{(2,2)} \left(\c_{(3,2)} +\c_{(2,1)}   \right)   +\c_{(3,2)} \left(\c_{(4,2)} +\c_{(2,2)}  + \c_{(3,1)}   \right) \nn 
&&\qquad+ \c_{(2,1)} \left(\c_{(3,1)} +\c_{(2,2)} +\c_{(2,0)}   \right)+\c_{(4,2)} \left(\c_{(5,2)} +\c_{(3,2)}  +\c_{(4,1)}  \right) \nn 
&& \qquad + \c_{(2,2)} \left(\c_{(3,2)} +\c_{(2,1)}   \right) + \c_{(3,1)} \left(\c_{(3,2)} +\c_{(4,1)}  +\c_{(3,0)} +\c_{(2,1)}  \right) \nn 
&&\qquad  + \c_{(2,0)} \left(\c_{(3,0)} +\c_{(2,1)}   \right) \Big]\nn
&&+\,q^{10}\,\Big[ \c_{(2,2)} ^2+\c_{(3,2)} ^2+\c_{(2,1)} ^2+\c_{(4,2)} ^2+\c_{(2,2)} ^2+\c_{(3,1)} ^2+\c_{(2,0)} ^2 \nn 
&& \qquad \quad +  \c_{(5,2)} ^{\odot 2}+\c_{(3,2)} ^{\odot 2} +\c_{(4,1)} ^{\odot 2} +\c_{(2,1)} ^{\odot 2} +\c_{(3,0)} ^{\odot 2}    \Big]\nn
&&+\,(\textrm{irrelevant terms})\,,
\ea
we find $M_2=19$ and $N_2= 12$.
From
\ba
\chi^{\odot 3}_{\cS(0,(2))}\eq q^9\,\Big[ \left(\c_{(5,2)} +\c_{(3,2)} +\c_{(4,1)} +\c_{(2,1)} +\c_{(3,0)}  \right)\, \c_{(2,2)}^{\odot 2}  \nn 
&& \qquad +\, \c_{(3,2)} ^{\odot 3} +  \c_{(3,2)} \,\c_{(2,1)} ^{\odot 2}  +\c_{(2,1)} ^{\odot 3} + \,\c_{(2,1)} \,\c_{(3,2)} ^{\odot 2}  \nn 
&&  \qquad +\c_{(2,2)} \left( \c_{(3,2)}  + \c_{(2,1)}  \right) \left( \c_{(4,2)} +\c_{(2,2)} +\c_{(3,1)} +\c_{(2,0)}  \right) \Big]\nn
&&+\,q^{10}\,\Big[\left( \c_{(6,2)} +\c_{(4,2)} +\c_{(5,1)} +\c_{(3,1)} \right)\,\c_{(2,2)} ^{\odot 2} \nn
&& \qquad \quad +\left( \c_{(4,0)} + \c_{(2,2)}  +\c_{(2,0)}  \right)\,\c_{(2,2)} ^{\odot 2}+ \c_{(2,2)} \,\c_{(4,2)} ^{\odot 2} \nn
&& \qquad \quad  +  \c_{(2,2)}  \left(    \c_{(3,1)} ^{\odot 2} +\c_{(2,2)} ^{\odot 2}+  \c_{(2,0)} ^{\odot 2}   \right) +\left( \c_{(4,2)} +\c_{(2,2)} +\c_{(3,1)} +\c_{(2,0)} \right) \, \c_{(3,2)} ^{\odot 2} \nn
&& \qquad \quad  +  \left( \c_{(4,2)} +\c_{(2,2)} +\c_{(3,1)} +\c_{(2,0)} \right)\, \c_{(2,1)} ^{\odot 2} \nn
&& \qquad + \c_{(3,2)} \, \c_{(2,1)} \left( \c_{(4,2)} +\c_{(2,2)} +\c_{(3,1)} +\c_{(2,0)}  \right) + \c_{(2,2)} \,\c_{(3,1)} \, \c_{(2,0)}  \nn
&& \qquad + \c_{(2,2)} \,\c_{(4,2)} \left( \c_{(2,2)} + \c_{(3,1)} +\c_{(2,0)}  \right) + \c_{(2,2)} ^2 \left( \c_{(3,1)}  + \c_{(2,0)}  \right) \nn
&& \qquad + \c_{(2,2)} \left( \c_{(3,2)} +\c_{(2,1)} \right) \left( \c_{(5,2)} +\c_{(3,2)} +\c_{(4,1)} +\c_{(2,1)} +\c_{(3,0)}  \right)  \Big]\nn
&&+\,(\textrm{irrelevant terms})\,,
\ea
we find $M_3=85$ and $N_3=62$.
From 
\ba
\chi^{\odot 4}_{\cS(0,(2))}\eq q^9\,\Big[ \left(\c_{(3,2)} +\c_{(2,1)}  \right) \,\c_{(2,2)} ^{\odot 3} \Big]\nn
&&+\,q^{10}\,\Big[ \left( \c_{(4,2)}  + \c_{(2,2)} +\c_{(3,1)}   + \c_{(2,0)}   \right) \,\c_{(2,2)} ^{\odot 3}  \nn
&&\qquad + \left(  \c_{(3,2)} ^{\odot 2}   +  \c_{(2,1)} ^{\odot 2}  \right)\,\c_{(2,2)} ^{\odot 2}  + \c_{(3,2)} \, \c_{(2,1)} \,\c_{(2,2)} ^{\odot 2}   \Big]\nn
&&+\,(\textrm{irrelevant terms})\,,
\ea
we find $M_4=46$ and $N_4=83$.

\bibliographystyle{JHEP}
\bibliography{biblio}

\providecommand{\href}[2]{#2}\begingroup\raggedright\begin{thebibliography}{10}

\bibitem{Fefferman:2007rka}
C.~Fefferman and C.~R. Graham, \emph{{The ambient metric}}, {\emph{Ann. Math.
  Stud.} {\bf 178} (2011) 1--128}, [\href{http://arxiv.org/abs/0710.0919}{{\tt
  0710.0919}}].

\bibitem{Henningson:1998gx}
M.~Henningson and K.~Skenderis, \emph{{The Holographic Weyl anomaly}},
  \href{http://dx.doi.org/10.1088/1126-6708/1998/07/023}{\emph{JHEP} {\bf 07}
  (1998) 023}, [\href{http://arxiv.org/abs/hep-th/9806087}{{\tt
  hep-th/9806087}}].

\bibitem{Wheeler:2013ora}
J.~T. Wheeler, \emph{{Weyl gravity as general relativity}},
  \href{http://dx.doi.org/10.1103/PhysRevD.90.025027}{\emph{Phys. Rev. D} {\bf
  90} (2014) 025027}, [\href{http://arxiv.org/abs/1310.0526}{{\tt 1310.0526}}].

\bibitem{Scholz:2017pfo}
E.~Scholz, \emph{{The unexpected resurgence of Weyl geometry in late 20-th
  century physics}},
  \href{http://dx.doi.org/10.1007/978-1-4939-7708-6_11}{\emph{Einstein Stud.}
  {\bf 14} (2018) 261--360}, [\href{http://arxiv.org/abs/1703.03187}{{\tt
  1703.03187}}].

\bibitem{Hobson:2021vzg}
M.~P. Hobson and A.~N. Lasenby, \emph{{Conformal gravity does not predict flat
  galaxy rotation curves}},  \href{http://arxiv.org/abs/2103.13451}{{\tt
  2103.13451}}.

\bibitem{Kehagias:2021smx}
A.~Kehagias and A.~Riotto, \emph{{Topological Early Universe Cosmology}},
  \href{http://arxiv.org/abs/2105.10669}{{\tt 2105.10669}}.

\bibitem{Duff:1993wm}
M.~J. Duff, \emph{{Twenty years of the Weyl anomaly}},
  \href{http://dx.doi.org/10.1088/0264-9381/11/6/004}{\emph{Class. Quant.
  Grav.} {\bf 11} (1994) 1387--1404},
  [\href{http://arxiv.org/abs/hep-th/9308075}{{\tt hep-th/9308075}}].

\bibitem{Bonora:1983ff}
L.~Bonora, P.~Cotta-Ramusino and C.~Reina, \emph{{Conformal Anomaly and
  Cohomology}},
  \href{http://dx.doi.org/10.1016/0370-2693(83)90169-7}{\emph{Phys. Lett. B}
  {\bf 126} (1983) 305--308}.

\bibitem{Bonora:1985cq}
L.~Bonora, P.~Pasti and M.~Bregola, \emph{{WEYL COCYCLES}},
  \href{http://dx.doi.org/10.1088/0264-9381/3/4/018}{\emph{Class. Quant. Grav.}
  {\bf 3} (1986) 635}.

\bibitem{Deser:1993yx}
S.~Deser and A.~Schwimmer, \emph{{Geometric classification of conformal
  anomalies in arbitrary dimensions}},
  \href{http://dx.doi.org/10.1016/0370-2693(93)90934-A}{\emph{Phys. Lett. B}
  {\bf 309} (1993) 279--284}, [\href{http://arxiv.org/abs/hep-th/9302047}{{\tt
  hep-th/9302047}}].

\bibitem{Boulanger:2007ab}
N.~Boulanger, \emph{{Algebraic Classification of Weyl Anomalies in Arbitrary
  Dimensions}},
  \href{http://dx.doi.org/10.1103/PhysRevLett.98.261302}{\emph{Phys. Rev.
  Lett.} {\bf 98} (2007) 261302}, [\href{http://arxiv.org/abs/0706.0340}{{\tt
  0706.0340}}].

\bibitem{Boulanger:2007st}
N.~Boulanger, \emph{{General solutions of the Wess-Zumino consistency condition
  for the Weyl anomalies}},
  \href{http://dx.doi.org/10.1088/1126-6708/2007/07/069}{\emph{JHEP} {\bf 07}
  (2007) 069}, [\href{http://arxiv.org/abs/0704.2472}{{\tt 0704.2472}}].

\bibitem{Parker:1987}
T.~Parker and S.~Rosenberg, \emph{{Invariants of conformal Laplacians}},
  \href{http://dx.doi.org/10.4310/jdg/1214440850}{\emph{Journal of Differential
  Geometry} {\bf 25} (1987) 199 -- 222}.

\bibitem{Boulanger:2004zf}
N.~Boulanger and J.~Erdmenger, \emph{{A Classification of local Weyl invariants
  in D=8}}, \href{http://dx.doi.org/10.1088/0264-9381/21/18/003}{\emph{Class.
  Quant. Grav.} {\bf 21} (2004) 4305--4316},
  [\href{http://arxiv.org/abs/hep-th/0405228}{{\tt hep-th/0405228}}].

\bibitem{Boulanger:2004eh}
N.~Boulanger, \emph{{A Weyl-covariant tensor calculus}},
  \href{http://dx.doi.org/10.1063/1.1896381}{\emph{J. Math. Phys.} {\bf 46}
  (2005) 053508}, [\href{http://arxiv.org/abs/hep-th/0412314}{{\tt
  hep-th/0412314}}].

\bibitem{Francois:2015pga}
J.~Francois, S.~Lazzarini and T.~Masson, \emph{Becchi-rouet-stora-tyutin
  structure for the mixed weyl-diffeomorphism residual symmetry},
  \href{http://dx.doi.org/10.1063/1.4943595}{\emph{J. Math. Phys.} {\bf 57}
  (2016) 033504}, [\href{http://arxiv.org/abs/1508.07666}{{\tt 1508.07666}}].

\bibitem{Curry:2014yoa}
S.~Curry and A.~R. Gover, \emph{{An introduction to conformal geometry and
  tractor calculus, with a view to applications in general relativity}},
  \href{http://arxiv.org/abs/1412.7559}{{\tt 1412.7559}}.

\bibitem{Erdmenger:1997gy}
J.~Erdmenger, \emph{{Conformally covariant differential operators: Properties
  and applications}},
  \href{http://dx.doi.org/10.1088/0264-9381/14/8/008}{\emph{Class. Quant.
  Grav.} {\bf 14} (1997) 2061--2084},
  [\href{http://arxiv.org/abs/hep-th/9704108}{{\tt hep-th/9704108}}].

\bibitem{sharpe}
R.~Sharpe, \emph{Differential Geometry: Cartain's generalization of Klein's
  Erlangen program}, vol.~166 of \emph{Graduate Texts in Mathematics}.
\newblock Springer-Verlag, 1997.

\bibitem{MacDowell:1977jt}
S.~W. MacDowell and F.~Mansouri, \emph{{Unified Geometric Theory of Gravity and
  Supergravity}},
  \href{http://dx.doi.org/10.1103/PhysRevLett.38.739}{\emph{Phys. Rev. Lett.}
  {\bf 38} (1977) 739}.

\bibitem{Stelle:1979aj}
K.~S. Stelle and P.~C. West, \emph{{Spontaneously Broken De Sitter Symmetry and
  the Gravitational Holonomy Group}},
  \href{http://dx.doi.org/10.1103/PhysRevD.21.1466}{\emph{Phys. Rev. D} {\bf
  21} (1980) 1466}.

\bibitem{CrispimRomao:1977hj}
J.~Crispim~Romao, A.~Ferber and P.~G.~O. Freund, \emph{{Unified Superconformal
  Gauge Theories}},
  \href{http://dx.doi.org/10.1016/0550-3213(77)90288-7}{\emph{Nucl. Phys. B}
  {\bf 126} (1977) 429--435}.

\bibitem{Kaku:1977pa}
M.~Kaku, P.~K. Townsend and P.~van Nieuwenhuizen, \emph{{Gauge Theory of the
  Conformal and Superconformal Group}},
  \href{http://dx.doi.org/10.1016/0370-2693(77)90552-4}{\emph{Phys. Lett. B}
  {\bf 69} (1977) 304--308}.

\bibitem{Crispim-Romao:1978zlo}
J.~Crispim-Romao, \emph{{CONFORMAL AND SUPERCONFORMAL GRAVITY AND NONLINEAR
  REPRESENTATIONS}},
  \href{http://dx.doi.org/10.1016/0550-3213(78)90099-8}{\emph{Nucl. Phys. B}
  {\bf 145} (1978) 535--546}.

\bibitem{Fradkin:1985am}
E.~S. Fradkin and A.~A. Tseytlin, \emph{{CONFORMAL SUPERGRAVITY}},
  \href{http://dx.doi.org/10.1016/0370-1573(85)90138-3}{\emph{Phys. Rept.} {\bf
  119} (1985) 233--362}.

\bibitem{Horne:1988jf}
J.~H. Horne and E.~Witten, \emph{{Conformal Gravity in Three-dimensions as a
  Gauge Theory}},
  \href{http://dx.doi.org/10.1103/PhysRevLett.62.501}{\emph{Phys. Rev. Lett.}
  {\bf 62} (1989) 501--504}.

\bibitem{Preitschopf:1998ei}
C.~R. Preitschopf and M.~A. Vasiliev, \emph{{Conformal field theory in
  conformal space}},
  \href{http://dx.doi.org/10.1016/S0550-3213(99)00087-5}{\emph{Nucl. Phys. B}
  {\bf 549} (1999) 450--480}, [\href{http://arxiv.org/abs/hep-th/9812113}{{\tt
  hep-th/9812113}}].

\bibitem{Aros:2013yaa}
R.~Aros and D.~E. Diaz, \emph{{AdS Chern-Simons Gravity induces Conformal
  Gravity}}, \href{http://dx.doi.org/10.1103/PhysRevD.89.084026}{\emph{Phys.
  Rev. D} {\bf 89} (2014) 084026}, [\href{http://arxiv.org/abs/1311.5364}{{\tt
  1311.5364}}].

\bibitem{Vasiliev:1986td}
M.~A. Vasiliev, \emph{{Free Massless Fields of Arbitrary Spin in the De Sitter
  Space and Initial Data for a Higher Spin Superalgebra}}, {\emph{Fortsch.
  Phys.} {\bf 35} (1987) 741--770}.

\bibitem{Vasiliev:1987hv}
M.~A. Vasiliev, \emph{{Linearized Curvatures for Auxiliary Fields in the De
  Sitter Space}},
  \href{http://dx.doi.org/10.1016/0550-3213(88)90325-2}{\emph{Nucl. Phys. B}
  {\bf 307} (1988) 319}.

\bibitem{Lopatin:1987hz}
V.~E. Lopatin and M.~A. Vasiliev, \emph{{Free Massless Bosonic Fields of
  Arbitrary Spin in $d$-dimensional De Sitter Space}},
  \href{http://dx.doi.org/10.1142/S0217732388000313}{\emph{Mod. Phys. Lett. A}
  {\bf 3} (1988) 257}.

\bibitem{Vasiliev:1988xc}
M.~A. Vasiliev, \emph{{Equations of Motion of Interacting Massless Fields of
  All Spins as a Free Differential Algebra}},
  \href{http://dx.doi.org/10.1016/0370-2693(88)91179-3}{\emph{Phys. Lett. B}
  {\bf 209} (1988) 491--497}.

\bibitem{Vasiliev:1988sa}
M.~A. Vasiliev, \emph{{Consistent Equations for Interacting Massless Fields of
  All Spins in the First Order in Curvatures}},
  \href{http://dx.doi.org/10.1016/0003-4916(89)90261-3}{\emph{Annals Phys.}
  {\bf 190} (1989) 59--106}.

\bibitem{DAuria:1982mkx}
R.~D'Auria, P.~Fre, P.~K. Townsend and P.~van Nieuwenhuizen, \emph{{Invariance
  of Actions, Rheonomy and the New Minimal $N=1$ Supergravity in the Group
  Manifold Approach}},
  \href{http://dx.doi.org/10.1016/0003-4916(84)90007-1}{\emph{Annals Phys.}
  {\bf 155} (1984) 423}.

\bibitem{Vasiliev:1990en}
M.~A. Vasiliev, \emph{{Consistent equation for interacting gauge fields of all
  spins in (3+1)-dimensions}},
  \href{http://dx.doi.org/10.1016/0370-2693(90)91400-6}{\emph{Phys. Lett. B}
  {\bf 243} (1990) 378--382}.

\bibitem{Vasiliev:2003ev}
M.~A. Vasiliev, \emph{{Nonlinear equations for symmetric massless higher spin
  fields in (A)dS(d)}},
  \href{http://dx.doi.org/10.1016/S0370-2693(03)00872-4}{\emph{Phys. Lett.}
  {\bf B567} (2003) 139--151}, [\href{http://arxiv.org/abs/hep-th/0304049}{{\tt
  hep-th/0304049}}].

\bibitem{Shaynkman:2004vu}
O.~V. Shaynkman, I.~Y. Tipunin and M.~A. Vasiliev, \emph{{Unfolded form of
  conformal equations in M dimensions and o(M + 2) modules}},
  \href{http://dx.doi.org/10.1142/S0129055X06002814}{\emph{Rev. Math. Phys.}
  {\bf 18} (2006) 823--886}, [\href{http://arxiv.org/abs/hep-th/0401086}{{\tt
  hep-th/0401086}}].

\bibitem{Vasiliev:2009ck}
M.~A. Vasiliev, \emph{{Bosonic conformal higher-spin fields of any symmetry}},
  \href{http://dx.doi.org/10.1016/j.nuclphysb.2009.12.010}{\emph{Nucl. Phys. B}
  {\bf 829} (2010) 176--224}, [\href{http://arxiv.org/abs/0909.5226}{{\tt
  0909.5226}}].

\bibitem{BGG}
J.~E. Humphreys, \emph{Representations of Semisimple Lie Algebras in the BGG
  Category O}, vol.~94 of \emph{Graduate Studies in Mathematics}.
\newblock American Mathematical Society, 2008.

\bibitem{Bekaert:2017bpy}
X.~Bekaert, M.~Grigoriev and E.~D. Skvortsov, \emph{{Higher Spin Extension of
  Fefferman-Graham Construction}},
  \href{http://dx.doi.org/10.3390/universe4020017}{\emph{Universe} {\bf 4}
  (2018) 17}, [\href{http://arxiv.org/abs/1710.11463}{{\tt 1710.11463}}].

\bibitem{Beccaria:2014jxa}
M.~Beccaria, X.~Bekaert and A.~A. Tseytlin, \emph{{Partition function of free
  conformal higher spin theory}},
  \href{http://dx.doi.org/10.1007/JHEP08(2014)113}{\emph{JHEP} {\bf 08} (2014)
  113}, [\href{http://arxiv.org/abs/1406.3542}{{\tt 1406.3542}}].

\bibitem{Basile:2018eac}
T.~Basile, X.~Bekaert and E.~Joung, \emph{{Conformal Higher-Spin Gravity:
  Linearized Spectrum = Symmetry Algebra}},
  \href{http://dx.doi.org/10.1007/JHEP11(2018)167}{\emph{JHEP} {\bf 11} (2018)
  167}, [\href{http://arxiv.org/abs/1808.07728}{{\tt 1808.07728}}].

\bibitem{Misuna:2019ijn}
N.~Misuna, \emph{{On unfolded off-shell formulation for higher-spin theory}},
  \href{http://dx.doi.org/10.1016/j.physletb.2019.134956}{\emph{Phys. Lett. B}
  {\bf 798} (2019) 134956}, [\href{http://arxiv.org/abs/1905.06925}{{\tt
  1905.06925}}].

\bibitem{Misuna:2020fck}
N.~G. Misuna, \emph{{Off-shell higher-spin fields in $AdS_{4}$ and external
  currents}},  \href{http://arxiv.org/abs/2012.06570}{{\tt 2012.06570}}.

\bibitem{Skvortsov:2006at}
E.~D. Skvortsov and M.~A. Vasiliev, \emph{{Geometric formulation for partially
  massless fields}},
  \href{http://dx.doi.org/10.1016/j.nuclphysb.2006.06.019}{\emph{Nucl. Phys. B}
  {\bf 756} (2006) 117--147}, [\href{http://arxiv.org/abs/hep-th/0601095}{{\tt
  hep-th/0601095}}].

\bibitem{Ponomarev:2010st}
D.~S. Ponomarev and M.~A. Vasiliev, \emph{{Frame-Like Action and Unfolded
  Formulation for Massive Higher-Spin Fields}},
  \href{http://dx.doi.org/10.1016/j.nuclphysb.2010.06.007}{\emph{Nucl. Phys. B}
  {\bf 839} (2010) 466--498}, [\href{http://arxiv.org/abs/1001.0062}{{\tt
  1001.0062}}].

\bibitem{Kuzenko:2019ill}
S.~M. Kuzenko and M.~Ponds, \emph{{Conformal geometry and (super)conformal
  higher-spin gauge theories}},
  \href{http://dx.doi.org/10.1007/JHEP05(2019)113}{\emph{JHEP} {\bf 05} (2019)
  113}, [\href{http://arxiv.org/abs/1902.08010}{{\tt 1902.08010}}].

\bibitem{Boulanger:2008up}
N.~Boulanger, C.~Iazeolla and P.~Sundell, \emph{{Unfolding Mixed-Symmetry
  Fields in AdS and the BMV Conjecture: I. General Formalism}},
  \href{http://dx.doi.org/10.1088/1126-6708/2009/07/013}{\emph{JHEP} {\bf 07}
  (2009) 013}, [\href{http://arxiv.org/abs/0812.3615}{{\tt 0812.3615}}].

\bibitem{Boulanger:2008kw}
N.~Boulanger, C.~Iazeolla and P.~Sundell, \emph{{Unfolding Mixed-Symmetry
  Fields in AdS and the BMV Conjecture. II. Oscillator Realization}},
  \href{http://dx.doi.org/10.1088/1126-6708/2009/07/014}{\emph{JHEP} {\bf 07}
  (2009) 014}, [\href{http://arxiv.org/abs/0812.4438}{{\tt 0812.4438}}].

\bibitem{Basile:2020gqi}
T.~Basile, E.~Joung, K.~Mkrtchyan and M.~Mojaza, \emph{{Dual Pair
  Correspondence in Physics: Oscillator Realizations and Representations}},
  \href{http://arxiv.org/abs/2006.07102}{{\tt 2006.07102}}.

\end{thebibliography}\endgroup

\end{document}